\documentclass[a4paper,titlepage,runningheads]{llncs}
\usepackage{hyperref}
\hypersetup{%
    hidelinks,
    pdftitle={SusTrainable: Promoting Sustainability as a Fundamental Driver in Software Development Training and Education},
    pdfauthor={Tihana Galinac Grbac, Csaba Szab\'o and João Paulo Fernandes (Eds.)},
    pdfsubject={2nd Teacher Training Proceedings},
}

\usepackage{url}
\usepackage{graphicx}

\usepackage{amsmath,amsfonts,xcolor}
\usepackage{xurl}

\newcommand{\diag}{\mathrm{diag}}

\usepackage{graphicx}
\usepackage{tikz}
\usepackage{pgfplots}
\usetikzlibrary{calc, shapes, arrows, decorations.markings}%
\usetikzlibrary{arrows,automata,calc,shapes, positioning, patterns, matrix}%
\usepackage{mathtools}
\usepackage{amssymb}
\usepackage{enumitem}
\usepackage{listings}
\usepackage{filecontents}
\usepackage{theorem}
\usepackage{comment}
\usepackage{subcaption}
\usepackage{graphicx}
\usepackage{comment}
\usepackage{amsmath,amsfonts,amssymb}
\usepackage{bbding}
\usepackage{caption}
\usepackage{graphicx}
\usepackage{listings}
\usepackage[T1]{fontenc}
\usepackage{varwidth}
\usepackage{xpatch}

\newcommand{\ignore}[1]{}

\usepackage{graphicx}
\usepackage{booktabs} 
\usepackage{stmaryrd} 
\usepackage{listings}  {} 
\lstdefinelanguage{Clean}{
	alsoletter={ABCDEFGHIJKLMNOPQRSTUVWXYZabcdefghijklmnopqrstuvwxyz_`1234567890},
	alsoletter={~!@\#\$\%^\&*-+=?<>:|\\},
	alsoother={.},
	alsodigit={1234567890},
	morekeywords={generic,implementation,definition,dynamic,module,import,where,in,of,case,let,infix,infixr,infixl,class,instance,with,if,derive},
	sensitive=true,
	morecomment=[l]{//},
	morecomment=[n]{/*}{*/},
	morestring=[b]", 
	morestring=[b]',
	emptylines=1,
	literate=   %
		{\\}{{$\lambda\:$}}1
		{A.}{{$\forall\;\,$}}2
		{E.}{{$\exists\;\,$}}2
		{<=}{{$\leq$}}1
		{<=.}{{\texttt{<=.}}}3
		{>=}{{$\geq$}}1
		{<>}{{$\neq$}}1
		{->}{{$\shortrightarrow$}}2
		{<-}{{$\shortleftarrow$}}1
		{=}{{$=$}}1
		{~}{{$\sim$}}1
		{\{|}{{$\{\!|$}}2
		{|\}}{{$|\!\}$}}2
		{:=}{{$:=$}}2
		{==}{{\tiny{$==$}}}2
		{===}{{\tiny{$===$}}}3
		{++}{{\tiny$+\!\!+$}}2
		{+++}{{\tiny$+\!\!+\!\!+$}}3
		{:==}{{$:==$}}3
		{\{|*|\}}{{$\{\!|\!\star\!|\!\}$}}4
		{<$>}{{\texttt{<\$>}}}3
		{<*>}{{\texttt{<*>}}}3
		{<|>}{{\texttt{<|>}}}3
		{>||>}{{$\triangleright\triangleright$}}2
		{>>>=}{{$>\!\!>\!\!>\!=$}}4
		{>>>|}{{$>\!\!>\!\!>\!|$}}4
		{>>=}{{$>\!\!>\!=$}}3
		{>>|}{{$>\!\!>\!\!|$}}3
		{?>>}{{\texttt{?>\!>}}}3
		{!>>}{{\texttt{!>\!>}}}3
		{@>>}{{\texttt{@>\!>}}}3
		{<<@}{{\texttt{<\!<@}}}3
		{-||-}{{\texttt{-\!|\!\!|\!-}}}3
		{-||}{{\texttt{-\!|\!\!|}}}3
		{||-}{{\texttt{|\!\!|\!-}}}3
		{.||.}{{\texttt{.\!|\!\!|\!.}}}3
		{.&&.}{{\texttt{.\&\&.}}}4
}
\newcommand{\CleanInline}[1]{\lstinline[language=Clean]!#1!}
\newcommand{\prog}[1]{\CleanInline{#1}}
\lstdefinestyle{numbers}{numbers=left, stepnumber=1, numberstyle=\tiny, numbersep=5pt}
\lstnewenvironment{CleanCode}{\lstset{language=Clean}}{}
\lstnewenvironment{CleanCodeN}{\lstset{language=Clean,style=numbers}}{}
\makeatletter
\lst@AddToHook{OnEmptyLine}{\vspace{-0.5\baselineskip}}
\makeatother
\usepackage{graphicx}
\usepackage{tikz}
\usepackage{pgfplots}
\usetikzlibrary{calc, shapes, arrows, decorations.markings}%
\usetikzlibrary{arrows,automata,calc,shapes, positioning, patterns, matrix}%
\usepackage{mathtools}
\usepackage{amssymb}
\usepackage{enumitem}
\usepackage{listings}
\usepackage{filecontents}
\usepackage{theorem}
\usepackage{comment}
\usepackage{subcaption}
\usepackage[T1]{fontenc}
\usepackage{graphicx}
\usepackage{url}

\usepackage{graphicx}
\usepackage{float}
\usepackage{tabularx}
\usepackage{color}
\usepackage{listings}
\usepackage{hyperref}

\lstdefinestyle{numbers}{numbers=right, stepnumber=1, numberstyle=\tiny, numbersep=-5pt}

\lstnewenvironment{Go}{\lstset{language=Go,frame=single,
  basicstyle=\ttfamily\scriptsize,
  keywordstyle=\color{black}\ttfamily,
  stringstyle=\color{black}\ttfamily,
  commentstyle=\color{black}\ttfamily}}{}

\usepackage{comment}
\usepackage{cite}
\usepackage{url}
\renewcommand{\orcidID}[1]{\href{https://orcid.org/#1}{\textsuperscript{\includegraphics[scale=.5, bb=0 0 20 1]{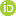}}}}

\title{
SusTrainable: Promoting Sustainability as a Fundamental Driver in Software Development Training and Education\\[2ex]
2nd Teacher Training\\
January 23--27, 2023, Juraj Dobrila University of Pula, Pula, Croatia\\
Revised lecture notes
}
\author{Tihana Galinac Grbac, Csaba Szab\'o and João Paulo Fernandes (Eds.)}
\date{June 2023}

\begin{document}


\begin{titlepage}
  \thispagestyle{empty}
  \let\footnotesize\small
  \let\footnoterule\relax
  \let \footnote \thanks
  \null\vfil
  \vskip 60pt
  \begin{center}%
    {\LARGE
        SusTrainable: Promoting Sustainability as a Fundamental Driver in Software Development Training and Education\\[2ex]
        \large
        2nd Teacher Training\\
        January 23--27, 2023, Juraj Dobrila University of Pula, \\
        Pula, Croatia\\
        Revised lecture notes
    \par}%
    \vskip 3em%
    {\large
     \lineskip .75em%
      \begin{tabular}[t]{c}%
        Tihana Galinac Grbac\and Csaba Szab\'o\and Jo\~ao Paulo Fernandes (Eds.)
      \end{tabular}\par}%
      \vskip 1.5em%
    {\large June 2023 \par}
  \end{center}\par
  \vfil\null
\end{titlepage}

\thispagestyle{empty}

\frontmatter

\chapter*{Preface}
\markboth{Preface}{Preface}

This volume exhibits the revised lecture notes of the second teacher training organized
as part of the project \emph{Promoting Sustainability as a Fundamental Driver in Software Development Training and Education},
and held at the Juraj Dobrila University of Pula, Croatia, in the week January 23--27, 2023.
The project is the Erasmus+ project \emph{No. 2020{-}1{-}PT01{-}KA203{-}078646 --- Sustrainable}.
More details can be found at the official project web site

\begin{center}
\url{https://sustrainable.github.io/}
\end{center}

\noindent but for convenience of the reader, the brief description of the project is quoted as follows.

``Sustainability as a key driver for the development of modern society and the future of the planet has never achieved as much consensus worldwide as today.

Ultimately, it is the hardware of ICT systems that consumes energy, but it is software that controls this hardware. Thus, controlling the software is crucial to reduce the ever-growing energy footprint of ICT systems. The Sustrainable project advocates the introduction of all facets of sustainability as a primary concern into software engineering practice.

This project aims to actively promote educating the next generation of software engineers to consider sustainability in all aspects of the software engineering process: SusTrainable means Training for Sustainability. We aim to provide future software engineers with essential skills to develop software that is not only functionally correct, but also easy to maintain and evolve, that is durable, has a low impact and uses the hardware it is running on in the most energy-efficient way.

Our objective is to train the future avant-garde software engineers for the sustainable software and ICT that the knowledge-based, environmentally concerned societies of the 21st century demand. The majority of summer school participants are close to their transition from learning to doing and will soon join the European software engineering workforce. They will carry the ideas, concepts and methods they learned at the summer schools into industrial software engineering practice across Europe and worldwide. Furthermore, they will act as facilitators and multipliers alike for a sustainable future software-driven Europe. Conversely, we provide the ubiquitous enthusiasm of the younger generation for sustainable development a constructive, meaningful and high-impact route forward.

For this purpose we assemble a broad and diverse consortium of researchers and educators from 10 selected universities and 7 countries from across Europe.''

\bigskip

One of the most important contributions of the project are two summer schools. The 2nd SusTrainable Summer School (SusTrainable -- 23) will be organized
at the University of Coimbra, Portugal, in the week July 10--14, 2023. The summer school
will consist of lectures and practical work for master and PhD students in the area of computing science and closely related fields.
There will be contributions from
Babe\textcommabelow{s}-Bolyai University,
E\"{o}tv\"{o}s Lor\'{a}nd University,
Juraj Dobrila University of Pula,
Radboud University Nijmegen,
Roskilde University,
Technical University of Ko\v{s}ice,
University of Amsterdam,
University of Coimbra,
University of Minho,
University of Plovdiv,
University of Porto,
University of Rijeka.

To prepare and streamline the summer school, the consortium organized a teacher training in Pula, Croatia.
This was an event of five full days in the week January 23--27, 2023, organized by Tihana Galinac Grbac and Neven Grbac at the Juraj Dobrila University of Pula, Croatia.
The Juraj Dobrila University of Pula is very concerned with the sustainability issues. The education, research and management is conducted with sustainability goals in mind. The university
participates in the UI GreenMetric World University Rankings.

The goal of this teacher training is to improve the efficiency and effectiveness of all the educational activities at the forthcoming
2nd SusTrainable Summer School at the University of Coimbra.

An effective and efficient summer school consists of a series of educational and teaching activities by a number of invited researchers and/or educators. It is important that these
activities are of high quality, but also that they are not totally unrelated and independent. The thing is that a summer school is usually very demanding for students, thus, a well
balanced combination of forms of activities and teaching methodologies would improve students' satisfaction with the knowledge and techniques acquired at the school. Besides that, it
is very convenient that different activities explain their part at the same or similar examples, the tasks are related and different aspects well explained.

At the teacher training, all the proposed summer school teachers are asked to present their plan for the activities at the summer school. This includes an outline and content for their series of
activities (course), the teaching methodology and the form of activities, prerequisites assumed the students will be familiar with (including their expected previous knowledge, practical
skills, programming tools).
The final outcome of the training should be a detailed plan for each teacher's series of activities at the summer school, orchestrated with all other teachers in the sense explained
above. The plans should include the list of prerequisites required by the students (with references and possibly some pre-school reading assignments), the required hardware and
software for the summer school (for the organizers), and an overview of the lecture material, examples and student assignments.

As in the first teacher training, the contents of the planned summer school lectures were discussed and fine-tuned.
The topics discussed in regard of each contribution are the following:
\begin{itemize}
\item
    the contents of the intended lecture, but not the entire lecture itself;
\item
    what is the intended audience;
\item
    what is the required knowledge and skills of the audience;
\item
    what are required materials (hardware and software) for the exercises;
\item
    what will students know after this lecture and can this be used in subsequent lectures;
\item
    what is the relation with other contributions, other dependencies and what are the lessons learned during the teacher training.
\end{itemize}

Apart from the discussion regarding the contributions of the summer school and other outcomes of the teacher training mentioned above and listed in the proceedings,
there were invited lectures focused on best practices in teaching IT-related topics in view of sustainability goals of the summer school:
\begin{itemize}
\item
    Dr.\ Patricia Lago from the Vrije Universiteit Amsterdam: \emph{VU Master-level education in software engineering, green IT, and digital sustainability}.
\item
    Dr.\ Darko Etinger and Dr.\ Sini\v{s}a Mili\v{c}i\'{c} from the Juraj Dobrila University of Pula: \emph{Virtual study of informatics: experiences and best practices}.
\end{itemize}
Those presentations provided useful background knowledge for the teaching team.
See

\begin{center}
\url{https://www.unipu.hr/en/international-cooperation/projects/sustrainable/teacher_training_pula}
\end{center}

\noindent
for the complete schedule of the teacher training.

The contributions in the proceedings were reviewed and provide a good overview of the range of topics that will be covered at the summer school.
The papers in the proceedings, as well as the very constructive and cooperative teacher training, guarantee the highest quality and
beneficial summer school for all participants. We are looking forward to contributing to a sustainable future.

\medskip

\begin{flushright}
\noindent June 2023\hfill
{Tihana Galinac Grbac\footnote{Juraj Dobrila University of Pula, Croatia}}\\
{Csaba Szab\'o\footnote{Technical University of Ko\v{s}ice, Slovakia}}\\
{Jo\~ao Paulo Fernandes\footnote{University of Porto, Portugal}}\\
\end{flushright}

\mainmatter%

\tableofcontents

%
\title{Code optimization with vectorization in data mining and machine learning}
\titlerunning{Code optimization for sustainable data mining}
%
\author{%
	Zalán Bodó\orcidID{0000-0003-1052-1849}  %
	\and
	Lehel Csató\orcidID{0000-0002-4857-878X} %
}
\authorrunning{Z. Bodó, L. Csató}
%
\institute{Faculty of Mathematics and Computer Science, Babeș-Bolyai University,\\
str. M. Kogălniceanu, nr. 1, RO-400084, Cluj\\
\email{\{zalan.bodo,lehel.csato\}@ubbcluj.ro}
}
\maketitle              
\begin{abstract}
%
%
Interpreted languages are preferred over compiled languages when considering scientific computing -- in this article, we focus on the MATLAB, Python, and Julia languages.
Whilst running interpreted code usually means longer runtime, the source code for interpreted languages is usually more compact and easier to understand.
A second significant advantage of interpreted languages is that they are easier to learn.
These languages, moreover, often provide built-in constructs or third-party libraries for efficient numerical computations.
In this course, we give a short introduction to basic machine learning algorithms and show how to implement these algorithms with vectorization in the different languages and modules -- with the scope of creating a more efficient and more understandable code.
Besides vectorization, the need for optimized data structures in scientific computing will also be discussed.

\end{abstract}
%

\section{Contents of the lecture}

The motivation of the lecture is the rise in popularity -- and indeed in the added value towards the teaching process -- of the interpreted languages, specifically when the focus of the application is an area related to data analysis, machine learning or -- more broadly -- artificial intelligence.
These languages are easier to code, easier to debug, but they are inefficient, for example, when executing loops, typically used when the same operations have to be applied to several elements -- mostly due to the higher level of abstraction due to the presence of the virtual machine that interprets the code \cite{birkbeck2007dimension}.\footnote{Loops are highlighted here because in vectorization we actually optimize cycles, but, of course, this inefficiency applies to almost all language constructs.}
In this respect, the compiled versions of the same code -- evidently written in a compiled language -- is at least one order of magnitude faster \cite{aho2007compilers}.

When numerical computations are required, however, some of these languages offer built-in features or libraries to support these.
For example, in MATLAB or more recently in the Julia language~\cite{bezanson2017julia,Sherrington2015}, the initial type for all variables is the multidimensional array\footnote{\url{https://www.mathworks.com/help/matlab/learn_matlab/matrices-and-arrays.html}}, and these languages provide specialized operations to make computations on these data structures more efficient than those realized by programmers and using loops.
Similarly, Python's NumPy\footnote{\url{https://numpy.org/}} package offers functions to perform optimized matrix computations \cite{van2011numpy}.

In this course, we will discuss the different approaches/methods to rewrite numerical computations using matrices to optimize the program code, if such a rewriting is possible/efficient.
This procedure is called \emph{vectorization}.
Besides vectorization, the need for using optimized data structures -- e.g. sparse vectors/matrices -- will also be discussed.

For the sake of example, we consider the problem of computing the Euclidean distance matrix of a set of items in a dataset -- usually needed to calculate nearest neighbors, the RBF kernel matrix when applying kernel methods \cite{scholkopf2018learning}, or in simple clustering algorithms -- i.e. we need to compute a matrix $\mathbf{D}$ of size $N\times N$, containing the pairwise Euclidean distances over data items:
$$
    D_{i,j} = \|\mathbf{x}_{i} - \mathbf{x}_{j}\|_{2}^{2}, \quad i,j\in \{1,2,\ldots,N\}
$$
Suppose the data $\mathbf{x}_{i}\in \mathbb{R}^d$, $i=1,2,\ldots,N$ -- represented as \emph{column} vectors -- is arranged into the $N\times d$ matrix $\mathbf{X}$, as below (left), then with only matrix operations we can compute $\mathbf{D}$ as below (right):
\begin{equation*}
  \mathbf{X} = \begin{bmatrix}
        \mathbf{x}_{1}^{T}\\
        \mathbf{x}_{2}^{T}\\
        \vdots\\
        \mathbf{x}_{N}^{T}
        \end{bmatrix}
  \qquad\qquad\qquad\qquad
  \begin{array}{ll}
    \mathbf{A} = \mathbf{X}\mathbf{X}^{T}\\
    \mathbf{B} = \diag(\diag(\mathbf{A}))\\
    \mathbf{D} = -2\mathbf{X}\mathbf{X}^{T} + \mathbf{B}\mathbf{1}_{NN} + \mathbf{1}_{NN}\mathbf{B},\\
  \end{array}
\end{equation*}
where $\diag(\cdot)$ is the diagonal function, implemented under the same name -- \verb!diag! -- in MATLAB/Octave\footnote{\url{https://www.mathworks.com/help/matlab/ref/diag.html}} which either keeps only the diagonal elements of a matrix -- the inner \verb!diag! function --, or makes a diagonal matrix out of a vector of elements -- the outer function\footnote{In the language Julia there is a distinction: \texttt{diag} takes a matrix and returns the vector of diagonals, for the other direction one uses \texttt{diagm}.}, and $\mathbf{1}_{NN}$ is the matrix containing all ones.

The above simplification of computing the pairwise distances highlights (1) the simplicity of the code -- which is extremely close to the mathematical notations in MATLAB and Julia -- and (2) the efficiency of the implementation, which originates in the optimized versions of the matrix multiplication and the matrix notation itself.
Using vectors and matrices in these methods allows the straightforward application of some formulae for more efficient computations when needed (e.g. Searle's set of identities) \cite{matrix_cookbook,golub2013matrix}.

We will explore the approaches to vectorization within the data processing and data analysis methods to
compute dot products, matrix products, weighted averages, etc., and use these in different algorithms.
The examples will be written for real-world data in order to keep the audience motivated.

\section{Intended audience}

The target audience of this course includes everybody who is interested in data mining, applied mathematics, numerical computing, and mathematical modeling within data analysis.
In the present, it is inevitable for someone to encounter applications of artificial intelligence, more specifically machine learning methods, since these can be found almost everywhere, from spam filtering, through intelligent image enhancement, to machine translation -- just to name a few.

Consequently, we recommend the course for everyone who is interested in artificial intelligence; specifically in understanding machine learning models and the crux of deep learning methodology.
They will acquire knowledge and skills that will help a quicker and ``deeper'' understanding of machine learning methods and their implementation details.
The use of vectorization and related concepts will help present and future PhD students improve the quality of their code and produce a more \emph{sustainable} code basis -- in every aspect of the word.
Sustainable coding/programming -- from the perspective of scientific computing -- reduces to the following three things: (i) using interpreted languages facilitating fast development and easy debugging; (ii) using optimized data representations, data structures; (iii) exploiting the benefits of vectorization.

\section{Prerequired knowledge}

The required knowledge to be able to follow the present course efficiently can be summarized as follows:
\begin{itemize}
    \item basic linear algebra knowledge (linear equations, vector spaces, matrix operations);
    \item basic probability theory (probability distributions, conditional probability, Bayes' theorem);
    \item programming skills, not necessarily in the languages mentioned above.
\end{itemize}
However, if some of the above-mentioned knowledge/skill is missing or is only partially available, the participants will be able to fill in the gaps and catch up during the theoretical and practical parts as explanations will be added upon questions from the audience.

\section{Required materials}

As a practical part, we will analyze basic machine learning/data mining algorithms (for example, centroid, $k$-means, kernel $k$-means, $k$-nearest neighbors, naive Bayes \cite{bishop2006pattern}, the PageRank  algorithm \cite{LangvilleMeyer2006}) and implement these using matrix operations, working in Python/Julia/Octave.
We will provide the participants with the materials needed for the course, including the presentation slides, bibliography, links to additional materials, datasets, as well as sample code snippets to facilitate the practice.
For effective cooperation, however, we would like to ask the participants to have Python\footnote{\url{https://www.python.org/}} and Julia\footnote{\url{https://julialang.org/}} installed on their computers, as well as an integrated development environment or code editor in which they can work comfortably -- we suggest the \emph{Visual Studio Code}\footnote{\url{https://code.visualstudio.com/}} system.

\section{Desired outcome}

After finishing the course the participants will have a basic knowledge of machine learning/data mining methods, will understand how to apply these and in which situations and they will gain practical knowledge of how to efficiently implement some of these algorithms using vectorization as well.

\section{Relation to other contributions and lectures}

This lecture is a continuation of the \emph{Sustainable functional programming and applications} lecture given by \emph{Lehel Csat\'o} at the Rijeka Summer school in 2022 -- where the Julia language was used to solve machine learning problems.
At the same summer school, the lecture \emph{Applying Soft Computing for Sustainability} by \emph{Goran Mauša} is also related to vectorization; since one computes the same functions for the whole dataset.
The 2023 Sust(r)ainable Summer School will host the \emph{Efficient evolutionary computing} lecture by \emph{Goran Mauša}, where the data mining problems require vectorization to function efficiently.

\section{Acknowledgments}

This work acknowledges the support of the ERASMUS+ project ``SusTrainable -- Promoting Sustainability as a Fundamental Driver in Software Development Training and Education'', no. 2020–1–PT01–KA203–078646.



\title{Sustainability through Green Gamification}%
\titlerunning{Sustainability through Green Gamification}
\author{Denitza Charkova\orcidID{0000-0003-0873-4415} \and Elena Somova\orcidID{0000-0003-3393-1058} }
\authorrunning{D. Charkova, E. Somova}
\institute {University of Plovdiv "Paisii Hilendarski", 24 Tzar Assen St., 4000 Plovdiv, Bulgaria,\\%
	\email{  dcharkova@uni-plovdiv.bg;eledel@uni-plovdiv.bg}\\%
	\url{https://www.uni-plovdiv.bg/} }%
\maketitle
\begin{abstract}%
Among the many trials that our contemporary society faces there is one which stands out as a priority to cultivate a healthy lifestyle on our planet. This issue is climate change and to tackle it, urgent actions are needed not only on the societal, but also on a personal level. The exponential growth of environmental problems stimulates us, as educators, to invent new strategies in the field of Green Education to teach our children how to maintain a more sustainable future. This article discusses the integration and application of sustainable practices through educating Green Gamification at the Plovdiv University “Paisii Hilendarski” (PU). The experiment was performed on first- and fourth-year students majoring in three different IT majors. The study aimed at raising awareness about climate change, the burning of fossil fuels and the crucial importance of developing green behaviours on a both personal and business level.%
\end{abstract}%

\section{Sustainable UI and UX}%
\par%
The economic, societal, and educational realities we currently face reveal how vulnerable communities, countries and the entire world can be in all aspects of life. Many alterations are needed for our future generations to live a healthier and more sustainable existence. These ideas should be implemented in the educational systems at an early age to have long lasting effects and change behaviours for a greener future.%
\par%
One interesting method for achieving these goals is teaching sustainability through Green Gamification. This method can be integrated in the educational systems and accompany students in their entire course of study. Furthermore, integrating practices such as developing sustainable UI (user interface) and sustainable UX (user experience) we aim to have a greener and cleaner experience when using the availability of the Internet. The design of the sustainable UI/UX will be the main topic of our SusTrainable Summer School’2023 lecture.%
\par%
There are four key areas where work can be done in terms of building a more sustainable UI/UX~\cite{BIB2}: findability, performance optimisation, design and UX, and green hosting. Findability focuses on making the content as accessible and easy to locate as possible, thus saving energy because users will have to load fewer pages in order to find the desired information. Performance optimization focuses on speeding up the website and thus using less processing power. The design and UX focus on approaches allowing websites to be accessible to all users, regardless of the hardware they use. In turn, green hosting reduces the carbon footprint of a website by switching to a green web hosting company.%

\section{Theories of action}\label{section:overview}%
\par%
There are existent theories of action which can further encourage users to practise green behaviours through models of conduct. Such examples are~\cite{BIB1}: Nudge theory, Hooked model, Mindful design, Theory of interpersonal behaviour, and Transtheoretical model of behaviour change. Likewise, the integration of Green Gamification in our software applications, classrooms and even community models is another method of stimulating and rooting sustainable behavioural practices. %
\par%
To begin with, the Nudge theory~\cite{BIB9} ~\cite{BIB14} suggests it is capable of affecting human behaviour - choices made by individuals can easily be affected in a predictable way without forbidding any options. Such a sustainable design example can be increasing the visibility to affect human behaviour - by giving one of two buttons (options) extra colour and contrast. %
\par%
The second theory – Hooked model~\cite{BIB10} is also called Model for building habit-forming products (sustainable habits), where the 4 key words are Trigger, Action, Variable reward and Investment. The model starts to act with some trigger (e.g. the need), then there is a need for action (to eliminate the trigger). For example, here the UI/UX designer can increase the probability for sustainable action. On the next step action is rewarded (e.g. through likes, levelling up or praising the choice of a sustainable option). The reward should not be predictable, since variability will keep the user’s interest over time. In the end, the reward encourages the user to invest in the choosen action/product in waiting of more rewards. %
\par%
The Mindful design~\cite{BIB11} ~\cite{BIB15} is another way to shape behaviours. Mindfulness is an optimal interaction between attention and awareness. There are 3 designated steps to Mindful design: 1. Identification of the lack of mindfulness (e.g. user intensively scrolling an app), 2. Identification of the mindful solutions (e.g. reminds the users of their sustainable intent), 3. Implementation of the mindful solutions (makes the users mindful they must be disrupted and forced to reflect). %
\par%
The Theory of Interpersonal Behaviour~\cite{BIB12}~\cite{BIB16} is a cognitive model that claims: the individual’s intention is the core factor behind how humans act. The intention is created by the individual's beliefs, social norms and emotions. Habits trigger the intention and facilitating conditions restrict what behaviours are possible. So, the designed solutions should focus on habits.%
\par%
The Transtheoretical model of behaviour change~\cite{BIB13} defines the stages people pass from the moment they start thinking about changing their behaviour to the moment the change is durably achieved – Precontemplation, Contemplation, Preparation, Action, Maintenance and Termination. After the user has gained insight, the design must empower the user to create an action plan: by facilitating information of what is sustainable action the design prepares the user to take action. The design should support the user to take correct actions: by only suggesting options that are considered sustainable. In addition, the design should encourage the user to keep up with the new behaviour and prevent relapse. In the end, the users have to sustain the new behaviour themselves.%
\section{Green Gamification}%
\par%
The second focal point in the theoretical models is the application of Green Gamification on a both community and educational level. By definition~\cite{BIB3}, Gamification is the application of game principles and design elements in non-game contexts. It uses game elements in non-gaming systems to improve user experience and user engagement, as it applies game design to make otherwise boring tasks more engaging~\cite{BIB4}. It hooks us by meeting our basic human needs for achievement, appreciation, reciprocity, and a sense of control over our little corner of life. In today’s competitive battle for attention, games are the most effective tool for leveraging technology, rising above marketing noise and engaging the socially networked consumer. The European Commission Acknowledges that the challenge is to mainstream the application of gaming technologies, design and aesthetics to non-leisure contexts, for social and economic benefits.%
\par%
Gamification consists of elements, techniques, and actions, which all come together to build an educational experience through games. Examples of game elements~\cite{BIB5}~\cite{BIB6} can be avatars, bonuses, badges, combos, rewards, leaderboards, progress, status, teams etc. The game techniques can also vary depending on the material, topic, class and aim of the specific class. Such examples~\cite{BIB5}~\cite{BIB6} can be identity shift, reward system, tracking progress, current status, rules, teamwork, time limit, hidden treasure, feedback etc. The game actions can also differ depending on the educational aim and circumstance e.g., role playing, receiving a bonus or award, gaining an advantage, retrieving resources, following progress level, completing missions etc. %
\par%
According to the Bartle Player quiz~\cite{BIB7}, the author designates 4 specific player types, which possess unique characteristics when engaging in a game: achiever, killer, explorer and socialiser. By nature, the achiever focuses on elements such as: level, badge, bonus, reward, resource, progress and techniques such as a reward system, feedback, challenge, mission, adventure and progress tracking. On the other hand, the profile of the socializer focuses on elements such as teamwork, the usage of avatars, chat and messaging and applies techniques like teamwork, communication etc.%
\par%
Taking a step further into the topic of Green Gamification~\cite{BIB8} we can define the term as the use of game mechanics to engage people and change behaviour, and apply it to sustainability issues.%

\section{PU green experiment}%
\par%
The Green Gamification experiment, performed at the university level consisted of two smaller experiments – an individual and team experiment. The experiment began with a lecture on Sustainability, Sustainable Development and Sustainable Education. After the acquaintance with the core concepts, students were introduced to the idea of Green Gamification. The aim was to highlight the importance of the issues and addressing them through meaningful education on the subject matter. Students were encouraged first to share their prior knowledge on Sustainability and were informed about legislations and sustainable goals both in Europe and worldwide. %
\par%
The first task was performed on an individual level and students were responsible for “creating a green log” each day. This was done using a shared file, where in the course of 1 week, students had to perform green actions, fill in the log which should not repeat previous logs (no matter if the logs were made by other students or were their own). They were in charge of discovering a green activity or behaviour that contributes to Environmental Sustainability. They also had to describe the essence of this action in 100 characters, using relevant vocabulary and explain why their action is sustainable. The game would be won by the one who manages to publish the most activities. This experiment showed that students were very actively motivated and most of them had multiple logs, explained their action’s impact and provided evidence. They shared that they found the task quite interesting and used a lot of humour in their explanations. Students were excited to share their actions in the logs and followed up in class. Students noted they were not so much motivated by the badges rather the new idea they had to generate. %
\par%
The second experiment was titled Team Green Business and was intended to be a group experiment. Students were in charge of building their own business which supports Green Behaviour and Sustainability. Students were encouraged to mind map their ideas in class, filter them and create a project of a business they thought would support green behaviour. Students were provided with helper questions to scaffold and guide them through the task. Such questions were: What does your company do? What kind of specialists do you have? What does your office environment look like? What green habits do you want to introduce in your company? Which unsustainable practices do you want to eliminate from your work environment? What is your position on fossil fuels? Would you work for clients who work for fossil fuel companies?%

\section{Conclusion}%
\par%
The core goals were to motivate sustainable behaviour through games and to provoke students to maintain the sustainable practices in their own urban environment. It is used to incentivise positive changes in behaviour and help young IT professionals to improve their overall quality of life. In addition, through the power of gamification we can make that experience predictable, repeatable, and financially rewarding. %
\par%
The plans for the SusTrainable Summer School'2023 are for the students to acquire and practice skills which will help them create a sustainable UI/UX design through the application of different approaches (incl. Green Gamification). The lecture will present suitable theories, models and approaches that can be used in sustainable UI/UX design. During the lab session students will design a green gamified mobile application that has to encourage/impact some type of sustainability, has to be sustainable from the technical point of view and has to include gamification techniques. In a few days of ongoing collaborative work in preset teams, students will present their ideas on UI/UX design. The training will finish with the evaluation of the ideas by all students and teachers and a subsequent award ceremony for the best works will be held. The evaluation questions for the students' work will be: Do they use Gamification (elements x techniques)?, Do they provide technical sustainable decisions? and Do they impact/encourage the user to adopt sustainable behavior (environmental/economic/social)?. Evaluators also will choose the most original and innovative idea. After completing the training, students will have gained knowledge on many approaches on how to design a sustainable UX/UI and will be able to develop sustainable websites and mobile applications.%

\par%
\textbf{Acknowledgements}%
\par%
This paper acknowledges the support of the Erasmus+ Key Action 2 (Strategic partnership for higher education) project No 2020-1-PT01-KA203-078646: “SusTrainable – Promoting Sustainability as a Fundamental Driver in Software Development Training and Education”. The information and views set out in this paper are those of the authors and do not necessarily reflect the official opinion of the European Union. Neither European Union institutions and bodies, nor any person acting on their behalf may be held responsible for the use which maybe made of the information contained therein.

%

%
\title{Large project sustainability: the role of code comprehension\thanks{Prepared with the professional support of the Doctoral Student Scholarship Program of the Co-operative Doctoral Program of Ministry of Innovation and Technology financed by the National Research, Development and Innovation Fund.}}
%
%
\author{Anett Fekete\orcidID{0000-0001-8466-7096} \and
Zoltán Porkoláb\orcidID{0000-0001-6819-0224}
}
\authorrunning{A. Fekete and Z. Porkoláb}
%
\institute{Eötvös Loránd University, Faculty of Informatics,\\ Pázmány Péter stny. 1/C, 1117 Budapest, Hungary\\
\email{\{afekete, gsd\}@inf.elte.hu}
}
\maketitle              
\begin{abstract}
Code comprehension is a daily challenge in a software developer's life. A substantial amount of time is spent with comprehension activities at the expense of productivity. Junior programmers are in the most dire need of proper code comprehension support, since they lack the amount of work experience which facilitates comprehension tasks. There is a wide variety of comprehension supporting software, however, developers are usually not aware of their options, and settle for the insufficient support in code editors. We aspire to raise awareness among programmers about the importance of proper code comprehension support by introducing the CodeCompass code comprehension framework to students. We expect that by showing the practical usage and benefits of a multi-purpose comprehension tool through solving a C++ programming task will instill the need in young programmers for such support who will later indicate their need in the workplace.

\end{abstract}
\section{Introduction}

Understanding the source code and architecture of a program is part of the daily challenges in a software developer's life, no matter how long they have been working as a programmer. Code comprehension is made difficult by several circumstances: incomplete or missing documentation, scattered knowledge among developers, insufficient comprehension support in code editors, etc. According to various studies, developers spend at least half of their working hours with code comprehension activities, and this amount of time has just been growing as computer science became a more and more paramount part of our lives \cite{corbi1989program, cherubini2007let, minelli2015know, siegmund2016program, xia2017measuring}. The statistics in these studies suggest that programmers are in need of more efficient code comprehension support. More time spent on comprehension activities mean slower task execution and more resource consumption. There are many resource types whose costs can be greatly reduced by using effective code comprehension support: working hours, money and actual energy used by computers are all included.

In this paper, we present the role of code comprehension in resource management by emphasizing the usage of comprehension tools in everyday programming tasks. We also present our plans for the summer school of 2023 in Coimbra, where we intend to popularize code comprehension tools among young developers with the hope that the culture of needing proper code comprehension support in the workplace will become more widespread.

\section{Code comprehension models}

There has been extensive research concerning the source code comprehension strategies of software developers. Program comprehension procedures form well-defined workflows that can be categorized into various code comprehension models which serve as unified abstractions of the comprehension process. Several models have been identified throughout the last couple decades, with two major categories that identify the code comprehension process based on the directions from where programmers approach the source code: the top-down and the bottom-up direction.

Some studies presented reviews of these categories \cite{storey2005theories}, with Mayrhauser and Vans also defining a new comprehension model \cite{von1995program}. Siegmund also summarized the categories in their study \cite{siegmund2016program}, along with an insight of the present and the possible future of code comprehension. We have also published a comprehensive review of the above mentioned code comprehension directions, putting the focus on classifying the function sets of modern code editors and integrated development environments into the categories \cite{fekete2020comprehensive}.

A top-down approach is built when the programmer first tries to build a mental image (hypothesis) of the purpose of the program, then moves on to finer details by comparing a the code to similar known programs. In this case, the programmer is initially familiar with the program domain. The finer details serve as further information which may help accepting, rejecting or modifying the hypothesis \cite{brooks1977towards}. The program domain elements (e.g. programming language syntac, algorithms, coding conventions etc.) provide a basis for the evaluation of the initial assumptions \cite{soloway1984empirical}.

A bottom-up strategy is applied when a programmer does not possess enough domain knowledge, so they will start understanding the code by searching for "pivot points" in the source code. The programmer relies on syntactic and semantic knowledge of a programming language to build a high level mental model of the program \cite{shneiderman1979syntactic}. In other cases, observation does not start at the code itself, but at the control flow of the program \cite{pennington1987stimulus}. Then the programmer considers the programming goals and the control flow to build a hypothesis of the program's operation.

There are comprehension models that include elements from both top-down and bottom-up approaches \cite{levy2019understanding} which are called integrated approaches. These models pay the most attention to the discursive way of human thinking. In this case, the programmer switches between the above discussed directions in an arbitrary way based on the previously existing and recently collected information~\cite{letovsky1987cognitive}.

\section{Code comprehension and sustainability}

In the realm of green computing and sustainability, code comprehension support holds paramount significance. As the world strives to mitigate the environmental impact of computing systems, it becomes imperative to develop software solutions that are energy-efficient, resource-conscious, and eco-friendly. However, achieving such objectives necessitates a comprehensive understanding of the underlying code base. Code comprehension support enables developers to navigate intricate software architectures, identify energy-intensive code segments, and optimize resource utilization. By comprehending the details of the code, developers can make informed decisions regarding algorithmic efficiency, memory management, and power consumption. Furthermore, code comprehension support facilitates the identification of areas for code refactoring, eliminating redundant or inefficient operations, and promoting cleaner, streamlined code. Ultimately, by enhancing code comprehension, developers can significantly contribute to the development of energy-efficient software systems, reducing the carbon footprint of computing and advancing the goals of sustainability in the digital age.

\section{CodeCompass}

CodeCompass \cite{porkolab2018codecompass} is an open-source\footnote{GitHub: \url{https://github.com/Ericsson/CodeCompass}} code comprehension framework developed by Eötvös Loránd University and Ericsson. It consists of a parser and a webserver binary. The CodeCompass parser applies static analysis to the given, ideally compiled source code and the corresponding build commands generated during compilation. Various information is stored afterwards about the project regarding structural data, code metrics, version control information, etc. This information is stored in the workspace database which is then accessed by the CodeCompass webserver. The webserver provides several different textual and graphic services, such as detailed searching, structural and code-level visualizations, and Git blame data.

CodeCompass is a pluginable framework. Each plugin is independent of every other, thus a certain plugin can be easily skipped from the parsing process if not needed. Plugins consist of a \emph{model}, a \emph{parser}, a \emph{service}, and a \emph{web GUI} component. Currently, CodeCompass fully supports C and C++ programs, and is in part capable of parsing C\#, Java, and Python projects. Apart from language parsing, CodeCompass provides code metrics analysis, advanced search functionalities, and Git repository visualizations.

\subsection{Role in education}

CodeCompass is mainly developed in the Model C++ Software Technology laboratory at Eötvös Loránd University, Faculty of Informatics. Students have the opportunity to join the project by solving issues (e.g. bug fixing, small improvements, refactoring), or by taking up a large subproject, such as creating a new plugin, or complementing an already existing one with various new features. Students who join the lab learn about working with various programming languages and APIs, using version control, reviewing other people's code, designing software architecture, etc. \cite{fekete2022building}

The software is integrated with the TMS assignment management system which is developed in our faculty and is used to handle programming assignments. This feature is supposed to facilitate the work of teachers: every submitted assignment can be parsed by CodeCompass in a containerized environment, and the teacher is able to browse the code by launching the webserver. This way the teacher does not have to download each assignment to their computer.

\section{Course information}

In our part of the summer school we plan to show the participants the importance of proper code comprehension support. For this purpose, during the lecture we will introduce the theoretical background of code comprehension along with the feature set of CodeCompass. This way, the students will be familiar with the wide variety of code and program comprehension supporting features that the tool offers; this lecture also serves as a thought provoking session in which we motivate the students to identify their specific needs in code comprehension tasks, e.g. come up with new functionalities based on real life examples from school or workplace.

In the practical session we will present a small C++ project, \emph{TinyXML2}\footnote{GitHub: \url{https://github.com/leethomason/tinyxml2}} which is readily parsed and available on the demo site\footnote{Demo website: \url{https://codecompass.net/}} of CodeCompass. TinyXML2 is a distinguished project that we use to test and present CodeCompass. It consists of three source files which contain hundreds of source code which makes it a perfect test project for code comprehension activities. After a short walk-through of all available functions, we will give the students a minor programming task to students in TinyXML2: they will have to find and modify one specific line in the source code in order to make the tag parsing feature of TinyXML2 case-insensitive. The students will get approximately 60 minutes to solve the task. Based on the results of a previous experiment which we conducted in the spring semester of 2022 with the participation of 27 Computer Science MSc students \cite{fekete2022report}, cc. 60 minutes should be enough for everyone to solve the task.

During the practice the students will be only allowed to use CodeCompass for code browsing and understanding. The software will be available on the demo site with TinyXML2 already parsed. We will anonymously collect information about the students' activities in CodeCompass via Google Analytics. They will be able to modify the code using Visual Studio Code, and compile the code with CMake and Make. The focus of this task is to correctly identify the exact line which has to be modified instead of writing new code; for this reason, we will readily provide the C++ method which has to be invoked to make the parsing case-insensitive. The students need will need fair understanding of C++ or any other imperative language, and a basic computer with 2 to 4 GB of RAM and Internet access. They will not need any knowledge in build systems as the exact commands will be given to them beforehand.

After our session we will analyze and evaluate the activity data. We will also publish various statistics for the students: best, average and medium solution times, most used functionalities in CodeCompass, etc.

\section{Conclusion}

As we discussed, an average programmer spends at least half of their work hours with code comprehension activities. Proper and effective code comprehension support is a very important aspect of software development: it reduces development costs in human resources, finances, and energy consumption. However, standalone comprehension supporting software is not widespread enough to make significant difference in resource usage. Our purpose in the summer school is to popularize code comprehension tools by showing the various functionality they serve. Our plan is to give the participating students a small C++ programming task which they have to solve by using a code comprehension supporting software. Our demo comprehension tool is CodeCompass which is equipped with several features which may help reducing the time spent with comprehension activities, and thus, resource consumption.

\section{Acknowledgement}

This paper acknowledges the support of the Erasmus+ Key Action 2 (Strategic partnership for higher education) project No. 2020-1-PT01-KA203-078646: “SusTrainable - Promoting Sustainability as a Fundamental Driver in Software Development Training and Education”.

The information and views set out in this paper are those of the author(s) and do not necessarily reflect the official opinion of the European Union. Neither the European Union institutions and bodies nor any person acting on their behalf may be held responsible for the use which may be made of the information contained therein.


%
\title{Energy Consumption and Optimization of Software}

%
%
\author{João Paulo Fernandes\inst{1}\orcidID{0000-0002-1952-9460} \and
Bernardo Santos\inst{1}\orcidID{0000-0002-6172-3830} \and
Maja H. Kirkeby\inst{2}\orcidID{0000-0003-0033-2438} \and
Luís Paquete\inst{3}\orcidID{0000-0001-7525-8901}
}
\authorrunning{J. P. Fernandes et al.}
%
\institute{LIACC \& DEI-FEUP, University of Porto, Porto, Portugal\\
\email{\{jpaulo,up201706534\}@fe.up.pt}
\and
Roskilde University, Roskilde, Denmark\\
\email{majaht@ruc.dk}
\and
University of Coimbra, Coimbra, Portugal\\
\email{paquete@dei.uc.pt}}
\maketitle              
\begin{abstract}
We describe a tutorial to be delivered in the second Summer School organized with the Sustrainable - Promoting Sustainability as a Fundamental Driver in Software Development
Training and Education, an Erasmus+ Strategic Partnership.

The tutorial aims to provide software engineers with the knowledge and tools to measure software energy consumption. This is regarded as a first step towards the more ambitious goal of optimizing such consumption. Reducing energy consumption within ICT systems is paramount in realizing a more sustainable exploration of the technology that ever more rules the world.

\end{abstract}
\section{Introduction}

Sustainability is a crucial driver for the development of modern society and the planet's future.
While groundbreaking progresses make of our age an exciting period for humanity, there exists a growing awareness that progress needs to consider the resources the world can provide.
If we continue over-exploiting such resources, the ability of the next generations to inhabit the planet is endangered.

However, the challenges towards sustainability are tremendous, as they are complex and multifaceted and require expertise from a wide range of disciplines to be addressed. For example, addressing climate change requires expertise in atmospheric science, ecology, engineering, economics, and policy analysis. Similarly, addressing social sustainability challenges such as poverty, inequality, and social justice requires sociology, psychology, political science, and public policy expertise.

Sustainability also entails understanding the interconnectedness of social, environmental, and economic systems. For example, environmental degradation can significantly impact human health and well-being, and economic development can significantly impact environmental sustainability. Understanding and addressing these interconnections requires interdisciplinary collaboration and a holistic approach to problem-solving.

Our work focuses on addressing and promoting sustainability within software development and engineering.

As a broad discipline, Sustainable Software Development/Engineering encompasses various aspects, ranging from economic to social, including technological, technical, and environmental factors~\cite{lago15}.
The social perspective explores the use of software to enhance people's quality of life, while the economic perspective emphasizes the development of products that can endure for an extended period. Additionally, from a software development standpoint, technical sustainability encourages the production of high-quality software, which in turn promotes reusability and reduces future development efforts and resource consumption.

One significant aspect that must be considered is environmental sustainability. It focuses on producing and utilizing software with minimal environmental impact, aiming to minimize resource use and maximize energy efficiency. This is a critical concern.

The emergence of large-scale cloud deployment models has led to the establishment of massive data centers with significant energy consumption, raising environmental sustainability concerns.
A 2020 study by the EU Commission revealed that data centers in the former 28 EU countries experienced a substantial increase in energy consumption. In 2010, they consumed 53.9 terawatt-hours, which escalated to 76.8 terawatt-hours by 2018. This significant rise accounted for 2.7\% of the total electricity demand in the EU.

While, in practice, the hardware components of ICT systems consume energy, their software counterparts govern how to run the hardware and how and when energy is consumed. This means that achieving the goals of Sustainability, particularly under its environmental lenses, is only possible by targeting Sustainable Software Development.

The relevance of addressing research in this line is confirmed by Software Developers themselves: it has been shown that Software Developers are indeed keen on developing energy-efficient software~\cite{pinto2014mining,pang2016programmers}. And studies have shown that different design patterns~\cite{DBLP:books/sp/21/Feitosa00F0S21}, sorting algorithms~\cite{bunse2009exploring}, software version changes~\cite{DBLP:journals/ese/Hindle15}, refactorings and transformations~\cite{park2014investigation,DBLP:conf/wcre/0001SF20}, and different Java based collections~\cite{Pereira:2016:IJC:2896967.2896968,hasan2016energy} have a statistically significant impact on energy usage.
Studies have also shown how even the choice of the programming languages to use influences energy usage~\cite{pereira2017sle,DBLP:journals/scp/PereiraCRRCFS21}.

While there is promising evidence of the feasibility of improving energy efficiency by targeting software components of ICT systems, the fact is that this topic is clearly under-represented in the education of modern software engineers.
This paper describes a summer school tutorial that aims to provide junior software engineers with the motivation, knowledge, and tools to measure, analyze and ultimately analyze the energy consumption of software. The structure and content of the tutorial we propose are described in detail in the next section.

\section{Structure and Contents of the Proposed Tutorial}

The tutorial will cover the essential knowledge and skills needed to understand and optimize software with energy consumption in mind. It is divided into two parts, a theoretical lecture to introduce key concepts and a hands-on laboratory session to become familiar with collecting and comparing energy consumption of software.
At the end of this tutorial, the participants will have acquired:

\begin{enumerate}
    \item [$\mathbf{O1}$] knowledge about what affects the energy consumption of software;
    \item [$\mathbf{O2}$] skills in measuring the energy consumption of software.
\end{enumerate}

\subsection{Theoretical Lecture}
A great way of demonstrating the importance of keeping energy efficiency in mind when developing software is to look at the energy consumption trends of the ICT sector, of which software is a very large part. Thus, showing the participants this sector's past consumption and forecasts for the near future is the first matter addressed in the lecture.

In order to successfully design and develop energy-efficient software, it is crucial to possess more than just a basic familiarity with energy consumption and efficiency concepts. It is essential to have a deep and comprehensive understanding of the various factors that influence consumption. Furthermore, one must be well-versed in the techniques and tools that are at our disposal for accurately measuring and optimizing energy efficiency in software development.

In software applications, energy consumption refers to the amount of energy the application consumes during runtime. Regarding energy efficiency, it is defined in~\cite{energy_efficiency_2} as "(...) using less energy for the same output or producing more with the same energy input (...)". Ultimately energy efficiency is the ratio between energy consumed and outputs produced, as it can be improved by reducing consumption and maintaining the output, maintaining consumption and increasing the output, or simultaneously reducing consumption and increasing the output.

\begin{figure}
\includegraphics[bb=0 0 600 400, scale=0.45]{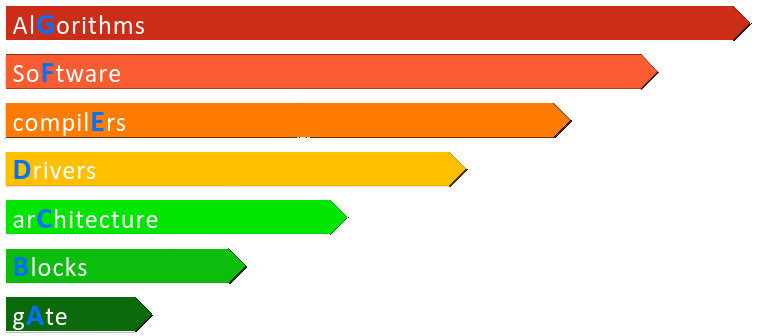}
\caption{Computer abstraction, adapted from~\cite{diagram-eder}}
\label{fig:stack}
\end{figure}

To better understand how different factors influence the energy consumption of software we can abstract a computer into 7 layers, presented in Figure~\ref{fig:stack}, providing us with a framework to understand the different levels of software and hardware involved in computing. At the top of the stack, algorithms represent the highest level of abstraction and are the most energy-agnostic. As we move down the stack, we encounter more specific and concrete layers, which can be optimized with some effort. And at the bottom of the stack are the most hardware-specific layers where energy efficiency can be most effectively optimized. Understanding this stack, and relating it to the factors identified as influential in the energy consumption in~\cite{cloud_survey}, can help identify opportunities to improve efficiency at various system levels.

The participants will be introduced to energy measurement and modeling, two different approaches to studying the energy consumption of software, each with its advantages and limitations. While energy measuring is considered the ground truth, as it provides accurate and precise measurements, it requires specialized hardware which may be expensive to acquire.
On the other hand, energy modeling approaches estimate energy consumption, resulting in values that may not be totally accurate but instead estimates that are relatively accurate, i.e., the relative distance between two measurements is correct. Still, the actual values may differ from the ground truth. Energy estimators are a more cost-effective approach that does not require specialized hardware.

The lecture will introduce Intel's Running Average Power Limit, or RAPL, an energy estimator that has proven itself to be reliable and accurate~\cite{pl:rank2017,tec:rapl1,tec:rapl2}, and has been used in multiple studies~\cite{pl:rank2017,ds:haskell,pereira2017sle,DBLP:journals/scp/PereiraCRRCFS21,Pereira:2016:IJC:2896967.2896968}. However, this tool has its limitations and the authors of~\cite{rapl_in_action} do a great job of pointing them out and presenting mitigation strategies when possible. Some of these include a lack of granularity, register overflows, and non-atomic updates. Therefore, it is important to be aware of some good practices when taking measurements, like disabling non-required services or WiFi connection, to reduce the noise in the system. Alternatively, one could use other tools, such as pTop~\cite{tec:ptop} or PowerTop\footnote{\url{https://github.com/fenrus75/powertop}, accessed 13/06/2023}, which overcome some of RAPL's limitations even though they might introduce some of their own. Nonetheless, for this tutorial, an overview of RAPL will be provided, so that the participants may realize their own experiments in the hands-on session.

\subsection{Practical Lecture}
The purpose of the hands-on lecture is to provide participants with an opportunity to put their recently acquired knowledge into practice. Each participant will be furnished with a laptop equipped with all the necessary materials and software for seamless execution.

The initial task involves two parts. Firstly, participants need to choose between programs with diverse functionality, such as sorting, provided in various programming languages that encompass different programming paradigms, including functional, object-oriented, and imperative. This allows participants to select programs and languages they are familiar with. Secondly, participants are required to utilize RAPL to measure the energy consumption of these selected programs. Once completed, the participants will have their first experience in gathering information about energy consumption.

In the second task, participants will be required to modify the chosen program while preserving its functionality, aiming to improve performance or make changes to implementation details. Following these modifications, participants will once again measure energy consumption. The number of measurements to be performed is up to the participant. Still, they should collect enough samples to assess whether or not the modifications influenced energy consumption.

\subsection{Reaching the Learning Outcomes}
To ensure that the participants acquire the learning outcomes of this tutorial, we have designed a comprehensive program that integrates theory and practice.

Regarding $\mathbf{O1}$, the theoretical lecture will introduce the key concepts of energy consumption in the software context, as well as some factors that affect energy consumption. To consolidate these concepts, the participants will be given the opportunity to experiment with software and modify them to observe the resulting energy consumption, on the hands-on laboratory session. This will enable the participants to develop a deeper understanding of how the various factors affect energy consumption in software.

When it comes to $\mathbf{O2}$, the participants will learn the difference between energy modeling and measuring, including the advantages and drawbacks of each method, during the theoretical lecture. Furthermore, they will also be introduced to RAPL and good practices for using tools such as this. In the practical session, participants will be required to measure the energy consumption of some programs, putting into practice what they have learned in the theoretical lecture.

\section{Conclusion}

This paper describes a summer school tutorial that aims to promote sustainability in software development and engineering. The tutorial provides participants with the knowledge and skills to optimize software with energy consumption in mind. Through a theoretical lecture and a hands-on laboratory session, participants will gain an understanding of the factors affecting energy consumption in software and how to measure it effectively.

By the end of the tutorial, participants will have acquired knowledge about what affects the energy consumption of software and the skills needed to measure and optimize it. This tutorial has the potential to make a significant impact in promoting sustainability in the field of software development and engineering. By equipping junior software engineers with the tools they need to make informed decisions about energy consumption, we can contribute to a more sustainable future.

\section*{Acknowledgements}

This work acknowledges the support of the ERASMUS+ project “SusTrainable —
Promoting Sustainability as a Fundamental Driver in Software Development
Training and Education”, no. 2020–1–PT01–KA203–078646.

%
%
%
%


%
\title{Sustainable and adaptive structures in networked software systems}
%
%
\author{Tihana Galinac Grbac\orcidID{0000-0002-4351-4082}
\and
Neven Grbac\orcidID{0000-0001-6657-6297}
}
\authorrunning{T. Galinac Grbac and N. Grbac}
%
\institute{Juraj Dobrila Univeristy of Pula,
Zagreba\v{c}ka 30, HR-52100 Pula, Croatia \\
\email{\{tgalinac,neven.grbac\}@unipu.hr}
}
\maketitle              
\begin{abstract}
Sustainability is a globally shared goal and is used to drive technology evolution for the prosperity of people and the planet in the future. Software is central to modern technology solutions. The way how we develop, design, and deliver software solutions significantly impacts technology operation sustainability. Therefore, in this lecture, we aim to describe the challenges to reaching sustainable goals while developing, designing, and delivering software in the new network architecture, and explain the relationship between software autonomy and technology sustainability. In the central focus of the lecture are design principles for building autonomous software architectures as a key driver for sustainable technology evolution. Finally, we discuss practical examples and software engineering practices in relation to these challenges. Note that our particular interests are related to software delivered in Clouds, as blockchains of microservices within future networks.

\end{abstract}

\section{Introduction}
\label{sect:Intro}

Software is interconnected and highly integrated within telecommunication networks and is viewed as a key enabler for future digitalization, which is one of the main prerequisites for achieving sustainable goals in various domains (e.g. e-health, e-government, smart water, smart city, etc) \cite{DigitSust}. Future networks are preparing for autonomous machines (cars, drones, robots) and their continuous communication, in order to accommodate the continuous flow of massive amounts of data and their use for effective management solutions \cite{Ericsson5G}. Furthermore, there is a progressive development of applications that can make decisions with the help of artificial intelligence models. The idea is to converge to a global and highly interconnected network of things that continuously share massive amounts of data and autonomously adapt to environmental conditions \cite{AInetwork}. However, future networks should be designed not only to meet technical requirements for meeting people's purposes and achieving economic goals in various domains, but network designs should also look beyond these technical issues to foresee their impacts on environmental conditions aiming to co-exist on Earth over a long time.

Further evolution of information-communication technologies (ICT) should secure its sustainable operation in every domain of application. Network, as a central part of modern ICT technologies, is evolving by developing network autonomy in which sustainable network designs would be based on dynamic network adaptations fulfilling the aforementioned goals. In our previous lectures \cite{GalinacRole1,GalinacRole2}, we presented the main enabler technologies for network autonomy and its sustainable behavior, such as software-defined networking and virtual network functions. In order to achieve its full potential of optimized infrastructure and time-effective response to customer, networks have implemented these main architectural elements as enablers for future networks. The network has prepared its technology for rising its level of autonomy. More precisely, network autonomy rises with the wide implementation of self–management functions such as self–configuration, self–healing, self–optimization, self–protection, \cite{Autonomicbook}. Recent research trends focus on developing adequate software algorithms and solutions to support the aforementioned self-management functions, \cite{AInetwork,AICloudRM}. In this lecture, we will introduce key design principles that are necessary to guide the designs of self-management functions.

Evolution is driven by the idea to standardize and automatize processes within the systems so that we can systematically monitor and measure their behavior and minimize human intervention. One of the main goals of autonomic systems is to manage system complexity and introduce the system's self-organized capability. For such a purpose IBM has introduced a simple control loop frequently referred as Monitor, Analyze, Plan, and Control, MAPE, \cite{IBMMAPE}. The MAPE loop model assumes the existence of sensors within the system that can measure operations and the effectors that can issue system commands on the system's managed entity. This sequential autonomic control loop makes two key assumptions. Firstly, it assumes the existence of directly controlled system entities, and secondly, it assumes the standardized interfaces for system entity monitor and control. However, in most technical systems these conditions are not satisfied. Nowadays, software is often delivered and deployed over the underlying network infrastructure, in Cloud environments and it is often impossible to achieve direct monitoring and control of particular system entities. There is currently a vast amount of different software engineering frameworks and technologies that support developers in their system engineering activities to deliver software products within Network and Cloud environments (e.g. various Web application development frameworks like for example React, Angular, Vue, etc.). Software products are concurrently assessing the shared pool of network and Cloud resources.  The main problems arise with performance aspects under their concurrent operation when software products are offered to the numerous end users,\cite{PerfomanceReact,PerformanceWeb}. In such environments, the software product business goals are balanced between the number of users concurrently using these services and service performances achieved during concurrent software access to shared resources. Thus, in order to optimize their business goals, some in-service monitor and measure solutions are usually proposed, \cite{Webinserviceperf}. However, it is often impossible to implement in-service control of the underlying shared resource pool. From the network perspective, the coexistence of a variety of these technologies in the runtime environment makes significant operational problems in terms of their autonomic and sustainable management. It becomes challenging to automatically manage numerous technologies and find the balance between network and service layer self-adaptive behaviors. One of the main problems for future network evolution would be to find new architectural models, in which compromise between network and service operation would be achieved by satisfying their combined sustainable behavior. In the next decade, our goal should be to move from the \emph{Cloud computing} paradigm to the \emph{Responsive computing} paradigm, in which we should develop software engineering skills and tools that would enable engineering of sustainable technology systems.

One of the main problems is that we still engineer these systems from the technological vision perspective and not from the operational behavior, reliability, or sustainability perspective. We still lack software engineering technology for engineering these so-called *ility attributes into our systems. The same is valid for recently identified vital software characteristic that is related to its sustainable operation. Software architecture is the main artifact we engineer during the software development process and the results of our engineering approach is reflecting the operational characteristics of our final software product. For example, the way how we design software parts and their interactions (that we refer to as system structure design) has a significant impact on system performance, reliability, security, etc. Our approach is to reverse the traditional top-down deductive systems engineering approach into a bottom-up inductive way, and towards framing inside-out design patterns \cite{Design4finSus}. We would like here to bring attention to considering autonomic networking design principles when studying software structure designs for software sustainable operation. Architecture designs should introduce light system simulations that would observe system behavior through a sequence of accumulating events and concurrent executions. Moreover, designing systems architecture should aim to implement mechanisms that would be able to adapt at runtime. The essence of sustainability design is to prioritize internal needs over external needs while keeping the system execution within its limits. Networking design principles are biologically inspired principles that try to translate these nature's adaptivity driven by the instinct to survive into the network. Here sustainable skills would refer to students' abilities of systems thinking in these adaptive designs.

In this lecture, we will teach students about design principles as key skills that would be needed to shape future network and software delivery architectures. The lectures would be supported by the tools that we developed to enable the design of sustainable software structures. Therefore, in the next section Sect.~\ref{sect:dp}, we introduce the main sustainable design principles that are used to implement autonomic network behavior. In Sect.~\ref{sect:ss} we explain our approach to teaching and promoting sustainable design principles within software architecture design.

\section{Design principles for autonomic systems}
\label{sect:dp}

As a starting point of discussion, we need to clarify the basic terms and try to make clear distinctions among them. There are three frequently mixed terms: automatic, autonomous, and autonomic, \cite{Autonomicbook}.

The term automatic refers to an action that is a spontaneous reaction of the system not related to any guided rule but some stimulants are always resulting in the same automatically produced action. From the computer science and programming perspective, we can interpret this term in the sense of functional programming in which we want that functions are consistent and always return the same result on the given input.

An autonomic system is defined as a system that exhibits some degree of self-governance, however, the self--governance is achieved solely from the system's inside processes and has no relation to outside stimuli. The concept of the autonomic system is reused from biology (e.g. autonomic nervous system) and is representing only a part of the nervous system that is responsible for the control of spontaneous or autonomic functions of the human body that are preconditions to survive, like breathing, heart beating etc. These processes are executed outside the human conscience and represent continuous background human activity. It is important to understand that the human body spends a lot of energy to execute solely these processes, without any additional activity.

The autonomous system is viewed as a self-governed system based on some internally system-defined policies and principles. Moreover, this system is self-controlled and functionality independent of the outside world. Note that this concept assumes the system sensing ability which enables the autonomous system to develop its own knowledge and thus make further decisions and make control over the system. All living systems have some autonomous behavior and inherently built-in mechanisms to support such autonomous behavior.

The aforementioned concept of the autonomic system is derived into the communication network context with the help of autonomic design principles. These autonomic design principles have been widely studied while planning the network evolution and as a core philosophy for developing new network architectures with high network management autonomy. Although there are numerous viewpoints on sustainability principles, here we will reflect on sustainable principles for autonomic network behavior. The following principles were recognized as valid in networks context \cite{Autonomicbook} that we will briefly discuss in our lecture.
\begin{itemize}
    \item Living systems inspired design
    \item Policy-based design
    \item Context awareness design
    \item Self–similarity
    \item Adaptive design
    \item Knowledge-based design
\end{itemize}

\subsection{Living system inspired design}
\label{subsect:livingsys}
Software design should follow the behavior of living systems. All living systems exhibit high levels of autonomy and the designs of systems we engineer can be highly inspired from these biological examples. In particular, there are two perspectives to explore in the context of living systems. These are survivability and collective behavior.

Observation of the instinct to survive in the living systems has always concluded with the fact that living systems work toward keeping the equilibrium state and any deviations caused by environmental conditions are forcing the system to bring back to this initial system state \cite{LivingSyst}. There are numerous adaptation mechanisms of living systems that can be reused also for human engineered systems such as software architecture.

The other interesting mechanism of living systems is their collective behavior, which is about system social characteristics while aiming to adapt to a group to which it belongs. There are some interesting studies on the influence of networking technologies like Facebook on social collective behavior, \cite{Loc2glob}. It is interesting how some local information has gained significant importance from the numerous local pieces of information and taken dominance in guiding system global behavior. Some of these concepts may be useful while engineering sustainable software system structures.

\subsection{Policy based design}
\label{subsect:policy}
Policy-based design means that there exists a predefined rule that governs the system's behavior. This design principle has been already used in software system design in which system behavior must adhere to different rules of behavior and these behaviors are specified during the system design phase. Then, based on some parameter settings within the system, the system behavior is chosen at system runtime.

This approach has two drawbacks. Firstly, it requires policy definition during the design time, thus limiting system dynamic adaptation to environmental conditions. Furthermore, we have to secure adequate conflict resolution mechanisms that may be needed in the intersection of system behaviors driven by different policies. One example of such design is software that can serve the same purpose in different conditions, i.e., some standard protocol implementation that has variant implementations in different markets and for these markets, some standard protocol behavior involves some specific feature.

\subsection{Context awareness design}
\label{subsect:context}
Context-awareness has been already used within computer science and is a concept related to the ability of the system to characterize the environment and adapt its behavior to the current situation parameters and knowledge gained from its historic behavior under the same conditions. The main challenge in applying this design principle is in the selection of appropriate variables that may be measured, and sampling strategies to adequately model and capture all relevant states and their interaction in order to effectively and efficiently recognize relevant environment and situation conditions.

\subsection{Self–similarity}
\label{subsect:similarity}
Self-similarity refers to the similarity of system organization on different scales. More precisely, the self–similarity design principle is related to a characteristic that system organization persists as the various system scales and thus guarantees its global properties. Here, it is the main point to develop system functions that preserve system scalability. The same functions may be used as the system is growing in hierarchy and functionality. Thus, further system evolution may be systematically enabled by a proper system architecture that minimizes the need for system change during the system evolution. Moreover, the system properties and system behavior at a large scale have to be the same as the system properties and system behavior at a low scale.

\subsection{Adaptive design}
\label{subsect:adaptive}
Adaptive design is related to the ability of the system to adapt its inner behavior as a reaction to various environmental conditions. Such a system is able to learn from its experience in operation and react accordingly by adapting its actions based on collected information and knowledge gained. Over time the system acquires new knowledge and is better adapting to the local environmental conditions of its operation.

\subsection{Knowledge-based design}
\label{subsect:knowledge}

Design based on the knowledge extracted from big data gathered from the complex system (using AI and other models) differs from the previous adaptive design concept in terms that it assumes some global data collection and artificial global intelligence that can be used to guide local decisions.

\section{Content of the summer school lecture and educational goals}
\label{sect:ss}

Education in software architecture should prepare students for their profession in a new sustainable world. One of the primary steps is to integrate sustainability principles into architecture design courses.

We will formally start the lecture by revising traditional system design skills, and traditional teaching approaches. Here we will introduce some basic definitions. System design is concerned with system decomposition into its components. The design decisions are fundamental for successful implementation and evolution of the system, \cite{Vliet}. The main output of the design phase is the software structure that is defined as a set of components and their mutual dependencies, \cite{Vliet}. The system structure is usually depicted by the graph that represents 'the uses' relation among the system components, which is called the call graph. During the system design phase, there are no clear rules on how to perform successful system designs. The system design rather offers design principles that we reviewed in \cite{GalinacRole1} such as abstraction, modularity, information hiding, layering and hierarchy. Traditional designing system architecture involves exploring various system structure designs in terms of requirements defined in the requirements phase. However, here we aim to rethink this approach and explore widening this narrow design principle list with design principles defined in the autonomous systems concept and explained in Sect.~\ref{sect:dp}. We will face here two challenging issues.

Firstly, we need to introduce novel learning approaches to engage students' abstraction abilities. Some studies \cite{RoleofFluid} have identified that fluid intelligence has a significant role in biological phenomena abstraction and that understanding of bioinspired design is an essential part for design creativity.

Secondly, we promote here a system thinking design approach \cite{systdynamics} that is reflecting on migration from a top-down system design analysis to a bottom-up approach, in which the system structure is examined from the system dynamics perspective. The system designer has to take here a more active role, more as a system experimenter than just a passive critical structure analyst's observer. In such a changed role, the system designer should develop skills that are related to the following: modeling of system simulation by formulating simulation formulas and models, coping with the understanding of system behavior, evaluation of system policies by the development of a control group, development and selection of appropriate treatment cases, establishing a hypothesis, ability to monitor and compare the results for different treatments, ability to judge the results, develop models that may represent intervention actions. Software structure designs are usually governed by the technical vision and usually rely solely on human unreliable intuition. In this lecture, we will try to present our ideas on a simple toy example by using popular among the Academy Python technology.

At the summer school, we plan to structure lectures as follows:

\begin{itemize}
    \item Introduction to System Thinking and System Dynamics
    \item Software as a concurrent system: Python skeleton
    \item Simulating software behavior
    \item Development of simulation scripts
    \item Observing system dynamics within system boundaries
    \item Python GUI for user monitor
    \item Autonomic system design principles
    \item Implementation of the student version of sustainable design principle
\end{itemize}
Firstly, we plan to explain the meaning of control MAPE loop and present its implementation within the Python environment. Some implementations of MAPE loop in Python environment are available publicly\footnote{\texttt{https://github.com/elbowz/PyMAPE}}. We will demonstrate how to implement Python functions for monitoring and control of software behavior by using public Python libraries.

In the second part of the lecture, we will describe system thinking approaches and system dynamics perspective. As a practical example, we plan to ask students to implement a simplistic version of a Python application. Then, we will explain how to implement a Python simulation script that can be used to monitor Python application behavior. Furthermore, we present several practical software implementations of bioinspired design principles and provoke students to play with simulations and observe software behavior. Finally, we motivate students to develop their own solutions, so that students can easily experiment with their versions of implementation of autonomic design principles in a predefined context.

The intended audience for the lectures is undergraduate or graduate students of Computing study programs that have previous experience in Python programming. For active exercising, it is required that students have PC or laptops with installed Python IDE.

The key learning outcomes will be on understanding how to measure software behavior, the ability to develop its own simulation models for capturing software behavior and thus understand how to use a bottom-up approach to system design, and understanding how bioinspired design principles may improve self-system management and thus affects sustainable goals.

\section{Conclusion}
\label{sect:conc}
In this lecture we provide a bottom--up approach to software design integrating system thinking and system dynamics skills into the software engineering curriculum. We focus our system design on the implementation of autonomic system concepts within software applications and provide autonomic design principles to as guidance for students. Throughout the lecture, we undertake an exercise-driven approach to learning autonomic design principles and orientation to system thinking and system dynamics design.

\section*{Acknowledgments}

This paper acknowledges the support of the Erasmus+ Key Action 2 (Strategic partnership for higher education) project No. 2020–1–PT01–KA203–078646: “SusTrainable - Promoting Sustainability as a Fundamental Driver in Software Development Training and Education” and the support of the Croatian Science Foundation under the project HRZZ-IP-2019-04-4216. The information and views set out in this paper are those of the author(s) and do not necessarily reflect the official opinion of the European Union. Neither the European Union institutions and bodies nor any person acting on their behalf may be held responsible for the use which may be made of the information contained therein.


\title{Teaching Sustainable IoT programming}
\author{Pieter Koopman \orcidID{0000-0002-3688-0957} \and
Mart Lubbers \orcidID{0000-0002-4015-4878}}
\institute{Institute for Computing and Information Sciences, Radboud University Nijmegen, The Netherlands\\
\email{\{pieter,mart\}@cs.ru.nl}}
\maketitle
\begin{abstract}
This paper explains how we want to teach sustainable programming for the Internet of Things, IoT, in the upcoming SusTrainable summer school.
Two aspects of sustainable IoT programming are discussed.
There is green computing, limiting the energy consumption of IoT applications.
Another important aspect is the efficient creation and maintenance of such IoT applications.
The lecture will show that both aspects can be served by Task-Oriented programming, TOP.
The students will experience the better maintenance of TOP based IoT applications by extending a given program in the practical session.


\end{abstract}

\section{Introduction}

This paper discusses how to teach two aspects of sustainable Internet of Things, IoT, programming.

The first aspect is known as green computing.
It concerns the energy consumption of the IoT, especially the power used by the edge nodes of the IoT.
This is important since many edge nodes are battery powered, and we want to stretch the interval between battery charges.
Due to the enormous amount of edge nodes,  also reducing the energy consumption of other nodes is worthwhile.
These edge nodes together consume several percents of the worldwide total of electricity.

The second aspect of sustainable IoT computing is the creation and maintenance of IoT system software.
The rapidly increasing number of applications for the IoT makes it important to produce and update this software efficiently and reliable.
However, traditional tiered IoT applications are notoriously hard to produce and maintain because they are built using a layered architecture.
The IoT system consists typically of a user interface in the presentation layer.
This is often a web page, but it can also be an app tailored for smartphones or a wall mounter tablet like devices.
Next, there is an application layer containing web servers, various data storages and collectors to fill these storages with fresh data from the edge nodes.
The communication between application layer and the edge nodes is done in the network layer.
Apart from the well-known TCP connections, there are also tailored network services like MQTT~\cite{mqtt} and Protobuf~\cite{protobuf} used.
Finally, the bottom layer contains the edge devices.
These are single board computers, like a Raspberry Pi, or microcontrollers, like an ESP8266, equipped with sensors and actuators to measure and control phenomena in the real world.

Each of the subsystems in an IoT application is programmed in its own programming language.
By various communication protocols, these subsystems co-operate to establish the goals of the IoT application.
This makes IoT software a multi-lingual distributed heterogeneous system.
The various languages and protocols used give rise to semantic friction~\cite{ireland_classification_2009}.
There is limited support to ensure that the diverse subsystems cooperate smoothly.
Combined with the physical distribution of the edge nodes, this makes the development and maintenance of IoT applications a challenging and error-prone endeavour.

The lecture shows that Task-Oriented Programming, TOP, provides solutions for those problems.
In the iTask system, a TOP system for distributed interactive applications, the programmer specifies tasks to be done by users and the system itself at a high level of abstraction~\cite{plasmeijer_task-oriented_2012}.
All communication between subsystems, the storage of data, and even a web interface tailored to the task of specific users is generated automatically.
This single program replaces the various components of an IoT system mentioned above.
The iTask system is embedded in the strongly typed functional programming language Clean~\cite{Clean:language}.
The strong type-system prevents run time errors by checking programs at compile time.
Hence, run time type errors cannot occur in iTask programs.

Recent results show that TOP programs for the IoT are considerably shorter and simpler than their traditional counterparts~\cite{LubbersTIOT}.
The TOP programs use less programming languages and paradigms.
Together, this makes those programs easier to develop and maintain.

\section{Green Computing}

To reduce the energy consumption of IoT nodes, several strategies are applied.

The communication between the edge node and the server can consume a significant fraction of the energy consumption.
Edge computing limits this energy consumption by processing as much data as possible on the nodes where it is collected, instead of sending all collected data to the server, processing it there, and communicating the results.

Another strategy to reduce the energy consumption of the IoT is by using energy efficient microcontrollers, like a ESP8266, instead of single board computers, like a Raspberry Pi, to drive the edge nodes.
This reduces the energy consumption of such a device by another of magnitude, but this comes at a price.
The microcontrollers have a very limited amount of memory and processing power.
This often forces us to use domain-specific programming languages, like MicroPython, on these devices.
These nodes typically run either no operating system, or a limited real-time operating system.
This implies that the user program has to take care of the interleaving of multiple subtasks.
The iTask ecosystem offers the mTask language to run tasks on such microcontrollers~\cite{lubbers_writing_2019}.

Most microcontrollers have some sleeping modes on top of the basic low energy consumption.
The system temporarily shuts down part of the system to save energy in such a sleeping mode.
Obvious candidates for parts to switch off are the Wi-Fi radio and peripheral controllers that are not needed for some time.
Even the main processor and the RAM can be put in sleep mode when it is known that nothing has to be done for some time.
The system wakes up and starts working again, and hence consuming energy, after a specified sleeping time, or when an external interrupt wakes the system.
In the 2022 Sustrainable summer school in Rijeka we have shown how the mTask system can achieve such energy savings automatically~\cite{TFP22}.
For the upcoming summer school, some results are reused, but do not require any knowledge of this lecture.

\section{Task-Oriented IoT Programming by Example}

A complete introduction to TOP is outside the scope of this paper.
We illustrate the basic concepts with a few variations of a program reading a temperature sensor and showing this value in a web browser.
This example is concise, but contains all layers of an IoT application mentioned in the introduction.

The examples use a Shared Data Source, SDS, to store the latest temperature.
Such an SDS is a memory location that various tasks can read and update.
Within a single program, this behaves like a global variable.
Between various programs, an SDS is more like a database.
Here, the SDS \prog{tempSDS} contains just a real number to represent the temperature.
But, SDSs can contain values of any first order datatype, including for instance large lists of records.

\subsection{Local Temperature Sensor}

In the first version of our program, all tasks run on the same server.
The sensor is connected directly to this server, the SDS lives here and the web-server also runs on this machine.

\begin{lstlisting}[language=Clean,caption={An iTask program to read a local temperature sesnsor and display the value.},label={lst:itask1}]
tempSDS :: SimpleSDSLens Real
tempSDS = sharedStore "tempSDS" -273.15 // brrr

localSensor :: Task Real
localSensor =
    withDHT TempID \dht ->
    get tempSDS >>- \old ->
        devTask dht old -|| viewSharedInformation [] tempSDS <<@ Label "temperature"

devTask :: DHT Real -> Task Real
devTask dht old =
    temperature dht >>~ \new ->
        if (old <> new)
            (set new tempSDS >-| devTask dht new)
            (waitForTimer False delayTime >-| devTask dht old)

Start world = doTasks localSensor world

delayTime = 5
\end{lstlisting}

The \prog{localSensor} task first creates a sensor object, \prog{dht}, with \prog{withDHT}.
The argument \prog{TempID} tells how to access the physical sensor, we skip details.
Next, the task reads the current value of the SDS by \prog{get}.
The combinator \prog{>>-} bind the result of this task to the lambda function with \prog{old} as argument.
This function uses the combinator \prog{-||} for the parallel composition of the tasks \prog{devTask} and the \prog{viewSharedInformation}.
The view of the SDS is automatically updated when a change occurs.
This task keeps displaying the latest temperature as long as it runs.
The \prog{devTask} controls the sensor.
With \prog{temperature dht} we read the current value of the sensor.
Next, we check if the \prog{new} value is differs from the \prog{old} one.
When they are different, we update the SDS with the new value and call the function recursively.
Otherwise, we wait some time and start this task recursively.

This simple program contains two energy optimizations.
First, we update the SDS only when the temperature is changed.
This prevents that the web page is updated to display the same value with all associated computations and network traffic.
Future versions will execute this computation on the edge node to obtain edge-computation.
Next, the \prog{waitForTimer} tasks impose a delay.
During this time, the task does not require energy.

\subsection{Remote Temperature Sensor}

The temperature sensor is connected to a remote node in a more realistic IoT example.
We assume that the remote machine is sufficiently powerful to run an iTask program and that we have access to it.
A Raspberry Pi is perfectly suited for this job.

Our example program remains largely unchanged.
We only have to indicate which tasks runs on which machines and where the SDS is stored.
We chose to store the SDS on the remote machine.
We replace \prog{localSensor} by \prog{remoteSensor}.
The differences with the previous task is that it asks the user of the program for the address of the remote machine to run \prog{devTask}.
The function \prog{asyncTask} takes care of moving the task to the remote machine.
We tell the \prog{viewSharedInformation} that it has to display the value of this remote SDS instead of a local SDS by \prog{remoteShare}.
All other code is unchanged.

\begin{lstlisting}[language=Clean,caption={The iTask task to read a remote sensor and display its value.},label={lst:itask2}]
remoteSensor :: Task Real
remoteSensor =
    withDHT TempID \dht ->
    get tempSDS >>- \old ->
    enterInformation [] <<@ Title "device" >>? \dev ->
        asyncTask dev.domain dev.port (devTask dht old)
    -|| viewSharedInformation [] (remoteShare tempSDS dev) <<@ Label "temperature"
\end{lstlisting}

In general, it is hard to decide which part of the code is executed where.
The code executed remotely can use all functions and data types specified in the program.
To ensure that the desired code is available, the iTask system makes the entire code available on every machine executing some part of the program.
This holds even for a browser displaying the user interface.

\subsection{Remote Temperature Sensor on Microcontroller}

The introduction explains that microcontrollers are attractive hardware for edge nodes since they consume less energy and are cheaper.
Unfortunately, we cannot execute the code from the previous section on common microcontrollers.
The generated code is just too large and requires too much processing power.

The mTask system allows us to identify what parts of the program must be executed on the microcontroller.
The simple data structures and first-order strict evaluation of mTask ensure that these tasks can be executed on the restricted hardware.
The mTask system is powered by an embedded domain-specific language, DSL, offering task definitions very similar to iTask.
In contrast to standalone DSLs, embedded DSLs are expressed in terms of the host language~\cite{hudak_modular_1998}.
The strong type system of the host language ensures that the mTask parts of a program will co-operate smoothly within iTask programs.

In \prog{mTaskSensor}, the user enters the remote device.
Next, it starts the mTask task on the specified device with \prog{withDevice} in parallel with displaying the value of the temperature SDS.

All code of the \prog{devTask} is compiled dynamically to byte code that is executed by the mTask run time system running on that device.
First, the \prog{liftsds} makes a copy of the \prog{tempSDS} on the device.
The mTask machinery takes care of synchronization between those shares.
The \prog{DHT} definition creates a sensor object.
The \prog{fun} primitive defines a mTask function called \prog{measure}.
This function has the old temperature as its argument.
Just like the iTask version of this task, it reads the temperature sensor and updates the SDS when the \prog{new} value differs from the \prog{old} one.
The \prog{main} expression reads the old temperature from the SDS and calls the \prog{measure} function.

\begin{lstlisting}[language=Clean,caption={An iTask\slash{}mTask program to read a remote sensor and display its value.},label={lst:mtask}]
mTaskSensor :: Task Real
mTaskSensor =
    enterInformation [] <<@ Title "device" >>? \dev ->
        withDevice dev (liftmTask devTask)
    -|| viewSharedInformation [] tempSDS <<@ Label "temperature"

devTask :: v Real | mTask v
devTask =
  liftsds \rSDS -> tempSDS In
  DHT DHT_I2C \dht ->
  fun \measure = (\old ->
    temperature dht >>~. \new ->
    If (new  !=. old) (setSds rSDS new >>|. measure new) (measure old) In
  {main = getSds rSDS >>~. measure}
\end{lstlisting}

The reader might notice that this mTask code does not contain an explicit delay.
The mTask system is equipped with a scheduler that will automatically put the device into sleep mode in orde to save energy when there are no tasks that need execution~\cite{TFP22}.

\section{Sustainable Programming}

To verify that the TOP code is indeed more sustainable than traditional tiered code, we implemented a real world IoT example~\cite{smartSensors} in four ways~\cite{lubbers20tiered,LubbersTIOT}.
For the traditional implementation, we use a Raspberry Pi 3 as edge node and Python, JSON, HTML, PHP, Redis and MongoDB to construct the software.
This is called Python Raspberry Sensor, PRS.
Its direct counterpart uses just Clean and the iTask library and is called Clean Raspberry Sensor, CRS.
We made a variant with a WEMOS D1 mini microcontroller instead of the Raspberry Pi for both implementations.
The microcontroller runs MicroPython in the tiered approach.
This is called Python WEMOS Sensor, PWS.
The microcontroller runs mTask in the tierless TOP approach.
This variant is called Clean WEMOS Sensor, CWS.
Table~\ref{table:t1} give some key figures about these implementations.

\begin{table}[h!]
\centering
\caption{Key figures about the four implementations of the smart sensor.}%
\label{table:t1}
\begin{tabular}{l r r r r}
\toprule
   & PRS & CRS & PWS & CWS \\
 \midrule
 number of languages used & 6 & 1 & 7 & 2 \\
 total SLOC & 576 & 155 & 562 & 155 \\
 files & 38 & 3 & 35 & 3 \\
 memory residence (KiB) & 3557 & 2726 & 20 & 0.9 \\
 power consumption edge node (W) & 1-2 & 1-2 & 0.2 & 0.2 \\
 \bottomrule
\end{tabular}
\end{table}

The number of languages and files used in these TOP variants is significant smaller.
In addition, these are single source files that are checked by the strong static type system of Clean.
The Clean based versions use considerable less Source Lines of Code, SLOC, than the Python based variants.
Checking the details about the source code reveals that the TOP code is in most case 95\% shorter than their tiered counterpart.
The interface to MongoDB is only 26\% smaller and consumes almost half of the TOP code.
We can achieve a huge reduction of code by using a native SDS instead of the external MongoDB.

The code size differs not much by replacing the Raspberry Pi to a WEMOS microcontroller.
However, the restrictions of the microcontroller require an additional programming language and an appropriate distribution of tasks that obeys the limitations of the microcontroller languages.
This might be tricky for complex programs.

The microcontroller based software has the required small memory residence and achieve the desired power saving of about one order of magnitude.

\section{Teaching}

The lecture in the upcoming SusTrainable summer school will tell this story in some more detail.
We will focus on easier maintenance and its support in the TOP approach.
Other lectures in the 2023 summer school will argue the importance of maintenance by showing that up to 90\% of the total project effort is allocated to this phase of the software life cycle.

We will give the students an initial project that works with a microcontroller in the practical part of the tutorial.
The plan is to start with a project like the one in this paper.
Students will experience the support for changing such a project.
This project will be extended in steps by controlling the measurement interval dynamically, adding humidity measurements, storing historical data, and extending the user interface.


\section*{Acknowledgements}
This work acknowledges the support of the ERASMUS\raisebox{.25ex}{+} project ``SusTrainable---Promoting Sustainability as a Fundamental Driver in Software Development Training and Education'', no.\ 2020--1--PT01--KA203--078646.

%

\title{Distributed Systems Architectures with \\Green Labels}
\titlerunning{Distributed Systems Architectures with Green Labels}

\author{Jianhao Li \and Vikt\'oria Zs\'ok}

\authorrunning{J. Li \and V. Zs\'ok }

\institute{E\"otv\"os Lor\'and University, Faculty of Informatics\\
Department of Programming Languages and Compilers\\
H-1117 Budapest, P\'azm\'any P\'eter s\'et\'any 1/C., Hungary\\
\email{lijianhao288@hotmail.com, zsv@inf.elte.hu}\\
}
\maketitle
\begin{abstract}
Communication middleware is critical in distributed systems. However, many existing distributed middleware only focus on performance. This tutorial concentrates on distributed messaging middleware that balances sustainability, scalability, and performance. Through comparison, let students understand the energy-saving benefits each design detail can bring. Additionally, through the analysis of the overall architecture of the middleware, students will have a deeper understanding of sustainable distributed system architecture. The goal is to give students the ability and awareness to consider sustainability and energy efficiency when developing distributed systems.

\end{abstract}

\section{Introduction}
Distributed systems are increasingly widespread in our daily lives and play an important role.
2 \% of the global CO2 emissions are generated by the ICT (Information and Communication Technology) infrastructure~\cite{en10101470}. Therefore, besides the energy efficiency optimization of the hardware, more and more researchers realized that reducing software energy consumption (SEC) is also essential to improve the sustainability of distributed systems of various data centers.

The energy-aware programmers can use energy analysis and modeling techniques with software engineering tools to reduce the energy consumption of code~\cite{Gallagher17}. The measurement of the energy consumption of software is essential when we teach about green software implementation~\cite{GreenCurriculum}. However, finding the best measurement tool for the distributed system in \textsc{Go} is not easy. Hardware energy measurement tools are very precise and may involve an additional financial cost, while many studies have shown that software energy measurements are unstable~\cite{phdthesis}. The on-chip power sensor is a sub-category of hardware measurement tools that generally does not require extra boards or devices. The RAPL (Running Average Power Limit) is an on-chip power sensor that provides power-limiting features and accurate energy readings for CPUs and DRAM~\cite{rapl1}.

Many important energy consumption studies have been conducted with the help of RAPL~\cite{EnergyWar,PEREIRA2021102609} and Joulemeter~\cite{joulemeterR,GreenRefactorJoulemeterJ}. At the summer school, we will introduce the best practices of using RAPL to measure the energy consumption of distributed systems in \textsc{Go}, and we will also evaluate measurements of the distributed communication system.

The power model using CPU utilization, as the primary signal of machine-level activity, tracks the dynamic power usage behavior extremely well~\cite{cpuu}.
There are many built-in profilers of \textsc{Go} runtime. For example, the CPU profiler~\cite{GolangPerf} shows which functions consume what percent of CPU time. However, it is a software measurement tool that may need to be more stable and accurate, maybe it is only suitable for performance profiling. Therefore, we need to assess its usage in green computing, i.e. can it provide valuable information to locate the energy consumption problems of a program?

The software energy consumption can be reduced at different levels: execution environment level and code level~\cite{phdthesis}. At the code level, code refactoring can significantly impact the energy usage of an application~\cite{coder1,coder2}, so developing energy-efficient software through code refactoring is meaningful~\cite{refactor1}. Therefore, the energy consumption reduction of our distributed communication project starts with sustainable code refactoring.

The energy consumption of small code snippets of \textsc{Java}~\cite{url_Java} using RAPL is reported in~\cite{javaCode1}. It can guide the developer in building energy-efficient software in \textsc{Java}. The book~\cite{refactorBook} also guides the refactoring of object-oriented code. \textsc{Go} is a compiled programming language with efficient built-in concurrency constructs~\cite{url_Golang}. Based on the studies of refactoring, we research the energy consumption of the \textsc{Go} distributed communication system from the code refactoring point of view by evaluating the code snippets summarized from different implementation decisions. We will also evaluate the energy consumption of the essential elements of \textsc{Go}, such as channel, map, slice, and mutex.

There are general energy efficiency strategies that guide green computing implementations~\cite{Conceptual}. However, more practical details and suggestions are needed, which we will provide during the school to acquire green computing in \textsc{Go} distributed system.

Beside the sustainable code refactoring, we also study the sustainability of the distributed systems architecture. The author of~\cite{EnterArch} doubts the transparency of distributed objects based on performance considerations.

The Erlang distributed communication system achieved location transparency by unifying local and remote communication interfaces~\cite{url_Erlang}. It is convenient for the user that does not need to think about whether the destination actor is in the same node or a remote node when sending a message. However, we think location transparency is not a proper sustainability consideration because the users cannot intentionally reduce the number or size of the remote messages if they cannot tell the difference between local and remote sending. We suggest implementing different interfaces for local (in the same node) and remote communications (among different nodes) for sustainability so that users can make sustainable design choices. For example, in case of the \textsc{Go} programming language, the built-in channel construct deals with local communication among different lightweight threads, while other distributed communication systems focus on providing distributed remote communication interfaces.

In the lecture, we will share with students our considerations and measurements of sustainability when designing a distributed system.
We will split different parts of the implementation design of Uactor, such as various group joining algorithms, abstract them into standalone programs, and attempt to make more sustainable design choices for distributed systems by measuring the energy consumption of programs representing different design decisions. We refer to these programs that abstract the energy consumption of distributed systems as "distributed system energy evaluation abstraction programs", which have simplified the distributed system design implementations, not encompassing all the details but reflecting the overall design approach and highlighting some energy-relevant aspects. Then, a comparative analysis with controlled variables will be conducted. Ultimately, after the initial design, before the formal implementation, we can consider which design approach may be more environmentally friendly.

The accuracy of energy consumption evaluation for different designs mainly depends on the design of the abstract evaluation programs, which should contain certain features that can manifest the energy consumption differences among different design decisions or algorithms. We will discuss some experiences designing abstract energy evaluation programs with students in the course.

\section{Uactor project}

The Uactor is a distributed communication system that we are currently developing. We have presented and submitted a tutorial about it at the SusTrainable Summer School 2022~\cite{SummerSchool22}. It integrates our research on distributed systems, including exploring distributed communication models and the coordination of underlying protocols, investigating scalable peer-to-peer networks, and analyzing novel distributed group membership algorithms. Currently, we are designing this project's group membership protocol and connection mechanisms under sustainability and scalability considerations.

Uactor is a connectionless and brokerless distributed messaging middleware that balances sustainability, scalability, and performance. It follows a modified actor model and provides group communication service. The actors are organized in a dynamic overlay of connected non-hierarchical structured overlays. Additionally, this model has unified communication constructs, which means we can effortlessly use different communication mechanisms without connecting different constructs.

We are further improving this project's scalability, so we must redesign its group service mechanisms and architecture. In addition, we will design the architecture of this project with more sustainability features in the next versions. Finally, the practical green design and implementation experience will be shared with the students.

\section{Course information}

This course guides the students in implementing sustainable distributed systems by teaching energy consumption measurement and comparing the energy efficiency of different implementation code snippets. Additionally, we provide sustainable distributed system architecture suggestions and introduce sustainable distributed communication middleware.

Here are some prerequisites for students to understand the course properly: Go programming language, TLA+, and RaPL.

The students should have general knowledge of the essential elements of the programming language \textsc{Go}: variable, functions, if statement, for loop, for range loop, slice, struct, pointer, method, map, goroutine, wait group, mutex, select, and channel~\cite{goBook}.

TLA+ can assist system architects in modeling the designs or algorithms of distributed systems~\cite{tla}. To describe the design more explicitly, we will summarize a part of the Uactor design using TLA+ specification.

The lecture will first introduce how to measure the energy consumption of \textsc{Go} programs. For example, we teach students how to configure RAPL in the distributed system and evaluate the energy efficiency of their code. The built-in profilers of \textsc{Go} runtime will be used in cases where they provide essential information about energy consumption, such as locating the functions that cause the energy leakage.

After that, we will introduce the energy efficiency suggestions for a distributed system in \textsc{Go} and guide the students to compare the energy efficiency of different implementing decisions. Many decisions in distributed systems are summarized into small code snippets that are easier to understand and test.

Finally, we will introduce the green design of the Uactor project and provide suggestions for designing the green architecture of distributed systems.
The students can modify the prepared \textsc{Go} programs at the practice session according to the green computing suggestions and compare the energy consumption.

After this course, the students will gain more experience in energy consumption measurement and implementing energy-efficient distributed systems in \textsc{Go}. In addition, they will have an insight into the sustainable design of the distributed communication system.

\section*{Acknowledgements}
This  paper  acknowledges  the  support  of  the  Erasmus+ Key Action 2 (Strategic partnership for  higher education) project $No$ 2020-1-PT01-KA203-078646: “SusTrainable – Promoting Sustainability as a Fundamental Driver in Software Development Training and Education”. The information and views set out in this paper are those of the authors and do not necessarily reflect the official opinion of the European Union. Neither European Union institutions and bodies, nor any person acting on their behalf may be held responsible for the use which maybe made of the information contained therein.



%
\title{Efficient evolutionary computing}
%
%
\author{Goran Mau\v{s}a\inst{1,2}\orcidID{0000-0002-0643-4577}}
\authorrunning{G. Mau\v{s}a}
%
\institute{University of Rijeka, Faculty of Engineering, Vukovarska 58, 51000 Rijeka, Croatia \and
University of Rijeka, Center for Artificial Intelligence and Cybersecurity, Radmile Matejčić 2, 51000 Rijeka, Croatia \\
\email{gmausa@riteh.hr}\\
}
\maketitle              
\begin{abstract}

In computer science, evolutionary computing is a family of nature-inspired algorithms for solving complex search-based and optimization problems.
The idea of evolutionary computing is to find the best solution to a given problem in a smaller number of steps than traditional and computationally demanding approaches like exhaustive or grid search.
This tutorial opens the question whether this strategy can be made even greener and analyzes this issue through hyper-parameter tuning, selecting the appropriate optimization algorithm and programming language, building surrogate models and neuroevolution.
The goal is to present theoretical background of evolutionary computing and genetic algorithm, provide a practical exercise as a showcase of its potential and to demonstrate its usage in a sustainable manner.

\end{abstract}

\section{Introduction}

Evolutionary computation is a sub-field of artificial intelligence (AI) and it is used extensively in complex combinatorial problems and for continuous optimization.
In technical terms, it is a family of population-based trial and error problem solvers with a metaheuristic and stochastic character.
Although these algorithms can be used to solve modelling and simulation problems, their main application is optimization~\cite{eiben2015introduction}.

Evolutionary computation is used to solve problems that have too many variables for traditional algorithms and when optimal solution cannot be derived in (desirable) polynomial time \cite{borah2022applied}.
Although inherently efficient in solving complex problems, there are several efficiency enhancement techniques: parallelization, hybridization, time continuation, and evaluation relaxation~\cite{goldberg2002design}.
The task of parallelization is to distribute the computational load among processor units, hybridization combines local and global search techniques to reach a ballance between exploration and exploitation~\cite{sinha2005designing}, time continuation exploits the trade-off between the population size and the number of convergence epochs~\cite{srivastava2002time}, while evaluation relaxation promotes substitution of computationally expensive fitness with inexpensive approximation, also know as the surrogate model~\cite{smith1995fitness}.

In this tutorial, the aim is to provide the means of making evolutionary computing more sustainable, present a set of case studies designed to improve their energy efficiency and give a practical session for the second Summer School organized with the SusTrainable project.

\section{Evolutionary algorithms}

Darwinian's theory of evolution, governed by \textit{the survival of the fittest} principle, inspired automated problem solving family of algorithms that form evolutionary computing~\cite{fogel1998evolutionary}.
The most popular representative is the Genetic Algorithm (GA), named by Holland in the '70s~\cite{holland1973genetic}.
Individuals in a population, whose size is constrained by the limited amount of available resources, reproduce to prolong their species and the ones that are better adapted to environmental conditions have greater chances for survival.
In GA, individuals represent candidate solutions to a given problem, their genomes are sets of characteristics, i.e. parameters that define them and their quality is defined by a fitness function that plays the role of \textit{adaptation to the environment}~\cite{eiben2015introduction}.

\begin{figure}
\includegraphics[bb=0 0 1400 400, width=\textwidth]{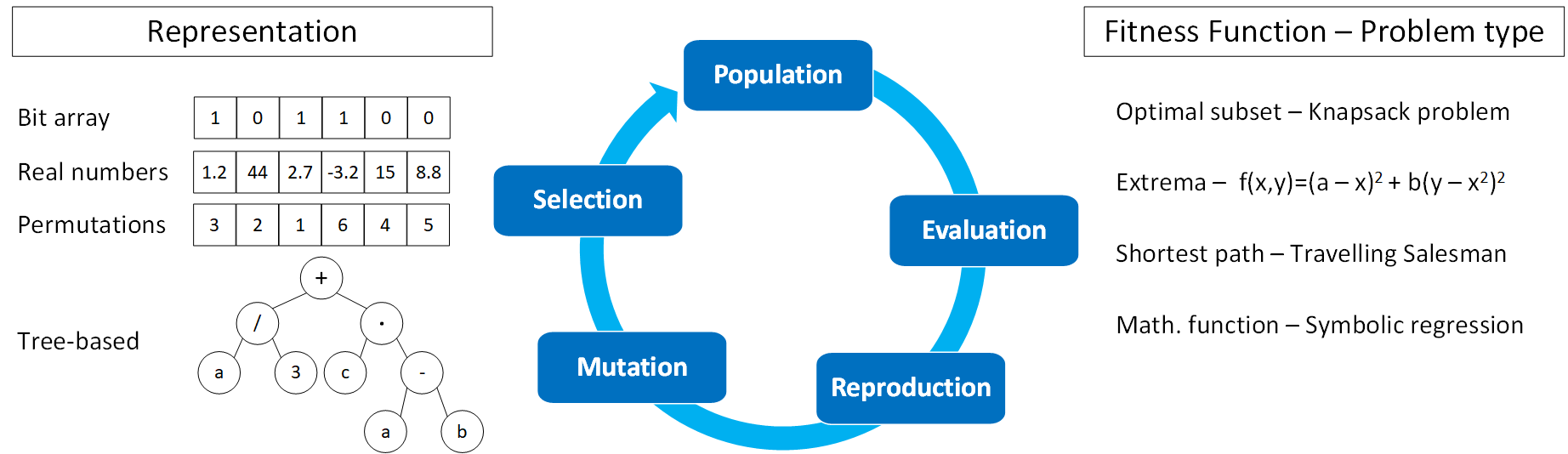}
\caption{Schematic representation of the main components and workflow for Genetic Algorithm, together with usual representation types accompanied by examples of fitness functions from classical deductive problem types.} \label{fig1:evolutionary}
\end{figure}

The main requirements to solve an optimization problem using a GA is to have an adequate \textbf{representation} for candidate solutions and a \textbf{fitness function} that can numerically evaluate their quality.
The usual representation types, given in Figure \ref{fig1:evolutionary}, include bit string for binary variables, vector of real numbers for continuous variables, permutations for arranging the order of events and tree-based structures for creating programs, mathematical functions or similar more complex entities.
Each of the classical deductive optimization tasks like knapsack problem, finding minima of a complex mathematical function, travelling salesman and symbolic regression require one of these representation types.
In knapsack problem we are selecting the best subset of items to maximize their weight without exceeding its carrying capacity and the best representation is the bit string.
Finding extrema of a mathematical function such as the Rosenbrock function~\cite{rosenbrock1960automatic}, presented in Figure \ref{fig1:evolutionary}, requires the candidate solutions to be expressed through a vector of real numbers.
A travelling salesman needs to find the shortest path to visit a number of predetermined number of places of his route only once and permutations are the obvious choice for describing the candidate solutions.

The schematic overview of GA workflow, as presented in Figure \ref{fig1:evolutionary}, stars from initialization of the population, usually by random generation of genomes.
The quality of each individual within the population is estimated numerically by the fitness function.
Better solutions have greater chances for becoming parents of new individuals produced in the reproduction phase by the operator of genome crossover.
To introduce small random changes in the genes inherited from their parents, a mutation operator is applied in the offspring population.
Final phase of this iterative process is the quality-oriented selection of the individuals that will form the next generation of individuals.
Most of these functions are of stochastic nature, and their overall goal is to gradually drive the evolution in the direction of the best solutions~\cite{eiben2015introduction}.

Drawing inspiration from swarm intelligence, biology, physics, chemistry or even social phenomena and grammar, nowadays there is a great number of evolutionary algorithms~\cite{fister2013brief}.
Several exotic algorithms, like the music-inspired Harmony Search Algorithm, demonstrated the ability to enhance computational efficiency with simpler implementation and a lower number of setting parameters~\cite{geem2009music}.
Although often used as black box solution, understanding the underlying subtleties of evolutionary computing may lead to their cost-effective and more sustainable utilization.

\section{Training Course}

The tutorial will start by introducing the basic concepts of evolutionary computing, explain the main operators and functions within the GA and what is its purpose.
Although GA is known to be a more efficient alternative to computationally demanding approaches like exhaustive or grid search, the tutorial will discuss the strategies which may be implemented to make it even more sustainable.
Besides the obvious hyper-parameter tuning of numerous selection, crossover and mutation operators, we will analyze which is the appropriate choice of programming language for the implementation of GA and compare the popular languages like java, R and python.
We will also compare several exotic optimization algorithms like Firefly, Artificial Bee Colony and Flower Pollination Algorithm in terms of execution time and accuracy of the prediction model after performing feature selection.
To present more elaborate ways of making GA greener, we will introduce the concept of surrogate models, whose aim is to replace expensive fitness function with simpler approximate solutions.
The practical exercise will demonstrate the potential of evolutionary computing applied to evolving neural network (NN)-based classifiers, known as the neuroevolution.

The key take-away messages of this course will be:
\begin{enumerate}
    \item Understanding the fundamental operators of evolutionary computing;
    \item Know-how for estimating and improving the efficiency of GA;
    \item Practical guidelines for developing a sustainable neuroevolution framework.
\end{enumerate}

\subsection{Prerequisite Knowledge and Skills}

This lecture is designed to fit the expertise of a master or PhD student enrolled in the study programme of computer science or related studies.
The intended audience should have theoretical knowledge of algorithms and data structures to follow the training materials on evolutionary computing.
To fully benefit from the training materials related to surrogate modelling and neuroevolution, a basic understanding of machine learning and its application is recommended, along with the following concepts:
\begin{itemize}
    \item Understanding the complexity of neural networks;
    \item Familiarity with prediction model training pipeline;
    \item Evaluation metrics for estimating the classification performance.
\end{itemize}

\subsection{Materials and Methods}

The sustainability aspect of the practical exercise will require energy measurements for executing the optimization of NN architecture and the training of the prediction model.
Running Average Power Limit (RAPL) and its Python version (pyRAPL) offer an estimated energy consumption of CPUs and DRAM~\cite{RAPL2019} with minimal performance overhead~\cite{khan2018rapl} and we will be used it for that purpose.
Energy consumed by the grid search and by the neuroevolution will be compared to demonstrate the benefit of using the evolutionary computing paradigm for the task of evolving low-complexity prediction models.

The software packages and libraries that are going to be needed for the implementation and execution of examples in the practical session are:
\begin{itemize}
    \item tensorflow - model construction and training;
    \item pyRAPL - energy consumption measurement;
    \item scikit-learn - machine learning library containing grid search implementation and various benchmark datasets;
    \item matplotlib - visualizing the results;
    \item numpy - supporting mathematical operations;
    \item pandas - manipulating data;
    \item neat-python - implementation of neuroevolution algorithm in python;
    \item seqprops - encoding for peptides dataset
    \item scikit-bio - various tools for dealing with peptides datasets
    \item scikit-optimize - hyperparameter optimization
    \item scipy - optimization, statistics
    \item dask - parallelization library
\end{itemize}

Datasets available in scikit-learn library, such as California Housing dataset, Diabetes dataset or Iris plants dataset and/or publicly available peptide datasets like DRAMP 2.0~\cite{kang2019dramp}, will be used to demonstrate the principles of training NN-based prediction models.
The goal is to examine the benefit of using neuroevolution for simultaneous arcitecture optimization and training process in terms of time and energy consumption.

\section{Conclusion}

This paper presents a tutorial on basic concepts of evolutionary computing, with the aim to minimize energy consumption when solving optimization tasks.
It is a continuation of our lecture entitled Soft Computing for Sustainability Science, which dealt with the problem of data pre-processing and feature selection in particular, its application on dimensionality reduction and impact on reducing the time and energy when training predictive models.
This tutorial tackles the issue of developing NN-based models, which are emerging as the most popular choice for solving complex classification problems.
A lack of clear guidelines for choosing the appropriate NN architecture impedes their wider applicability and often leads to implementation of computationally demanding optimization procedures like grid-search.
Not only do these methods present an unsustainable means of solving the problem, but they also often result with architectures whose complexity is higher than actually needed, which increases computational cost for training and using them.

Participants of this tutorial will go through a theoretical lecture and a practical exercise, after which they will understand the importance of evolutionary algorithms and gain skills required to train NN-based prediction model in a sustainable manner.
Aligned with the aim of our SusTrainable project, the tutorial strives to provide scientific and technical support in identifying new development approaches, promote sustainability as an important topic in AI and to disseminate these concepts to a wider audience.




\section*{Acknowledgement}

This paper acknowledges the support of the Erasmus+ Key Action 2 (Strategic partnership for higher education) project No. 2020-1-PT01-KA203-078646: “SusTrainable - Promoting Sustainability as a Fundamental Driver in Software Development Training and Education”.

The information and views set out in this paper are those of the author(s) and do not necessarily reflect the official opinion of the European Union. Neither the European Union institutions and bodies nor any person acting on their behalf may be held responsible for the use which may be made of the information contained therein.

%


\title{Energy- and Time-aware Coordination\\ for Multi-core Cyber-physical Systems}
\newcommand{\correspondingauthor}{ (\raisebox{-0.25em}{\Envelope})}
\author{Lukas Miedema\inst{1}\orcidID{0000-0002-7295-6568}
  \and
  Clemens Grelck\inst{1,2}\orcidID{0000-0003-3003-1388}\thanks{Corresponding author}
}
\authorrunning{L.~Miedema, C.~Grelck}
\titlerunning{Energy- and time-aware coordination for multi-core cyber-physical systems}
\institute{University of Amsterdam, Netherlands\\
  \email{\{l.miedema, c.grelck\}@uva.nl}
  \and
  Friedrich Schiller University Jena, Germany\\
  \email{clemens.grelck@uni-jena.de}
}
\maketitle
\begin{abstract}
We present the coordination language TeamPlay for the high-level specification of cyber-physical systems with strong requirements on non-functional properties such as execution time and energy consumption, both as a constraint (deadline, energy budget) and as an optimisation objective (save energy, run quickly).
TeamPlay makes trade-offs between time, energy and (optional) redundancy for fault-tolerance visible in early system design and implementation stages.

In this paper we describe the essential concepts of the TeamPlay coordination language, mostly in an abstract way.
Eventually, we sketch out our programme for the SusTrainable summer school to be held in Coimbra, Portugal, in July 2023.
\end{abstract}

\section{Introduction}
\label{sec:summary}

Cyber-physical systems (CPS) combine hardware/software (cyber) systems with the physical word seen through sensors and impacted through actuators.
In cyber-physical systems the functional correctness of software, thus, is of pa\-ra\-mount importance as erroneous behaviour is not ``just'' an annoyance, but may have disastrous consequences in the physical world including loss of life.
Beyond functional correctness, non-functional properties such as execution time and energy consumption play an equally important role.
Typically, reactions to sensor inputs must occur within some given deadline.
Energy consumption is not only relevant as a general environmental concern, but many cyber-physical devices are in fact battery-driven.
As such energy consumption is in a direct relation to the usefulness of a cyber-physical system.

From a software engineering point of view the question is how confidence in functional correctness, meeting of execution deadlines and energy consumption can all be addressed at an equal level of importance without creating an unmaintainable software mess.
We proposed the coordination language \emph{TeamPlay} \cite{RoedRouxAltm+20} to address these concerns.
TeamPlay embraces the concept of \emph{exogenous} coordination \cite{Arbab98,Arbab04} and organises a concurrent application as a set of independent, identifiable black-box \emph{components} (aka \emph{tasks}) that communicate with each other solely via data-flow channels.
Components are meaningful, sequential application building blocks, implemented in a general-purpose programming language.

TeamPlay enforces a stringent two-layer software architecture where components are simple enough to be implemented correctly for a given input-output relation.
Furthermore, components have a defined dynamic behaviour regarding worst-case/average-case execution time and energy consumption for any given execution platform under consideration.
These data could be obtained through static code analysis or through experimentation, depending on the compute architecture.
A system integrator then takes these (in coding terms) black-box components and synthesise a systems by means of the TeamPlay coordination language.

We primarily target high-performance embedded system architectures, like the Odroid or Jetson families.
These combine heterogeneous multi-core CPUs with DVFS and highly parallel GPUs for bulk computing.
Mapping TeamPlay coordination programs to high-performance embedded systems under time and energy constraints creates a complex scheduling (in time) and mapping (in space) problem that we address with various scheduling/mapping schemes \cite{RoedRouxAltm+20b,RoedRouxGrel21,RoedPimeGrelAIAI23} and a runtime middleware for said high-performance embedded systems \cite{RouxAltmGrel21}.

During the SusTrainable summer school in Coimbra, Portugal, in July 2023 we will motivate the use of exogenous coordination in the field of cyber-physical systems, introduce the essential features of the TeamPlay coordination language and explain the ins and outs of the scheduling and mapping problem in the presence of heterogeneous high-performance embedded systems.
During the labs students are supposed to implement a simple (mock) cyber-physical system using TeamPlay.
We will provide the latest implementation via a virtual machine to avoid installation hassle and uncertainties.


\section{Components and contracts}
\label{components}

\begin{figure}[htb]
	\centering
	\includegraphics[bb=0 0 750 200, width=0.8\linewidth]{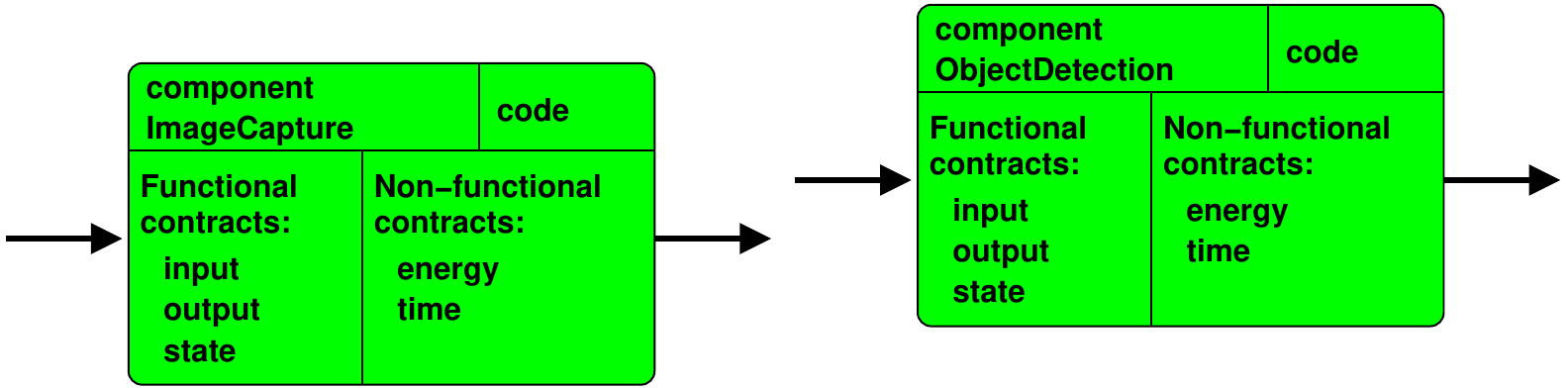}
	\caption{Two examples of TeamPlay components and their functional as well as non-functional contracts}
	\label{fig:components3}
\end{figure}

As pointed out before, a TeamPlay application is organised as a collection of interacting components.
Components are meaningful, sequential application building blocks, implemented in a general-purpose programming language.
As is common in the domain of cyber-physical systems, we exclusively work with the system programming language C in practice.
We illustrate TeamPlay components in Figure~\ref{fig:components3}.

\noindent
A TeamPlay component consists of four entities:
\begin{description}
\item[Name:] a unique name: \textit{ImageCapture} and \textit{ObjectDetection} in the example;
\item[Code:] a component implementation in the form of a callable, linkable C function object code;
\item[Functional contracts:] define the component's functional interaction with the outside world;
\item[Non-functional contracts:] define the component's non-functional properties, namely with respect to energy and time.
\end{description}
We further refine the functional contracts of a component into:
\begin{description}
\item[Input ports:] a set of pairs specifying type and name of each input port;
\item[Output ports:] a set of pairs specifying type and name of each output port;
\item[State ports:] a set of pairs specifying type and name of each internal state, which in this way is externalised to the coordination layer.
\end{description}
If the set of input ports is empty, we speak of a \textit{source component} that controls one or more sensors.
If the set of output ports is empty, we speak of a \textit{sink component} that controls one or more actuators.
And, if the set of state ports is empty, we speak of a \textit{state-less component}.
Information regarding the internal state of a component is crucial for its automatic replication or migration between multiple execution units.

We further distinguish two kinds of non-functional contracts:
\begin{description}
\item[Time contract:] the timing behaviour of the component, namely average and/or worst-case execution time;
\item[Energy contract:] the energy behaviour of the component, namely average and/or worst-case energy consumption.
\end{description}
Functional and non-functional contracts differ from each other in one crucial aspect: any implementation of a component must obey the functional contracts, but the non-functional contracts are specific to one particular execution platform (system architecture).
Hence the functional contracts of a component are not stored in the coordination source file, but in a separate data base.
In a simplistic view this data base could be seen as a matrix with all components of an application on one axis and all potential execution platforms and their individual processing units (e.g.~Arm big.LITTLE) on the other axis.
Determining a component's non-functional properties on a specific processing unit is beyond the scope of our work.
For relatively simple architectures this could be done with static analysis \cite{WegeNikoNune+23}, but for more complex architectures (deep caches, out-of-order execution, etc.) this is usually accomplished by dynamic profiling \cite{Seewald2019coarse} and some configurable safety margin.

\begin{figure}[htb]
\centering
\includegraphics[bb=0 0 750 200, width=0.8\linewidth]{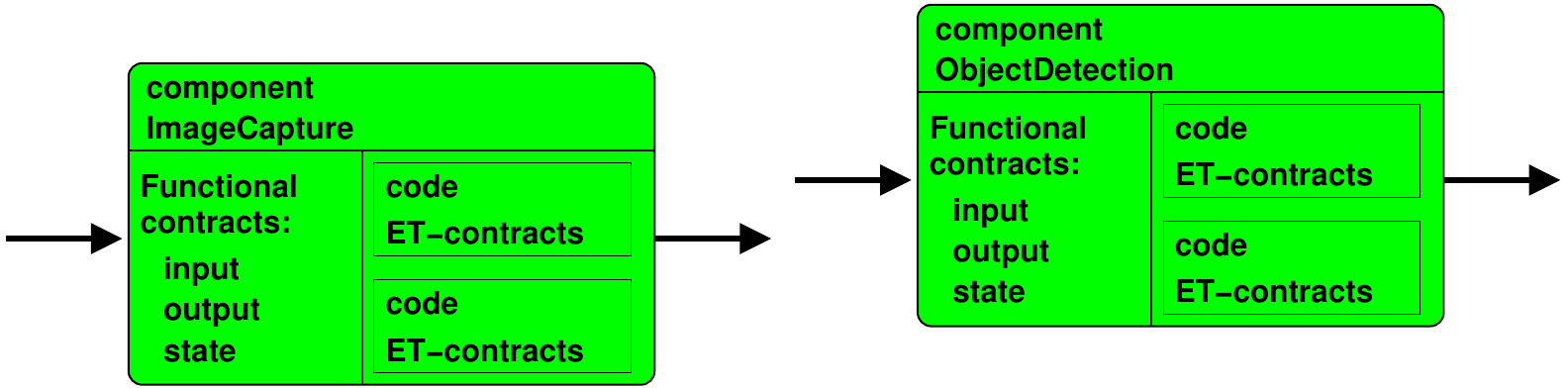}
\caption{Two examples of TeamPlay components with multiple alternative implementations and their implementation-specific non-functional contracts}
\label{fig:components4}
\end{figure}

Our observation is that usually a component can be implemented in various different ways.
Furthermore, component implementations can be compiled into executable code with different compilation objectives (e.g.~speed, energy, code size, etc.).
Hence, we further refine our component model to support \emph{multi-version components}, as illustrated in Figure~\ref{fig:components4}.
Multi-version components have multiple implementations that all share the component's functional contracts, but behave differently with respect to the non-functional contracts, even on the very same execution platform and processing unit.

The genuine rationale behind multi-version components is to offer a choice of different non-functional behaviours as described by the non-functional contracts.
Consequently, the coordination layer, or more precisely the component space/time scheduler, may choose among the various versions according to global application-specific objectives.
Multiple component versions can further be used to support multiple binary-incompatible execution units or to optimise one (source code) implementation of a component for one specific type of execution unit, even if they are binary-compatible in principle.
The coordination run-time system \cite{RouxAltmGrel21} may even react on changing environmental conditions, as for example the battery status of some device.


\ignore{
In essence, (non-functional) contracts define the ETS properties of the component. Hence, each component has a defined run-time behaviour, amount of energy required and a security level obtained from the analysis tools of the other partners. Additionally, each component can have multiple implementations with equivalent functional, but different extra-functional behaviour. Besides the guaranteed performance of the individual components, we need to specify the component-specific and system-wide ETS-constraints, expressed as, for instance, deadlines on the response-time of a component or energy budgets of the entire system or minimum security levels.
}

\ignore{
We use the very same annotations explained in Deliverable D7.4 to define the ETS-characteristics of each component, e.g.:\\
\textit{ \_\_teamplay\_\{qualifier\}\_metric(name\{,default, confidence\});stmt;} \\
These annotations form the central glue between the technical work packages: while some work packages are primarily concerned with deriving these non-functional properties from component implementations, others, among them work package 2, are concerned with interpreting and exploiting this additional information on the application level.
}

\section{Component interaction}

\begin{figure}[htb]
	\centering
	\includegraphics[bb=0 0 900 300, width=0.9\linewidth]{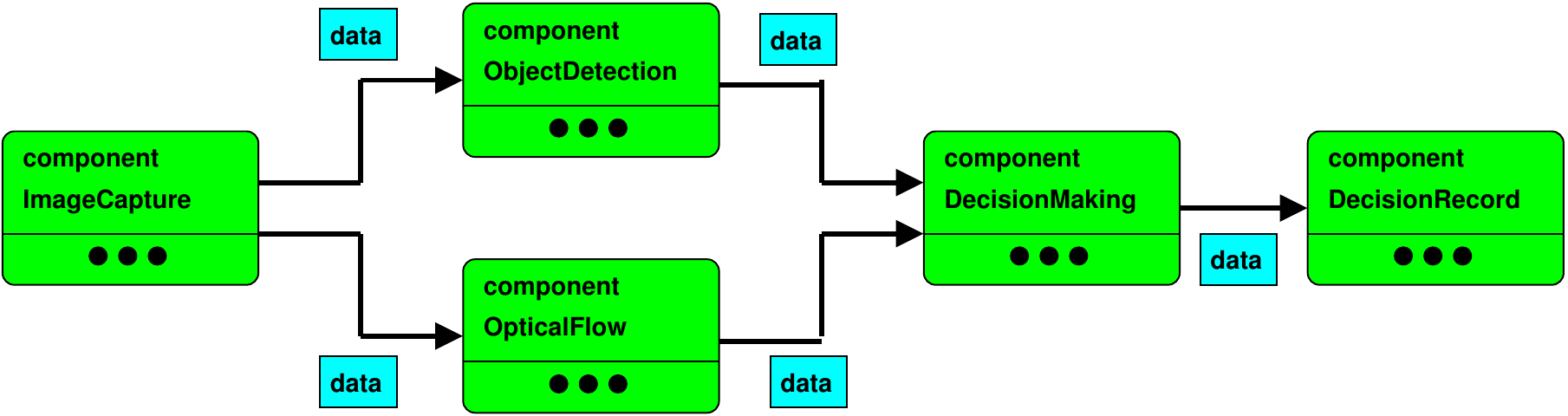}
	\caption{Workflow definition of the different components.}
	\label{fig:coordination}
\end{figure}

As illustrated in Figure~\ref{fig:coordination}, several components together form an application (or system).
They interact in the form of a streaming network or component workflow.
Such streaming networks may have arbitrary shape, but must be free of cycles.
In other words, our streaming networks form \emph{acyclic directed graphs}, or \emph{DAGs} for short.
In the example we see a source component \textit{ImageCapture} that feeds data into the streaming network.
The rationale behind the example is that this component is connected to some kind of camera, but in principle any sensor input is a typical case of such a source component.

The data is passed on to two subsequent components, called \textit{ObjectDetection} and \textit{OpticalFlow}. This could be characterised as a broadcast of one data item to two components, as is the intention behind the example, or the \textit{ImageCapture} component could send two different data items to \textit{ObjectDetection} and \textit{OpticalFlow}, respectively.
Components \textit{ObjectDetection} and \textit{OpticalFlow} are stateless and process the data received from \textit{ImageCapture} in different ways and output their results to the subsequent component \textit{DecisionMaking}.

The \textit{DecisionMaking} component in the given example synchronises the data received from both incoming
streams in a pair-wise manner, derives a decision (as the name suggests) and forwards it to the final component of the workflow \textit{DecisionRec} for recording.
The final component \textit{DecisionRec} is a sink component and as such does not have any output port.

The whole execution regime is entirely data-driven: components are activated by the availability of data items on their input ports. Components process the data received and in turn produce data items on their output ports, which trigger subsequent components further down the workflow or streaming network.
Both the external behaviour of components as well as their orderly interaction in a streaming network or work flow are described in the TeamPlay coordination language \cite{RoedRouxAltm+20}.
The (static) type system of TeamPlay ensures the soundness of streaming networks.

\section{Component scheduling and mapping}

\begin{figure}[htb]
\centering
\includegraphics[bb=0 0 900 600, width=0.9\linewidth]{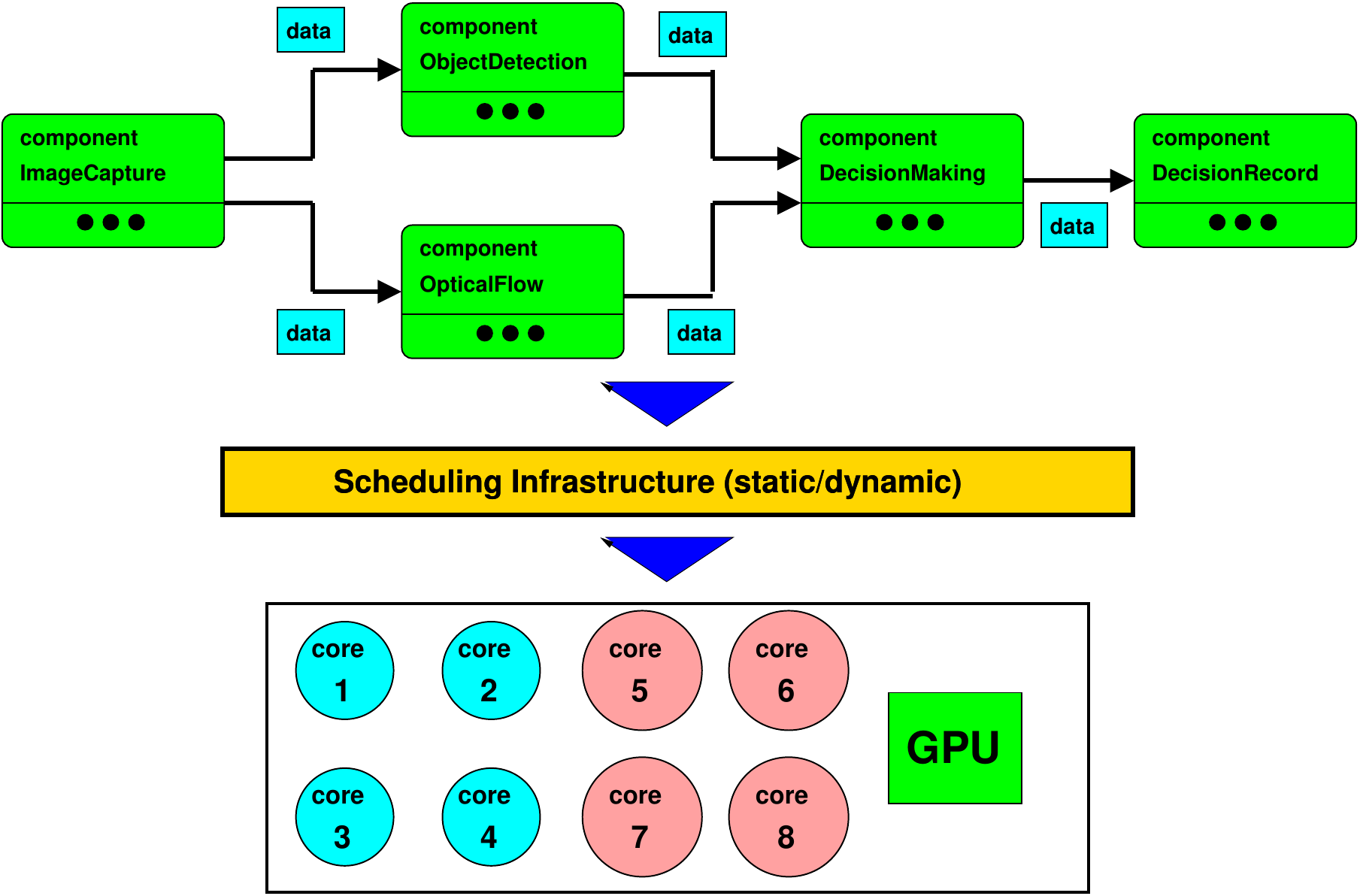}
\caption{Scheduling a TeamPlay data-flow coordination programme onto a heterogeneous multi-core platform. The target architecture is inspired by the ARM big.LITTLE architecture with a GPU accelerator.}
\label{fig:scheduling}
\end{figure}

Eventually, componentised applications and their coordination glue must be mapped to execution hardware.
We specifically target heterogeneous parallel systems with multiple different types of execution units.
These types of execution units may or may not be binary-compatible, but they (almost) always expose different energy/time trade-offs.
Together with our multi-version component model we have a created a rich design space for the crucial question where to run what and when to meet application-specific global constraints and to achieve application-specific global objectives.
We illustrate this plethora of options and opportunities in Figure~\ref{fig:scheduling}.

It is the combination of multi-version components with heterogeneous parallel hardware that in essence causes this explosion of options.
For example one may want to switch from a fast implementation to a more energy-efficient implementation as the battery status of the executing device degrades.
Different versions of the same component could implement different algorithms, could be compiled using different compiler flags (or even different compilers) use different parameters (frame-rate) or could make use of different execution units of a heterogeneous system, such as sketched out in Figure~\ref{fig:scheduling}.

Depending on application requirements we consider both energy and time as either a budget limitation or an optimisation problem.
For time both variations have classical role models and solutions: the real-time community works with time budgets while the high-performance computing community sees time as an object of optimisation.
Energy can be addressed analogously as a budget constraint or an optimisation objective.
The most typical combination in a our target application domain of cyber-physical systems, however, is to combine time budgets, i.e.~real-time deadlines, with energy optimisation objectives.
In other words we aim to run an application within the deadline with the least energy possible.
The energy reduction objective could be motivated by either environmental concerns or by battery life times, but that again is beyond the scope of our work.

The scheduling problem as such is NP-complete, and so optimal solutions can effectively only be computed for very small concrete problems.
Consequently, heuristics are used in practice, and we have integrated a number of such scheduling heuristics into our TeamPlay compiler \cite{seewald2019component,RoedRouxAltm+20b,RoedRouxGrel21,RoedPimeGrelAIAI23} that explore different aspects of the general subject matter.
Our energy-aware schedulers make use of an elaborate energy model that distinguishes between static and dynamic energy consumption and take full advantage of platform-specific DVFS capabilities (i.e.~dynamic voltage and frequency scaling).
Static energy consumption refers to the energy consumption of a processing unit in idle status, i.e.~without executing any code.
Static energy consumption, as the name suggests, is constant over time and only affected by application makespan, i.e.~the time it takes an application to run to completion or to complete one application cycle.
In other words, static energy consumption is a property of the type of execution unit, but application-independent.
Dynamic energy consumption depends on the instruction mix of an application component, the type of execution unit and the choice of voltage and frequency in case the hardware supports DVFS, which is a common case nowadays.

Execution time and energy consumption are not independent of each other, but they are not the same either.
In this sense, the fastest solution is typically not the most energy-efficient one.
Running an application faster cuts down the static energy consumption, but this is often over-compensated by running code on higher clock frequency and voltage or by even running it on a more powerful but likewise more energy-hungry execution unit.
For example, on the ARM big.LITTLE architecture we have used so far for illustration, moving a component from a LITTLE core to a big core typically increases platform energy consumption.
Should the compute power of a big core be required to meet the deadline, this additional energy is well invested, but otherwise most likely not.

\section{Related work}
\label{related}

Coordination itself is a well established computing paradigm with a plethora of languages, abstractions and approaches~\cite{CML+18}.
Yet, we are neither aware of any explicit adoption of the principle in the domain of cyber-physical systems, nor are we aware of energy- and time-aware approaches to coordination as a paradigm.

We have previously worked on the coordination language and component technology S-Net~\cite{GSS10,GrelJulkPencCCGRID12}, from which we draw both inspiration and experience for the design of TeamPlay.
However, like other coordination approaches S-Net merely addresses the functional aspects of coordination programming and has left out any non-functional requirements, not to mention energy or time specifically.

A notable exception in the otherwise pretty much uncharted territory of energy/time-aware coordination is the functional language Hume~\cite{HaM03} that was specifically designed with real-time systems in mind. Thus, guarantees on time (and space) consumption are key to Hume.
The main motivation behind Hume was to explore how far high-level functional programming features, such as automatic memory management, higher-order functions, polymorphism, recursion and currying, eventually up to full-scale Haskell, can be supported while still providing accurate real-time guarantees.

Pradhan et al.~\cite{Pradhan15} proposed \textit{CHARIOT}, a \textit{Domain Specific Language} for CPSoSes. While they do not describe their language as a \emph{coordination language}, it provides capabilities such that it could be considered a coordination language. The authors also do not explicitly describe utility for systems-of-systems and, instead, focus on \textit{fractionated} systems. A fractionated system is an architecture within which one system is composed of a set of networked systems at the same location (e.g.~a satellite swarm). The advantage of such a system is the ease of extendibility: a system providing a new capability (e.g.~sensor or processing platform) can be added to the general area of the other systems, thus enhancing the capabilities of the composite system.

More work published under the \emph{DSL} label shows considerable overlap with coordination languages. The contribution by Berger~\cite{Berger14} proposes a DSL for the integration of sensors into a cyber-physical system. The platform is supplied in a \textit{board specification file}. However, while the board can be changed, the work is limited to a single board. The DSL encodes the requirements for sensors attached to the board. Each sensor is attached by pins, which must be set to a particular mode (e.g.~analog or use i2c). Not all pins support all modes, which forms the basis of a \textit{constraint-satisfaction problem} to perform the assignment of sensors to pins. As such, this work is an example of using a language to formulate a problem that cannot be solved manually, similar to the TeamPlay language. However, the scope and focus on real-time of our coordination language remains unique.

\section{Summer school}
\label{school}

Our contribution to the planned summer school targets an audience of MSc and PhD students with a broad understanding of computer science in general and solid programming skills in particular.
We do not expect prior exposure to engineering cyber-physical systems or mapping/scheduling problems nor any prior knowledge of the topics described in this document.

The learning objectives for the summer school are as follows:
\begin{itemize}
\item obtain a basic understanding of the requirements of cyber-physical systems and resource-aware computing;
\item understand the concept of exogenous coordination;
\item familiarise with the TeamPlay coordination language;
\item solve a (simple) scheduling/mapping problem for a (mock) application;
\item optimise the energy consumption of said cyber-physical system by using the TeamPlay methodology \cite{RouxPaghAkes+20}
\end{itemize}

We will provide any subject-specific background required during the lecture.
However, we are convinced that just telling the story is not sufficient for a deeper and lasting learning experience.
Thus, a hands-on practical session is essential to meet the learning objectives.
Such practical sessions making use of research (prototypical) software incur the risk of losing considerable time in overcoming installation and compatibility issues.
We will address and avoid such problems by providing a docker container with a light-weight virtual machine and all required non-standard software pre-installed and tested.

The concrete task to be solved by the summer school participants has not yet been finalised entirely, but our working hypothesis circles around a fork-join component/task graph, where the fork component reads wifi signal strength (from some imaginary sensor) and forwards the data to a fixed number of tasks \emph{representing students in a room}.
These tasks determine the approximate locations of the students in the room, which are forwarded to a join task that takes some decision based on the locations, e.g.~raising an alarm should the 1.5m Covid-19 distance rule be violated.

We plan to provide the summer school participants with a monolithic solution to the problem and let them identify the logical components and their interaction using the TeamPlay coordination technology.
We provide non-functional properties, i.e.~time and energy, for the (rather obvious) building blocks of our application.
As a result the students will find out that the monolithic solution cannot be scheduled with the given deadline.
With the explicit componentisation and the TeamPlay methodology the application not only becomes schedulable, but as an extra incentive we will have a small competition who of the participants manages to solve the problem with the least energy demand.

Time-permitting, we can also illustrate the fault-tolerance features of TeamPlay \cite{LoevGrel20} that permit to easily run selected components under a variety of fault-tolerance regimes, such as double or triple modular redundancy.
Here, the students can experiment with how much redundancy we can afford on a given hardware platform before the application becomes unschedulable.
Moreover, the price in terms of energy consumption that needs to be paid for redundancy and fault-tolerance becomes evident.

\ignore{

  Lukas:

> = Sustrainable
> - Goal: play a bit with the coordination language. NFP and source code provided, slight changes to the configuration
>   - Needs: a container / VM with everything ready to go (Cecile on path, Yasmin, build-essentials)
>   - [!] this will be tricky on ARM macs (need an ARM mac)
>     -> solution: docker container with bash as start shell and all tools on path
> - Concrete plan:
>   - Application: fork-join esq graph where
>     Fork task reads wifi signal strength to a fixed number of students
>     Multiple process task locates each student
>     Join task makes some decision based on locations (e.g. 1.5m rule violations)
>   - Excercise: students get a monolithic version + code, nfp for parts of the code
>     Monolithic version is not schedulable
>     Students rewrite the task graph using SDF to enable hardware parallism
>     Run join task with fault tolerance (replicas = no. of replicas, (ignore) votingReplicas = no. of voters)
>     [!] Fix spelling mistake in homogenize (from homogeneise)
>     [!] Finish expand-ft-compiler-pass
> - Other:
>     [Clemens] will send slide deck regarding other instructions

After this I realized it actually needs to be about energy-aware scheduling -- the monolithic version should be schedulable as well, just use a lot of energy. But we can still take the same basic idea.
}

\section*{Acknowledgments}
\label{ack}

This project has received funding from the European Union's Horizon-2020 research and innovation programme under grant agreement No.~779882 (TeamPlay: Time, Energy and Security Analysis for Multi/Many-core Heterogeneous Platforms) and under grant agreement No.~871259 (ADMORPH: Towards Adaptively Morphing Embedded Systems).
Our work has received further funding from the European Union via the Erasmus Plus Key Action 2 (Strategic partnership for higher education) under grant agreement No.~2020-1-PT01-KA203-078646 (SusTrainable: Promoting Sustainability as a Fundamental Driver in Software Development Training and Education).
Last not least, this work has been supported by the COST Association through COST Action CA19135 (CERCIRAS: Connecting Education and Research Communities for an Innovative Resource Aware Society).

Special thanks go to all who have contributed to the design and implementation of the TeamPlay coordination language:
Julius Roeder,
Benjamin Rouxel,
Steven Swatman,
Wouter Loeve and
Sebastian Altmeyer.




%
\title{Sustainable Behaviour of IT Practitioners}%
\titlerunning{Sustainable Behaviour of IT Practitioners}
\author{Elena Somova\orcidID{0000-0003-3393-1058} \and Denitza Charkova\orcidID{0000-0003-0873-4415}  }
\authorrunning{E. Somova, D. Charkova}
\institute {University of Plovdiv "Paisii Hilendarski", 24 Tzar Assen St., 4000 Plovdiv, Bulgaria,\\%
	\email{ eledel@uni-plovdiv.bg; dcharkova@uni-plovdiv.bg}\\%
	\url{https://www.uni-plovdiv.bg/} }%
\maketitle
\begin{abstract}%
Environmental protection is a key topic nowadays. Many world and European organizations and bodies (with their respective documents) fight against the global negative impact of industries such as fossil fuels. The work of IT specialists could influence other sectors of sustainability in both positive and negative ways. The paper presents a questionnaire, exploring students' attitude to sustainability and the place of sustainability in higher education and future employment. The survey will be conducted during the Second SusTrainable Summer school.%
\end{abstract}%
\section{Introduction}%
\par%
Sustainability has many dimensions - environmental, economic, social, technical, educational, cultural, etc., as environmental sustainability will lead to the most long-term and positive global consequences. For example, the negative effects of burning fossil fuels (coal, oil and natural gas) are well-known: air pollution, water pollution, climate change and health problems.%
\par%
The work of IT specialists could also influence sustainability in other sectors. Something more, IT professionals can impact environmental sustainability, both positively and negatively – positively by reducing the impact of IT on the environment (Green IT) and applying IT to improve sustainability (\#Tech4Good), and negatively by supporting and working for fossil fuel companies and other unsustainable sectors (\#Tech4Bad)~\cite{1BIB1}.%
\par%
This work presents the survey aimed at exploring IT specialists (students and teachers)’ attitude towards sustainability (in particular to the fossil fuel industry) and the place of sustainability in higher education and future employment. The survey is going to be carried out with the participants of the Second SusTrainable Summer School.

\section{Global and European political and legislation issues}\label{section:overview2}%
\par%
The global and European policy is directed to environmental protection incl. first reduction and late cessation of the fossil fuel industry:%
\begin{itemize}%
\item The United Nations Framework Convention on Climate Change~\cite{1BIB2}, signed at the United Nations Conference on Environment and Development in 1992;
\item The Paris Agreement~\cite{1BIB3}, adopted at COP 21 in 2015;%
\item The United Nations 2030 Agenda for Sustainable Development, adopted in 2015, with its 17 Sustainable Development Goals (SDGs)~\cite{1BIB4};%
\item The Global Coal to Clean Power Transition Statement, signed at the 26th United Nations Climate Change Conference (COP 26) in 2021;%
\item The European Green Deal~\cite{1BIB5}, approved in 2020;%
\item The set of proposals to reduce net greenhouse gas emissions, adopted by the European Commission in 2021;%
\item The priority areas of the Energy Transition Council, established in 2020;%
\item The International Energy Agency (IEA) Report~\cite{1BIB6} in 2021, etc.%
\end{itemize}%
\par%
All these reductions in environmentally harmful industries will inevitably affect the IT sector.%
\section{Sustainable behavior of IT practitioners}
\par%
When IT practitioners make their behavioral decisions, the choose how sustainable their behavior is depending on some pragmatic and ethical issues. The paper~\cite{1BIB1} explores in detail these pragmatic and ethical concerns that IT practitioners have to take into account when they want to start working or already work in a fossil fuel company or its supplier company.%
\par%
From a pragmatic point of view, the fossil fuel sector is decreasing and will end in the next few decades, according to the global politics (see Section 2), therefore IT specialist should move to another sector. A lot of jobs will be lost according to IEA~\cite{1BIB6} very soon, because companies will not invest in new fossil fuel exploration and development. Because of the "net-zero emissions by 2050" pathway, some industries will end by 2050 or earlier, such an example being  the transport fossil fuel industry. In recent years, there has been a rapid increase in net zero emissions announcements by companies with target 2050 or earlier~\cite{1BIB6}. A common classification system – EU Taxonomy for sustainable activities~\cite{1BIB7}, establishing a list of environmentally sustainable economic activities has been created, where the fossil fuels are excluded. Some organizations provide lists that can be used for recognition of sustainable companies – for examples, MSCI launched Global Fossil Fuels Exclusion Indexes~\cite{1BIB8}, used by institutional investors to to eliminate or reduce some or all fossil fuel reserves exposure from their investments, and Investopedia’s Environmental, social, and governance (ESG) criteria~\cite{1BIB9}, giving a set of standards for a company’s behavior used by investors to screen potential investments based on corporate sustainable policies.%

\par%
From an ethical standpoint, to work for a company producing or supplying fossil fuels is not illegal, but the ethical point of view should be considered. Whether to work for an environmentally damaging sector or not, is a personal decision. Ethical principles, connected to the issue of sustainability, are reflected in the main documents of a number of IT organizations – IEEE Code of Ethics~\cite{1BIB10}, BCS Code of Conduct~\cite{1BIB11}, ACM Code of Ethics and Professional Conduct~\cite{1BIB12}, etc. The environmentally damaging fossil fuels sector directly hinders the two of the 17 SDGs of the UN – SDG13. Climate Action and SDG7. Affordable and Clean Energy, and indirectly – SDG3. Good Health and Well-Being, SDG11. Sustainable Cities and Communities, etc.%
\par%
All these pragmatic and ethical considerations~\cite{1BIB1} that IT practitioners should take into account when deciding which company to work for show that this decision is very important and with consequences.%
\section{Survey questionnaire}
\par%
The objectives of the questionnaire are:%
\begin{itemize}%
\item to investigate young IT professionals’ and teachers’ attitude to sustainability (incl. the fossil fuel industry);%
\item to explore IT students’ and teachers’ opinion about the place of sustainability in higher education;%
\item to study their opinion about future employment in relation to sustainability.%
\end{itemize}%
\par%
The survey consists of 20 questions (see the Survey Questionnaire): 5 demographic questions and 15 questions (presented as statements) on issues of interest. The survey uses 7 of the UK survey questions~\cite{1BIB13}~\cite{1BIB15}~\cite{1BIB16} to enable comparison with existing results. The statements are divided into 6 groups (see Table 1).%

\begin{table}[]
	\centering
	\begin{tabular}{ll}
		\textbf{Group} & \textbf{Questions}  \\
		Sustainability awareness & 1-4  \\
		Sustainability at university &   5-7  \\
		Sustainability in employers &   8-10 \\
		Ethical concerns &  11-13 \\
		Practical concerns &  14-16 \\
		Community behavior &   17-20
	\end{tabular}
\caption{\label{demo-table2}Survey question groups}
\end{table}

\par%
The majority of the questions in the survey use a 5-step Likert scale to give students a better opportunity to express their opinion to the fullest: Strongly Disagree, Disagree, Undecided, Agree and Strongly Agree. The remaining questions provide students with situational cases from which they need to choose the option best suited for them.%

\par%
\textbf{SURVEY QUESTIONNAIRE}%
\par%
\textbf{Part 1. Personal data}%
\par%
1. Sex – Male/Female%
\par%
2. Year of study/Teacher – 1/2/3/4/teacher%
\par%
3. Working now/have worked in the ICT sector? – Yes/No%
\par%
4. If yes, for how long?%
\par%
5. Country of study/teaching%
\par%
\textbf{Part 2. Personal opinion}%
\par%
1. I know of the United Nations’ Sustainable Development Goals.%
\par%
2. I know that globally we must reach net zero carbon emissions by 2050.%
\par%
3. Fossil fuels are at the root of the climate crisis.%
\par%
4. I personally carry out the skill “understand people’s relationship to nature”.%
\par%
5. The university practices and promotes good social and environmental skills.%
\par%
6. The University is obliged to develop students’ social and environmental skills as part of the courses.%
\par%
7. The university courses consider the ethical (in the direction of environmental and social aspects) implications of the course subject.%
\par%
8. My employer should consider the environmental and social impacts of its products and/or services.%
\par%
9. My employer should actively work for the reduction of carbon emissions (incl. by refusing to provide products/services to the fossil fuels sector).%
\par%
10. The employee’s skill of “understanding people’s relationship to nature” should be important for my employers (current or future).%
\par%
11. I am concerned about environmental climate change.%
\par%
12. I will refuse to do any work at my company that supports the fossil fuel industry.%
\par%
13. I will leave my company if I learn that it produces software/ provides services for the fossil fuel industry.%
\par%
14. I share my position about not supporting the Fossil Fuel Industry with other ICT practitioners.%
\par%
15. I participate in conversations/events about the negative impact of the ICT sector on the environment when supporting the Fossil Fuel Industry.%
\par%
16. I show support to others who are speaking up and saying no to working for the Fossil Fuel Industry.%
\par%
17. I know which organization (Trade union, Non Government Organizations) to contact with to help me find an employer who doesn’t support the fossil fuel industry or to help me leave a company that supports the environmentally damaging fossil fuel sector.%
\par%
18. Which option would you choose?
Assuming all other factors are equal, I would choose a graduate position with a starting salary of 100 Euro higher than average in a company with a poor environmental and social record.
Assuming all other factors are equal, I would choose a graduate position with a starting salary of 100 Euro lower than average in a company with a strong environmental and social record.%
\par%
19. Which option would you choose?
Assuming all other factors are equal, I would choose a graduate position with a starting salary of 300 Euro higher than average in a company with a poor environmental and social record.
Assuming all other factors are equal, I would choose a graduate position with a starting salary of 300 Euro lower than average in a company with a strong environmental and social record.%
\par%
20. Which option would you choose?
Assuming all other factors are equal, I would accept a graduate position with a starting salary of 300 Euro higher than average in a role that does not contribute to positive environmental and social change.
Assuming all other factors are equal, I would accept a graduate position with a starting salary of 300 Euro lower than average in a role that contributes to positive environmental and social change.%

\par%
The survey was completed by 260 young IT professionals – IT students from University of Plovdiv “Paisii Hilendarski”, Bulgaria.%
\par%
During the summer school, a survey will be conducted among IT students and teachers, and the results of the survey will be able to be compared across participating countries. The survey will be voluntary and anonymous.%
\section{Conclusion}
\par%
IT specialists have to take into account that a career in an environmentally damaging sector like the fossil fuel industry will be short-term and in near future they have to change their job in other sectors. Working for the fossil fuel industry, through supporting or developing software that continues the production and sales of fossil fuels, is a legal job, but it is incompatible with the ethical principles of the ICT professional organizations and ethical job decisions are very important.%
\par%
During the Second SusTrainable Summer School, a survey will be conducted among IT students and teachers to explore IT specialists’ attitude to sustainability (in particular to the fossil fuel industry) and the place of sustainability in higher education and future employment.%
\par%
\textbf{Acknowledgements}%
\par%
This paper acknowledges the support of the Erasmus+ Key Action 2 (Strategic partnership for higher education) project No 2020-1-PT01-KA203-078646: “SusTrainable – Promoting Sustainability as a Fundamental Driver in Software Development Training and Education”. The information and views set out in this paper are those of the authors and do not necessarily reflect the official opinion of the European Union. Neither European Union institutions and bodies, nor any person acting on their behalf may be held responsible for the use which maybe made of the information contained therein.


%

%
\title{Energy Consumption Related Questions of Developing and Evolving Dynamic Web Applications}
\titlerunning{Energy Consumption of Web Applications}
%
\author{Csaba Szab\'o\orcidID{0000-0001-5147-2452}}
\authorrunning{Cs. Szab\'o}
%
\institute{Technical University of Ko\v sice, Letn\'a 9, 042 00 Ko\v sice, Slovakia\\
\email{csaba.szabo@tuke.sk}}
\maketitle              
\begin{abstract}
The aim of this paper is to shortly introduce the need of considering energy consumption when developing dynamic web applications. The underlying problem is presented from the users' point of view, namely that a web page that consumes too much energy might be the main reason of unwanted browser restarts or worse responsibility. Later, similarities and differences to related work are shown to contrast specifics of web applications, in both contexts of development and evolution. Then, we define an example development project using React with Typescript as frontend technology and Tensorflow with Python as backend technology. Energy consumption measurement will be distributed between the Firefox browser running the frontend and PyRAPL library at backend. After the initial measurements based on selected scenarios, we focus on analysis of energy consumption changes caused by directed evolution of the web application. All presented content aims to contribute to the topic of software engineering education by focusing on sustainable software solutions.

\end{abstract}
\section{Motivation}
"Everything is on the Internet" is a very common phrase of the last two decades. All industries consider this medium as the most important one to inform (potential) customers, clients etc. Common types of Internet-based communication include e-mail, chat and web pages. Using amount of information as classification criteria, we could say that web pages aim to provide much more information and functionality than e-mail or chat\cite{similarweb}. Usual implementations often include both chat and e-mail option as a form of one-on-one communication. Current webs offer lots of functionalities in form of rich web applications.

Nowadays, web browsers obviously offer the possibility to open more than one web page at the same time. So, the user does not need to run many browsers. With increasing number of open web pages, memory and CPU load from browsing also increases. Modern web browsers are optimised for a huge number of simultaneously opened pages, but the browser developers used selected webs when they optimised these products\cite{mdntesting}.

Despite this kind of usage focused optimisations, sometimes the users experience browser failures. Opened pages require restart (become slow) due to high resource usage, servers stop responding or more precisely the request is timed out. Mobile devices such as laptops and phones are more sensitive to such intensive resource usage since they are not continuously charged. Servers might have a shorter lifetime under higher workload.

Web pages also have their creators: designers and developers. In a world close to perfect, systematic testing process\cite{mdntesting} is also executed before deployment. Testing can uncover design and implementation weaknesses, but some types of weaknesses are usage specific, which is impossible to completely cover by testing. Users will also report some system failures they experience during application use.

User feedback\cite{swevolutionbook} points to further application improvements. This is what we call evolution of web applications. We have two sides. On the one, there are the users requesting changes, on the other one, there are the developers trying to implement the best improvement.

In this paper, we focus on a kind of web application evolution that is addressing energy consumption optimisation. First, we present related work and tools. Then, we introduce an example project using selected (measurable) technologies\cite{raplzhang,raplinaction}. To improve the measured stats of the example web application, we implement selected evolution techniques. To make it more interesting, we take a look at a different type of evolution: extending functionality; and its impact on the energy consumption profile of the application. Result analysis and what-if discussions will close the teaching class. We only present the process in this paper.

\section{Server/Client-side Green Computing and Evolution}
Web applications usually follow the same architecture. The logic and data of the application are distributed into a number of components, which are also distributed and/or grouped into specific virtual or physical locations\cite{architecture}.

The roles of these system modules are clear: presentation of information, storing underlaying data and implementation of the logic of the application. Logical distribution is much simpler in terms that there are only two parts: the modules run from within the browser (a.k.a. frontend) and everything else (a.k.a. backend). As the application is growing during its evolution, the architecture might be subject of change as well, but there also exist generic architectural patterns preparing the system for long-term evolution that do not require significant changes in the system architecture, such as the microservices' or microfrontends' architecture\cite{architecture}. In all of these architectures, front- and backends are clearly defined. Thus, the architecture comprehension could be minimised to understanding of the historical client/server architecture, considering frontend the client and backend the server.

Evolution of software is necessary\cite{swevolutionbook}. Since there is almost impossible to provide a final software from the first try so that software will not require any further modifications. The literature defines these necessary and/or user-requested modifications as software evolution. Sure, there exist solutions that do not require any improvements after implementation as they are complete and perfect (S-type software by Lehman, perfection is not subject of change over time) or, at least there is no better solution (P-type software by Lehman, where a minimal chance of further one-time modification exists, but that requires progress in theory and practical state of the art).

Majority of software is of E-type, meaning we expect a kind of evolution. This evolution is guided by the feedback from the system usage, which includes direct user feedback, usage data analysis etc. Part of this is profiling in different production environments.

With web applications, the problem of profiling a distributed application raises\cite{architecture}. But, a partial profiling can be done by separately analysing the selected nodes the system is being distributed to.

When we focus on energy profiling\cite{balkishanprofiling}, a significant problem is the limited ability to separate the effects of the artefacts under profiling from the side-effects of measuring and effects of other system components. Invasive and non-invasive measurements are used at different level of error\cite{raplzhang,raplinaction}. Non-invasive is the way, when one is measuring the whole system, and the artefact selection is done by scenario design (test setup in automation). An invasive way is to introduce changes into the system to be able to measure with a lower error. Some techniques emit signals to external measuring only, other might be derived from Unit-testing techniques\cite{unit}.

RAPL\cite{raplzhang,raplinaction} supports both, since there exists its language library integrations such as PyRAPL integration, but it could be also used as an external measurement tool such as the power gadget. Getting inside interpreters is much more complicated, therefore browser or operating system level measurements are kind different. The browser itself needs to include a development support interface for data collection, which data could be then used in further analysis, e.g. using RAPL again. Firefox web browser includes energy profiling ability since version 104. The energy model is not included in the browser, collected data are sent to a web service that processes them to a readable form.

\begin{figure}
\includegraphics[bb=0 0 725 300, width=\textwidth]{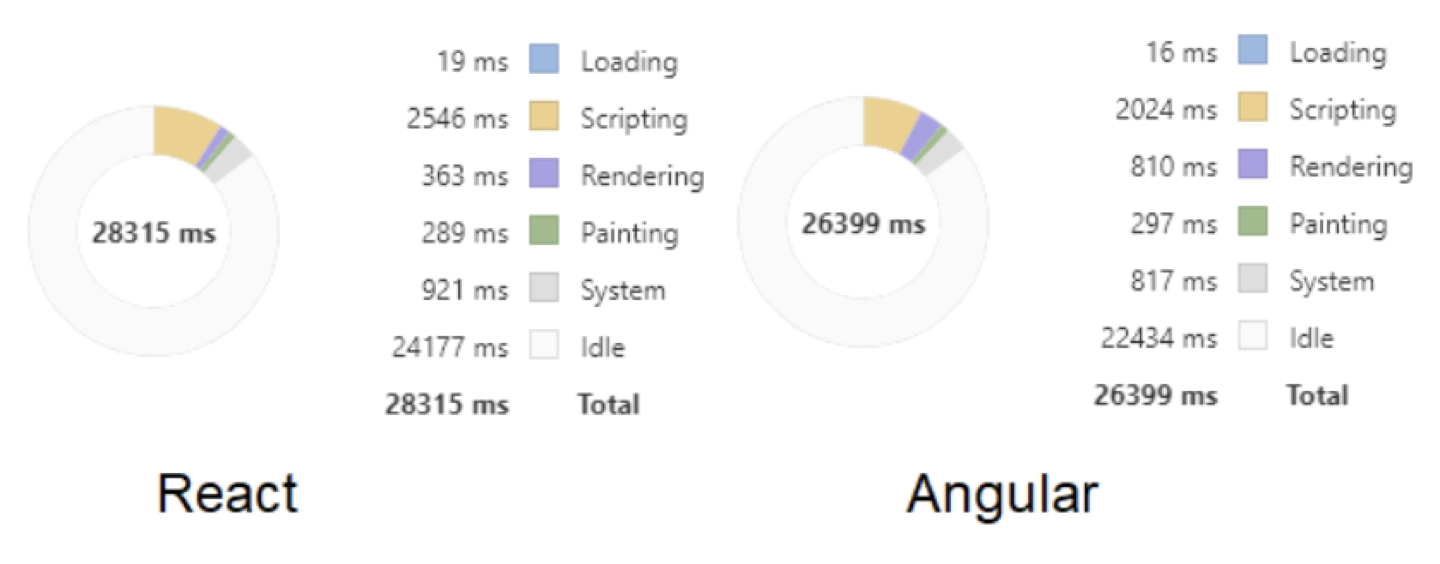}
\caption{Total application runtime destructured into Loading, Scripting, Rendering, Painting, System and Idle time for the two different implementations of the example news application. Angular implementation was a little bit faster than the React one in our experiments.} \label{RvA1}
\end{figure}

\begin{figure}
\includegraphics[bb=0 0 725 300, width=\textwidth]{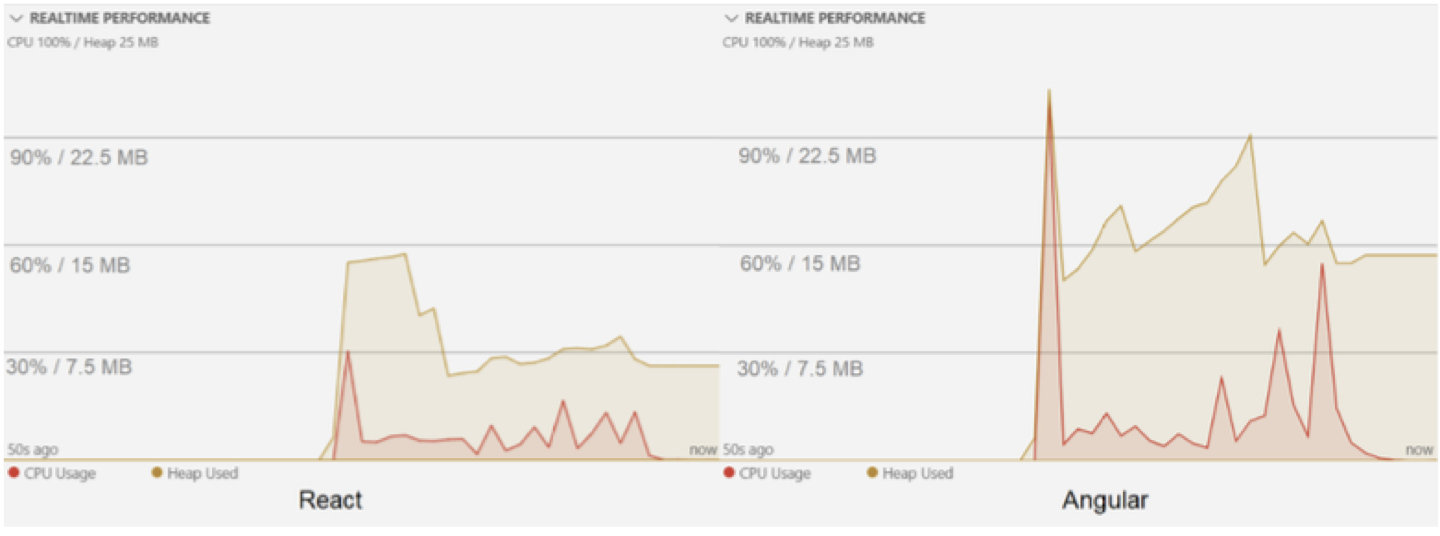}
\caption{A comparison of system workload (CPU and Heap usage) for the two different implementations of the example news application. Angular implementation was using significantly more resources than the React one in our experiments.} \label{RvA2}
\end{figure}

Our preliminary profiling results in technologies show that even React and Angular web application frameworks are already significantly different when developing standard news applications (fetching news data from a database and displaying them in an equally formatted form). While React seems to be slower in execution (see Fig.~\ref{RvA1}), Fig.~\ref{RvA2} shows a significantly lower workload.

\section{Our Example Web Application}
We propose a web application containing user profiles and allowing chat messaging between these users. We select React and Typescript as frontend technologies, while a very basic backend will be implemented using Python. Two development projects as building blocks of one application.

Repair and evolution tasks for the example:
\begin{enumerate}
\item In the application, there will be a programming error causing increased energy consumption of the browser (too many re-renders), which will be eliminated "based on user feedback".
\item According to energy saving design guides, we will evolve the style of the application to decrease frontend energy consumption (i.e. moving from light to dark theme style).
\end{enumerate}

Later, we will focus on the Python backend, measuring it using the pyRAPL library:
\begin{verbatim}
import pyRAPL

pyRAPL.setup( ... )

@pyRAPL.mesureit()
def Application(params):
\end{verbatim}

Here, the task will be to extend the backend. E.g. by an infantry filter, translator or even more advanced AI NLP procedure (supported by the Tensorflow library).

\section{Expected outputs}
We expect the increase of understanding of RAPL reports and profiles, especially their usage during software evolution. Practical experience with using the selected technologies in web application development will point out the similarity between different programming languages when using the same (object or component oriented) pragmas.

Certain level of competition between student groups (we plan to apply pair programming) will be set up based on comparing level of sustainability of the provided versions of the web application. Applications covering the same subset of functionalities will be sorted according to their energy requirements. The more classic competition in functional size will be also applied, thus, the ultimate winners will be the members of the team that implements the most functionalities by reaching a local minimum in energy consumption. We consider this factor as the evaluation of sustainability of the provided software solution:
\begin{itemize}
\item best functionality,
\item lowest energy consumption,
\item best of both worlds.
\end{itemize}

Practically, we expect that the teams will measure the frontends in Firefox and, separately, the Python backends as well. Then, the sum of the measurement results over a 5 minute interval will serve as the value entering the evaluation. Proof of correctness and function coverage of the measurements will be checked based on the submitted source codes and measurement logs.

\section*{Conclusion}
This paper aimed to present topics of the identically entitled lecture at the SusTrainable Summer School in Coimbra. These topics are web application architecture, development and evolution, with an emphasis on energy consumption measurement of selected components of the application architecture. Topics' presentation is enriched by using the illustrative example described in Section 3, more precisely its setup and first evolutionary steps.

The following practical session will provide discussion and competition environment to the attendees to evolutionary improve both frontend and backend functionality and energetic performance.

To be able to fulfil all described tasks, the attendees must have at least basic programming skills in the presented technologies (Typescript, React, Python), and basic soft skills to "survive" in a team of three developers. From technical perspective, there would be the usage of low-edge configurations pointing to more significant improvements, since top developer computers will be not running out of any resources in the lab exercises.

\subsubsection{Acknowledgements} This work acknowledges the support of the ERASMUS+ project “SusTrainable -- Promoting Sustainability as a Fundamental Driver in Software Development Training and Education”, no. 2020–1–PT01–KA203–078646.

%
%
%

\begin{thebibliography}{10}
\providecommand{\url}[1]{\texttt{#1}}
\providecommand{\urlprefix}{URL }
\providecommand{\doi}[1]{https://doi.org/#1}

\bibitem{aho2007compilers}
Aho, A.V., Sethi, R., Ullman, J.D.: Compilers: principles, techniques, and
  tools. Addison-Wesley, 2nd edn. (2007)

\bibitem{bezanson2017julia}
Bezanson, J., Edelman, A., Karpinski, S., Shah, V.B.: Julia: A fresh approach
  to numerical computing. SIAM review  \textbf{59}(1),  65--98 (2017)

\bibitem{birkbeck2007dimension}
Birkbeck, N., Levesque, J., Amaral, J.N.: {A dimension abstraction approach to
  vectorization in Matlab}. In: International Symposium on Code Generation and
  Optimization (CGO'07). pp. 115--130. IEEE (2007)

\bibitem{bishop2006pattern}
Bishop, C.M.: Pattern recognition and machine learning. Springer (2006)

\bibitem{golub2013matrix}
Golub, G.H., Van~Loan, C.F.: Matrix computations. JHU Press (2013)

\bibitem{LangvilleMeyer2006}
Langville, A.N., Meyer, C.D.: Google's PageRank and Beyond: The Science of
  Search Engine Rankings. Princeton University Press (2006)

\bibitem{matrix_cookbook}
Petersen, K.B., Pedersen, M.S.: The matrix cookbook (nov 2012),
  \url{https://www2.imm.dtu.dk/pubdb/edoc/imm3274.pdf}

\bibitem{scholkopf2018learning}
Schölkopf, B., Smola, A.J.: Learning with kernels: support vector machines,
  regularization, optimization, and beyond. MIT Press (2018)

\bibitem{Sherrington2015}
Sherrington, M.: Mastering Julia. Pact Publishing (2015)

\bibitem{van2011numpy}
Van Der~Walt, S., Colbert, S.C., Varoquaux, G.: {The NumPy array: a structure
  for efficient numerical computation}. Computing in science \& engineering
  \textbf{13}(2),  22--30 (2011)

\end{thebibliography}

\begin{thebibliography}{10}
\providecommand{\url}[1]{\texttt{#1}}
\providecommand{\urlprefix}{URL }
\providecommand{\doi}[1]{https://doi.org/#1}

\bibitem{BIB13}
The transtheoretical model (stages of change).
  \url{https://sphweb.bumc.bu.edu/otlt/mph-modules/sb/behavioralchangetheories/behavioral-
  changetheories6.html}, accessed: 2023-06-16

\bibitem{BIB9}
What is nudge theory?
  \url{https://www.imperial.ac.uk/nudgeomics/about/what-is-nudge-theory/},
  accessed: 2023-06-16

\bibitem{BIB2}
Sustainable ux design. \url{https://www.rossul.com/2019/blog/
  sustainable-ux-design/} (2019), accessed: 2023-06-16

\bibitem{BIB7}
Bartle, R.: Hearts, clubs, diamonds, spades: Players who suit muds. MUD
  research  \textbf{1} (1996)

\bibitem{BIB8}
Bates, H.: 5 businesses using green gamification to save the world.
  \url{https://www.givingforce.com/blog/5-businesses-using-green-gamification-to-save-the-world},
  accessed: 2023-06-16

\bibitem{BIB1}
Birgersson, J.F.: 5 approaches for sustainable ux.
  \url{https://uxplanet.org/5-approaches-for-sustainable-ux-53eb4eb672b4}
  (2021), accessed: 2023-06-16

\bibitem{BIB5}
Charkova, D., Somova, E., Gachkova, M.: Gamification in cloud-based
  collaborative learning. Mathematics and Informatics  \textbf{63},  471--483
  (2020)

\bibitem{BIB3}
Deterding, S., O’Hara, K., Sicar, M., Dixon, D., Nacke, L.: Gamification:
  Using game design elements in non-gaming contexts. International Conference
  on Human Factors in Computing Systems, Vancouver, Canada p. 2425–2428
  (2011)

\bibitem{BIB10}
Eyal, N.: Hooked: How to build habit-forming products hardcover. Portfolio
  (2014)

\bibitem{BIB4}
Hommes, V.: Using green gamification for fun and fame. https://www.
  greenbiz.com/article/using-green-gamification-fun-and-fame

\bibitem{BIB11}
Niedderer, K.: Mindful design as a driver for social behaviour change. 5th
  International Congress of International Association of Societies of Design
  Research pp. 4561--4571 (2013)

\bibitem{BIB15}
Riley, S.: Mindful Design: How and Why to Make Design Decisions for the Good of
  Those Using Your Product. Apress (2018)

\bibitem{BIB12}
Salonen, A., Fredriksson, L., Järvinen, S., Korteniemi, P., Danielsson, J.:
  Sustainable consumption in finland – the phenomenon, consumer profiles, and
  future scenarios. Marketing Studies  \textbf{6},  59--82 (2014)

\bibitem{BIB6}
Somova, E., Gachkova, M.: Strategy to implement gamification in lms.
  Next-Generation Applications and Implementations of Gamification Systems, IGL
  Global pp. 51--72 (2021)

\bibitem{BIB16}
Sung, K., Cooper, T., Kettley, S.: Factors influencing upcycling for uk makers.
  Sustainability  \textbf{11},  1--26 (2019)

\bibitem{BIB14}
Zhang, Y., Liu, J., Song, S.: The design and evaluation of a nudge-based
  interface to facilitate consumers' evaluation of online health information
  credibility. Journal of the Association for Information Science and
  Technology  \textbf{74},  828--845 (2023)

\end{thebibliography}

\begin{thebibliography}{10}
\providecommand{\url}[1]{\texttt{#1}}
\providecommand{\urlprefix}{URL }
\providecommand{\doi}[1]{https://doi.org/#1}

\bibitem{brooks1977towards}
Brooks, R.: Towards a theory of the cognitive processes in computer
  programming. International Journal of Man-Machine Studies  \textbf{9}(6),
  737--751 (1977). \doi{https://doi.org/10.1016/S0020-7373(77)80039-4}

\bibitem{cherubini2007let}
Cherubini, M., Venolia, G., DeLine, R., Ko, A.J.: Let's go to the whiteboard:
  how and why software developers use drawings. In: Proceedings of the SIGCHI
  conference on Human factors in computing systems. pp. 557--566 (2007).
  \doi{https://doi.org/10.1145/1240624.1240714}

\bibitem{corbi1989program}
Corbi, T.A.: Program understanding: Challenge for the 1990s. IBM Systems
  Journal  \textbf{28}(2),  294--306 (1989).
  \doi{https://doi.org/10.1147/sj.282.0294}

\bibitem{fekete2022building}
Fekete, A., Gera, Z., Porkol{\'a}b, Z.: Building bridges between academy and
  industry: Improving crucial development skills through participating in open
  source projects. In: INTED2022 Proceedings. pp. 1716--1722. IATED (2022)

\bibitem{fekete2020comprehensive}
Fekete, A., Porkol{\'a}b, Z.: A comprehensive review on software comprehension
  models. In: Annales Mathematicae et Informaticae. vol.~51, pp. 103--111.
  L{\'\i}ceum University Press (2020)

\bibitem{fekete2022report}
Fekete, A., Porkol{\'a}b, Z.: {Report on a Field Experiment of the
  Comprehension Strategies of Computer Science MSc Students}. In: 2022 IEEE
  16th International Scientiﬁc Conference on Informatics - Proceedings. pp.
  73--81. IEEE (2022)

\bibitem{letovsky1987cognitive}
Letovsky, S.: Cognitive processes in program comprehension. Journal of Systems
  and software  \textbf{7}(4),  325--339 (1987).
  \doi{https://doi.org/10.1016/0164-1212(87)90032-X}

\bibitem{levy2019understanding}
Levy, O., Feitelson, D.G.: Understanding large-scale software: a hierarchical
  view. In: Proceedings of the 27th International Conference on Program
  Comprehension. pp. 283--293. IEEE Press (2019).
  \doi{https://doi.org/10.1109/ICPC.2019.00047}

\bibitem{minelli2015know}
Minelli, R., Mocci, A., Lanza, M.: I know what you did last summer-an
  investigation of how developers spend their time. In: 2015 IEEE 23rd
  International Conference on Program Comprehension. pp. 25--35. IEEE (2015)

\bibitem{pennington1987stimulus}
Pennington, N.: Stimulus structures and mental representations in expert
  comprehension of computer programs. Cognitive psychology  \textbf{19}(3),
  295--341 (1987). \doi{https://doi.org/10.1016/0010-0285(87)90007-7}

\bibitem{porkolab2018codecompass}
Porkol{\'a}b, Z., Brunner, T., Krupp, D., Csord{\'a}s, M.: Codecompass: an open
  software comprehension framework for industrial usage. In: Proceedings of the
  26th Conference on Program Comprehension. pp. 361--369 (2018).
  \doi{https://doi.org/10.1145/3196321.3197546}

\bibitem{shneiderman1979syntactic}
Shneiderman, B., Mayer, R.: Syntactic/semantic interactions in programmer
  behavior: A model and experimental results. International Journal of Computer
  \& Information Sciences  \textbf{8}(3),  219--238 (1979).
  \doi{https://doi.org/10.1007/BF00977789}

\bibitem{siegmund2016program}
Siegmund, J.: Program comprehension: Past, present, and future. In: 2016 IEEE
  23rd International Conference on Software Analysis, Evolution, and
  Reengineering (SANER). vol.~5, pp. 13--20. IEEE (2016)

\bibitem{soloway1984empirical}
Soloway, E., Ehrlich, K.: Empirical studies of programming knowledge. IEEE
  Transactions on software engineering (5),  595--609 (1984).
  \doi{https://doi.org/10.1109/TSE.1984.5010283}

\bibitem{storey2005theories}
Storey, M.A.: Theories, methods and tools in program comprehension: Past,
  present and future. In: 13th International Workshop on Program Comprehension
  (IWPC'05). pp. 181--191. IEEE (2005).
  \doi{https://doi.org/10.1109/WPC.2005.38}

\bibitem{von1995program}
Von~Mayrhauser, A., Vans, A.M.: Program comprehension during software
  maintenance and evolution. Computer  \textbf{28}(8),  44--55 (1995).
  \doi{https://doi.org/10.1109/2.402076}

\bibitem{xia2017measuring}
Xia, X., Bao, L., Lo, D., Xing, Z., Hassan, A.E., Li, S.: Measuring program
  comprehension: A large-scale field study with professionals. IEEE
  Transactions on Software Engineering  \textbf{44}(10),  951--976 (2017)

\end{thebibliography}

\begin{thebibliography}{10}
\providecommand{\url}[1]{\texttt{#1}}
\providecommand{\urlprefix}{URL }
\providecommand{\doi}[1]{https://doi.org/#1}

\bibitem{energy_efficiency_2}
\url{https://www.eea.europa.eu/en/topics/in-depth/energy-efficiency}

\bibitem{bunse2009exploring}
Bunse, C., H{\"o}pfner, H., Mansour, E., Roychoudhury, S.: {Exploring the
  energy consumption of data sorting algorithms in embedded and mobile
  environments}. In: Mobile Data Management: Systems, Services and Middleware,
  2009. MDM'09. Tenth International Conference on. pp. 600--607. IEEE (2009)

\bibitem{pl:rank2017}
Couto, M., Pereira, R., Ribeiro, F., Rua, R., Saraiva, J.: {Towards a Green
  Ranking for Programming Languages}. In: Proceedings of the 21st Brazilian
  Symposium on Programming Languages. pp. 7:1--7:8. SBLP 2017, ACM (2017).
  \doi{10.1145/3125374.3125382},
  \url{http://doi.acm.org/10.1145/3125374.3125382}, best Paper

\bibitem{DBLP:conf/wcre/0001SF20}
Couto, M., Saraiva, J., Fernandes, J.P.: {Energy Refactorings for Android in
  the Large and in the Wild}. In: Kontogiannis, K., Khomh, F., Chatzigeorgiou,
  A., Fokaefs, M., Zhou, M. (eds.) 27th {IEEE} International Conference on
  Software Analysis, Evolution and Reengineering, {SANER} 2020, London, ON,
  Canada, February 18-21, 2020. pp. 217--228. {IEEE} (2020).
  \doi{10.1109/SANER48275.2020.9054858},
  \url{https://doi.org/10.1109/SANER48275.2020.9054858}

\bibitem{tec:ptop}
Do, T., Rawshdeh, S., Shi, W.: ptop: A process-level power profiling tool (Jan
  2009)

\bibitem{diagram-eder}
Eder, K.: {Whole Systems Energy Transparency: More Power to Software
  Developers!} Electronic Proceedings in Theoretical Computer Science
  \textbf{248}, ~2--3 (2017). \doi{10.4204/eptcs.248.2}, lecture slides

\bibitem{DBLP:books/sp/21/Feitosa00F0S21}
Feitosa, D., Cruz, L., Abreu, R., Fernandes, J.P., Couto, M., Saraiva, J.:
  {Patterns and Energy Consumption: Design, Implementation, Studies, and
  Stories}. In: Calero, C., Moraga, M.{\'{A}}., Piattini, M. (eds.) Software
  Sustainability, pp. 89--121. Springer (2021).
  \doi{10.1007/978-3-030-69970-3\_5},
  \url{https://doi.org/10.1007/978-3-030-69970-3\_5}

\bibitem{tec:rapl2}
H\"{a}hnel, M., D\"{o}bel, B., V\"{o}lp, M., H\"{a}rtig, H.: {Measuring Energy
  Consumption for Short Code Paths Using RAPL}. SIGMETRICS Perform. Eval. Rev.
  \textbf{40}(3),  13–17 (Jan 2012)

\bibitem{hasan2016energy}
Hasan, S., King, Z., Hafiz, M., Sayagh, M., Adams, B., Hindle, A.: Energy
  profiles of java collections classes. In: Proceedings of the 38th
  International Conference on Software Engineering. pp. 225--236. ACM (2016)

\bibitem{DBLP:journals/ese/Hindle15}
Hindle, A.: Green mining: a methodology of relating software change and
  configuration to power consumption. Empirical Software Engineering
  \textbf{20}(2),  374--409 (2015)

\bibitem{rapl_in_action}
Khan, K.N., Hirki, M., Niemi, T., Nurminen, J.K., Ou, Z.: Rapl in action:
  Experiences in using rapl for power measurements. ACM Transactions on
  Modeling and Performance Evaluation of Computing Systems  \textbf{3}(2),
  1–26 (Jun 2018). \doi{10.1145/3177754}

\bibitem{lago15}
Lago, P., Ko{\c{c}}ak, S.A., Crnkovic, I., Penzenstadler, B.: Framing
  sustainability as a property of software quality. Commun. {ACM}
  \textbf{58}(10),  70--78 (2015). \doi{10.1145/2714560},
  \url{http://doi.acm.org/10.1145/2714560}

\bibitem{cloud_survey}
Li, Z., Tesfatsion, S., Bastani, S., Ali-Eldin, A., Elmroth, E., Kihl, M.,
  Ranjan, R.: A survey on modeling energy consumption of cloud applications:
  Deconstruction, state of the art, and trade-off debates. IEEE Transactions on
  Sustainable Computing  \textbf{2}(3),  255--274 (2017).
  \doi{10.1109/TSUSC.2017.2722822}

\bibitem{ds:haskell}
Melfe, G., Fonseca, A., Fernandes, J.P.: {Helping Developers Write Energy
  Efficient Haskell through a Data-Structure Evaluation}. In: Proceedings of
  the 6th International Workshop on Green and Sustainable Software. p. 9–15.
  GREENS '18, ACM (2018)

\bibitem{pang2016programmers}
Pang, C., Hindle, A., Adams, B., Hassan, A.E.: What do programmers know about
  software energy consumption? IEEE Software  \textbf{33}(3),  83--89 (2016)

\bibitem{park2014investigation}
Park, J.J., Hong, J.E., Lee, S.H.: Investigation for software power consumption
  of code refactoring techniques. In: SEKE. pp. 717--722 (2014)

\bibitem{pereira2017sle}
Pereira, R., Couto, M., Ribeiro, F., Rua, R., Cunha, J., Fernandes, J.P.,
  Saraiva, J.: {Energy Efficiency Across Programming Languages: How Do Energy,
  Time, and Memory Relate?} In: Proceedings of the 10th ACM SIGPLAN
  International Conference on Software Language Engineering. pp. 256--267. SLE
  2017, ACM, New York, NY, USA (2017). \doi{10.1145/3136014.3136031},
  \url{http://doi.acm.org/10.1145/3136014.3136031}

\bibitem{DBLP:journals/scp/PereiraCRRCFS21}
Pereira, R., Couto, M., Ribeiro, F., Rua, R., Cunha, J., Fernandes, J.P.,
  Saraiva, J.: Ranking programming languages by energy efficiency. Sci. Comput.
  Program.  \textbf{205},  102609 (2021). \doi{10.1016/j.scico.2021.102609},
  \url{https://doi.org/10.1016/j.scico.2021.102609}

\bibitem{Pereira:2016:IJC:2896967.2896968}
Pereira, R., Couto, M., Saraiva, J., Cunha, J., Fernandes, J.P.: {The Influence
  of the Java Collection Framework on Overall Energy Consumption}. In:
  Proceedings of the 5th International Workshop on Green and Sustainable
  Software. pp. 15--21. GREENS '16, ACM (2016)

\bibitem{pinto2014mining}
Pinto, G., Castor, F., Liu, Y.D.: Mining questions about software energy
  consumption. In: Proceedings of the 11th Working Conference on Mining
  Software Repositories. pp. 22--31. ACM (2014)

\bibitem{tec:rapl1}
Rotem, E., Naveh, A., Ananthakrishnan, A., Weissmann, E., Rajwan, D.:
  {Power-Management Architecture of the Intel Microarchitecture Code-Named
  Sandy Bridge}. IEEE Micro  \textbf{32}(2),  20--27 (2012)

\end{thebibliography}

\begin{thebibliography}{99}

\bibitem{Autonomicbook}
N. Agoulmine (ed.), \textit{Autonomic Network Management Principles: from Concepts to Applications}, Elsevier Inc. (2011). \doi{10.1016/C2009-0-62958-5}.

\bibitem{Webinserviceperf}
Y. Amannejad, D. Krishnamurthy, B. Far, Managing performance interference in Cloud-based web services,
\textit{IEEE Transactions on Network and Service Management} 12 (2015), no. 3,
320-333. \doi{10.1109/TNSM.2015.2456172}.

\bibitem{LagoSurvey}
N. Condori-Fernandez, P. Lago,
Characterizing the contribution of quality requirements to software sustainability,
\textit{Journal of Systems and Software} 137 (2018), 289-305.

\bibitem{TDA}
H. Edelsbrunner, J. L. Harer,
\textit{Computational Topology: An Introduction},
American Mathematical Society, Providence RI, 2010.

\bibitem{systdynamics}
J. W. Forrester, System dynamics, systems thinking, and soft OR,
\textit{Syst. Dyn. Rev.} 10 (1994), 245-256. \doi{10.1002/sdr.4260100211}.

\bibitem{GalinacRole1}
T. Galinac Grbac, The Role of Functional Programming in Management and Orchestration of Virtualized Network Resources Part I. System structure for Complex Systems and Design Principles. CoRR abs/2107.12136 (2021)

\bibitem{GalinacRole2}
T. Galinac Grbac, N. Domazet, The Role of Functional Programming in Management and Orchestration of Virtualized Network Resources Part II. Network Evolution and Design Principles. CoRR abs/2107.12227 (2021)

\bibitem{ICT4S2}
L. M. Hilty, B. Aebischer,
ICT for sustainability: an emerging research field,
In: L. M. Hilty, B. Aebischer (eds.),
\textit{ICT Innovations for Sustainability}, pp. 3-36,
\textit{Advances in Intelligent Systems and Computing}
vol. 310,
Springer, Cham, 2015.

\bibitem{ICT4S1}
L. M. Hilty, P. Arnfalk, L. Erdmann, J. Goodman, M. Lehmann, P. A. W\"{a}ger,
The relevance of information and communication technologies for environmental sustainability: A prospective simulation study,
\textit{Environmental Modelling \& Software} 21 (2006), 1618–1629.

\bibitem{IBMMAPE}
IBM Corporation,
An architectural blueprint for autonomic computing,
Technical Report, 2005.

\bibitem{PerfomanceReact}
A. Javeed, Performance optimization techniques for ReactJS, in: 2019 IEEE International Conference on Electrical, Computer and Communication Technologies (ICECCT), Coimbatore, India, 2019, pp. 1-5. \doi{10.1109/ICECCT.2019.8869134}.

\bibitem{FramingSustainability}
P. Lago, S. A. Ko\c{c}ak, I. Crnkovi\'{c}, B. Penzenstadler,
Framing sustainability as a property of software quality,
\textit{Commun. ACM} 58 (2015), 70–78.

\bibitem{LagoGreen}
P. Lago, N. Meyer, M. Morisio, H. A. M\"{u}ller, G. Scanniello,
Leveraging ``energy efficiency to software users'': summary of the second GREENS workshop,
\textit{SIGSOFT Softw. Eng. Notes}
39 (2014), no. 1, 36–38.

\bibitem{Petric}
J. Petri\'{c}, T. Galinac Grbac,
Software structure evolution and relation to system defectiveness,
In: EASE 2014, pp. 34:1-34:10.

\bibitem{Design4finSus}
J. Thomas, P. Mantri, Design for financial sustainability,
\textit{Patterns} 3 (2022), no. 9, Art. no. 100585 (32 pages).

\bibitem{SoftwareGraph}
S. Valverde, R. Sole,
Network motifs in computational graphs: A case study in software architecture,
\textit{Physical review. E, Statistical, nonlinear, and soft matter physics} 72 (2005), Art. no. 026107 (8 pages).

\bibitem{Vliet}
H. van Vliet,
\textit{Software Engineering: Principles and Practice},
John Wiley \& Sons, 2008.

\bibitem{PerformanceWeb}
Y. Yao and J. Xia, Analysis and research on the performance optimization of Web application system in high concurrency environment, in: 2016 IEEE Information Technology, Networking, Electronic and Automation Control Conference, Chongqing, China, 2016, pp. 321-326. \doi{10.1109/ITNEC.2016.7560374}.

\bibitem{Ericsson5G}
Dohler, M., ; Nakamura, T. (2016). 5G Mobile and Wireless Communications Technology (A. Osseiran, J. Monserrat, P. Marsch, Eds.). Cambridge: Cambridge University Press.

\bibitem{DigitSust}
Mondejar, Maria; Avtar, Ram; Baños Diaz, Heyker ;Dubey, Rama ; Esteban, Jesús ; Gómez-Morales, Abigail, Hallam, Brett, Mbungu, Nsilulu; Okolo, Chukwuebuka ; Kumar, Arun ; She, Qianhong ; Garcia-Segura, Sergi. (2021). Digitalization to achieve sustainable development goals: Steps towards a Smart Green Planet. Science of The Total Environment. 794. 148539. 10.1016/j.scitotenv.2021.148539.

\bibitem{AInetwork}
B. Mao, F. Tang, Y. Kawamoto and N. Kato, "AI Models for Green Communications Towards 6G," in IEEE Communications Surveys and Tutorials, vol. 24, no. 1, pp. 210-247, Firstquarter 2022

\bibitem{AICloudRM}
Shreshth Tuli, Sukhpal Singh Gill, Minxian Xu, Peter Garraghan, Rami Bahsoon, Schahram Dustdar, Rizos Sakellariou, Omer Rana, Rajkumar Buyya, Giuliano Casale, Nicholas R. Jennings, HUNTER: AI based holistic resource management for sustainable cloud computing, Journal of Systems and Software, Volume 184, 2022, 111124, ISSN 0164-1212.

\bibitem{LivingSyst}
James G. Miller, I: The nature of living systems, Biosystems, Volume 4, Issue 2, 1972, Pages 55-77, ISSN 0303-2647.

\bibitem{Loc2glob}
Carmen Leong, Isam Faik, Felix T.C. Tan, Barney Tan, Ying Hooi Khoo,
Digital organizing of a global social movement: From connective to collective action, Information and Organization, Volume 30, Issue 4, 2020, 100324, ISSN 1471-7727.

\bibitem{RoleofFluid}
Hung-Hsiang Wang, Xiaotian Deng,
The role of fluid intelligence in creativity: The case of bio-inspired design,
Thinking Skills and Creativity, Volume 45, 2022, 101059, ISSN 1871-1871.

\end{thebibliography}

\begin{thebibliography}{10}
\providecommand{\url}[1]{\texttt{#1}}
\providecommand{\urlprefix}{URL }
\providecommand{\doi}[1]{https://doi.org/#1}

\bibitem{Clean:language}
{Clean team}: Clean 3.0 language report (2020), \url{https://cloogle.org/doc},
  [Online; accessed 05-April-2022]

\bibitem{TFP22}
Crooijmans, S., Lubbers, M., Koopman, P.: Reducing the power consumption
  of iot with task-oriented programming. In: Swierstra, W., Wu, N. (eds.)
  Trends in Functional Programming. pp. 80--99. Springer, Cham (2022)

\bibitem{protobuf}
{Google}: Protocol buffers documentation (2023), \url{https://protobuf.dev/},
  [Online; accessed 13-February-2023]

\bibitem{smartSensors}
Hentschel, K., Jacob, D., Singer, J., Chalmers, M.: Supersensors: Raspberry pi
  devices for smart campus infrastructure. In: 2016 IEEE 4th International
  Conference on Future Internet of Things and Cloud (FiCloud). pp. 58--62
  (2016). \doi{10.1109/FiCloud.2016.16}

\bibitem{hudak_modular_1998}
Hudak, P.: Modular domain specific languages and tools. In: Proceedings.
  {Fifth} {International} {Conference} on {Software} {Reuse} ({Cat}.
  {No}.{98TB100203}). pp. 134--142 (1998). \doi{10.1109/ICSR.1998.685738}

\bibitem{ireland_classification_2009}
Ireland, C., Bowers, D., Newton, M., Waugh, K.: A {Classification} of
  {Object}-{Relational} {Impedance} {Mismatch}. In: First {International}
  {Conference} on {Advances} in {Databases}, {Knowledge}, and {Data}
  {Applications}. pp. 36--43. IEEE, Cancun, Mexico (2009).
  \doi{10.1109/DBKDA.2009.11}

\bibitem{lubbers_writing_2019}
Lubbers, M., Koopman, P., Plasmeijer, R.: Writing {Internet} of {Things}
  applications with {Task} {Oriented} {Programming}. In: Central {European}
  {Functional} {Programming} {School}: 8th {Summer} {School}, {CEFP} 2019,
  {Budapest}, {Hungary}, {July} 17–21, 2019, {Revised} {Selected} {Papers},
  p.~51. Springer, Budapest, Hungary (in-press)

\bibitem{lubbers20tiered}
Lubbers, M., Koopman, P., Ramsingh, A., Singer, J., Trinder, P.: Tiered versus
  tierless iot stacks: Comparing smart campus software architectures. In:
  Proceedings of the 10th International Conference on the Internet of Things.
  IoT '20, ACM, New York, NY, USA (2020). \doi{10.1145/3410992.3411002}

\bibitem{LubbersTIOT}
Lubbers, M., Koopman, P., Ramsingh, A., Singer, J., Trinder, P.: Could tierless
  languages reduce iot development grief? ACM Trans. Internet Things  (nov
  2022). \doi{10.1145/3572901}, \url{https://doi.org/10.1145/3572901}, just
  Accepted

\bibitem{mqtt}
{OASIS MQTT Technical Committee}: Mqtt: The standard for iot messaging (2023),
  \url{https://mqtt.org/}, [Online; accessed 13-February-2023]

\bibitem{plasmeijer_task-oriented_2012}
Plasmeijer, R., Lijnse, B., Michels, S., Achten, P., Koopman, P.: Task-oriented
  programming in a pure functional language. In: Proceedings of the 14th
  Symposium on Principles and Practice of Declarative Programming. p.
  195–206. PPDP '12, ACM, New York, NY, USA (2012).
  \doi{10.1145/2370776.2370801}

\end{thebibliography}

\begin{thebibliography}{10}
\providecommand{\url}[1]{\texttt{#1}}
\providecommand{\urlprefix}{URL }
\providecommand{\doi}[1]{https://doi.org/#1}

\bibitem{Conceptual}
Ardito, L., Procaccianti, G., Torchiano, M., Vetrò, A.: Understanding green
  software development: A conceptual framework. IT Professional
  \textbf{17}(1),  44--50 (2015). \doi{10.1109/MITP.2015.16}

\bibitem{en10101470}
Avgerinou, M., Bertoldi, P., Castellazzi, L.: Trends in data centre energy
  consumption under the european code of conduct for data centre energy
  efficiency. Energies  \textbf{10}(10) (2017). \doi{10.3390/en10101470},
  \url{https://www.mdpi.com/1996-1073/10/10/1470}

\bibitem{goBook}
Donovan, A.A., Kernighan, B.W.: The Go Programming Language. Addison-Wesley
  Professional, 1st edn. (2015)

\bibitem{Gallagher17}
Eder, K., Gallagher, J.P.: Energy-aware software engineering. In: Fagas, G.,
  Gammaitoni, L., Gallagher, J.P., Paul, D.J. (eds.) ICT - Energy Concepts for
  Energy Efficiency and Sustainability, chap.~5. IntechOpen, Rijeka (2017).
  \doi{10.5772/65985}, \url{https://doi.org/10.5772/65985}

\bibitem{cpuu}
Fan, X., Weber, W.D., Barroso, L.A.: Power provisioning for a warehouse-sized
  computer. SIGARCH Comput. Archit. News  \textbf{35}(2),  13–23 (jun 2007).
  \doi{10.1145/1273440.1250665}, \url{https://doi.org/10.1145/1273440.1250665}

\bibitem{EnterArch}
Fowler, M.: Patterns of Enterprise Application Architecture. Addison-Wesley
  Longman Publishing Co., Inc., USA (2002)

\bibitem{refactorBook}
Fowler, M., Beck, K., Brant, J., Opdyke, W., don Roberts: Refactoring:
  Improving the Design of Existing Code. Addison-Wesley Longman Publishing Co.,
  Inc., USA (1999)

\bibitem{url_Erlang}
Index - erlang/otp. \url{https://www.erlang.org/}, accessed: 2023-03-10

\bibitem{joulemeterR}
Jagroep, E., Procaccianti, G., van~der Werf, J.M., Brinkkemper, S., Blom, L.,
  van Vliet, R.: Energy efficiency on the product roadmap: An empirical study
  across releases of a software product. Journal of Software: Evolution and
  Process  \textbf{29}(2),  e1852 (2017).
  \doi{https://doi.org/10.1002/smr.1852},
  \url{https://onlinelibrary.wiley.com/doi/abs/10.1002/smr.1852}, e1852
  smr.1852

\bibitem{url_Java}
Java | oracle. \url{https://www.java.com/en/}, accessed: 2023-03-10

\bibitem{rapl1}
Khan, K.N., Hirki, M., Niemi, T., Nurminen, J.K., Ou, Z.: Rapl in action:
  Experiences in using rapl for power measurements. ACM Trans. Model. Perform.
  Eval. Comput. Syst.  \textbf{3}(2) (mar 2018). \doi{10.1145/3177754},
  \url{https://doi.org/10.1145/3177754}

\bibitem{javaCode1}
Kumar, M., Li, Y., Shi, W.: Energy consumption in java: An early experience.
  In: 2017 Eighth International Green and Sustainable Computing Conference
  (IGSC). pp.~1--8 (2017). \doi{10.1109/IGCC.2017.8323579}

\bibitem{tla}
Lamport, L.: Specifying Systems: The TLA+ Language and Tools for Hardware and
  Software Engineers. Addison-Wesley Longman Publishing Co., Inc., USA (2002)

\bibitem{EnergyWar}
de~Macedo, J.a., Alo\'{\i}sio, J.a., Gon\c{c}alves, N., Pereira, R., Saraiva,
  J.a.: Energy wars - chrome vs. firefox: Which browser is more energy
  efficient? In: Proceedings of the 35th IEEE/ACM International Conference on
  Automated Software Engineering. p. 159–165. ASE '20, ACM, New York, NY, USA
  (2021). \doi{10.1145/3417113.3423000},
  \url{https://doi.org/10.1145/3417113.3423000}

\bibitem{phdthesis}
Ournani, Z.: {Software eco-design : investigating and reducing the energy
  consumption of software}. Theses, {Universit{\'e} de Lille} (Nov 2021),
  \url{https://theses.hal.science/tel-03554712}

\bibitem{refactor1}
Park, J.J., Hong, J., Lee, S.: Investigation for software power consumption of
  code refactoring techniques. In: Reformat, M.Z. (ed.) The 26th International
  Conference on Software Engineering and Knowledge Engineering, Hyatt Regency,
  Vancouver, BC, Canada, July 1-3, 2013. pp. 717--722. Knowledge Systems
  Institute Graduate School (2014)

\bibitem{PEREIRA2021102609}
Pereira, R., Couto, M., Ribeiro, F., Rua, R., Cunha, J., Fernandes, J.P.,
  Saraiva, J.: Ranking programming languages by energy efficiency. Science of
  Computer Programming  \textbf{205},  102609 (2021).
  \doi{https://doi.org/10.1016/j.scico.2021.102609},
  \url{https://www.sciencedirect.com/science/article/pii/S0167642321000022}

\bibitem{GolangPerf}
Performance golang. \url{https://github.com/golang/go/wiki/Performance},
  accessed: 2023-03-10

\bibitem{coder1}
Sahin, C., Pollock, L., Clause, J.: How do code refactorings affect energy
  usage? In: Proceedings of the 8th ACM/IEEE International Symposium on
  Empirical Software Engineering and Measurement. ESEM '14, Association for
  Computing Machinery, New York, NY, USA (2014). \doi{10.1145/2652524.2652538},
  \url{https://doi.org/10.1145/2652524.2652538}

\bibitem{GreenCurriculum}
Saraiva, J.a., Zong, Z., Pereira, R.: Bringing green software to computer
  science curriculum: Perspectives from researchers and educators. In:
  Proceedings of the 26th ACM Conference on Innovation and Technology in
  Computer Science Education V. 1. p. 498–504. ITiCSE '21, Association for
  Computing Machinery, New York, NY, USA (2021). \doi{10.1145/3430665.3456386},
  \url{https://doi.org/10.1145/3430665.3456386}

\bibitem{GreenRefactorJoulemeterJ}
Sehgal, R., Mehrotra, D., Nagpal, R., Sharma, R.: Green software: Refactoring
  approach. Journal of King Saud University - Computer and Information Sciences
   \textbf{34}(7),  4635--4643 (2022).
  \doi{https://doi.org/10.1016/j.jksuci.2020.10.022},
  \url{https://www.sciencedirect.com/science/article/pii/S1319157820305164}

\bibitem{coder2}
Sehgal, R., Mehrotra, D., Nagpal, R., Sharma, R.: Green software: Refactoring
  approach. Journal of King Saud University - Computer and Information Sciences
   \textbf{34}(7),  4635--4643 (2022).
  \doi{https://doi.org/10.1016/j.jksuci.2020.10.022},
  \url{https://www.sciencedirect.com/science/article/pii/S1319157820305164}

\bibitem{SummerSchool22}
Sustrainable summer school 2022. \url{https://sustrainable.uniri.hr/},
  accessed: 2023-06-12

\bibitem{url_Golang}
The \textsc{Go} programming language. \url{https://go.dev/}, accessed:
  2023-03-10

\end{thebibliography}

\begin{thebibliography}{10}
\providecommand{\url}[1]{\texttt{#1}}
\providecommand{\urlprefix}{URL }
\providecommand{\doi}[1]{https://doi.org/#1}

\bibitem{borah2022applied}
Borah, S., Panigrahi, R.: Applied Soft Computing: Techniques and Applications.
  Apple Academic Press (2022)

\bibitem{eiben2015introduction}
Eiben, A.E., Smith, J.E.: Introduction to evolutionary computing. Springer
  (2015)

\bibitem{RAPL2019}
Fahad, M., Shahid, A., Manumachu, R.R., Lastovetsky, A.: A comparative study of
  methods for measurement of energy of computing. Energies  \textbf{12}(11),
  ~2204 (2019)

\bibitem{fister2013brief}
Fister~Jr, I., Yang, X.S., Fister, I., Brest, J., Fister, D.: A brief review of
  nature-inspired algorithms for optimization. arXiv preprint arXiv:1307.4186
  (2013)

\bibitem{fogel1998evolutionary}
Fogel~David, B.: Evolutionary computation: The fossil record (1998)

\bibitem{geem2009music}
Geem, Z.W.: Music-inspired harmony search algorithm: theory and applications,
  vol.~191. Springer (2009)

\bibitem{goldberg2002design}
Goldberg, D.E.: The design of innovation: Lessons from and for competent
  genetic algorithms, vol.~1. Springer (2002)

\bibitem{holland1973genetic}
Holland, J.H.: Genetic algorithms and the optimal allocation of trials. SIAM
  journal on computing  \textbf{2}(2),  88--105 (1973)

\bibitem{kang2019dramp}
Kang, X., Dong, F., Shi, C., Liu, S., Sun, J., Chen, J., Li, H., Xu, H., Lao,
  X., Zheng, H.: Dramp 2.0, an updated data repository of antimicrobial
  peptides. Scientific data  \textbf{6}(1), ~148 (2019)

\bibitem{khan2018rapl}
Khan, K.N., Hirki, M., Niemi, T., Nurminen, J.K., Ou, Z.: Rapl in action:
  Experiences in using rapl for power measurements. ACM Transactions on
  Modeling and Performance Evaluation of Computing Systems (TOMPECS)
  \textbf{3}(2),  1--26 (2018)

\bibitem{rosenbrock1960automatic}
Rosenbrock, H.: An automatic method for finding the greatest or least value of
  a function. The computer journal  \textbf{3}(3),  175--184 (1960)

\bibitem{sinha2005designing}
Sinha, A., Chen, Y.p., Goldberg, D.E.: Designing efficient genetic and
  evolutionary algorithm hybrids. Springer (2005)

\bibitem{smith1995fitness}
Smith, R.E., Dike, B.A., Stegmann, S.: Fitness inheritance in genetic
  algorithms. In: Proceedings of the 1995 ACM symposium on Applied computing.
  pp. 345--350 (1995)

\bibitem{srivastava2002time}
Srivastava, R.P.: Time continuation in genetic algorithms. Ph.D. thesis,
  University of Illinois at Urbana-Champaign (2002)

\end{thebibliography}

\begin{thebibliography}{10}

\bibitem{RoedRouxAltm+20}
Roeder, J., Rouxel, B., Altmeyer, S., Grelck, C.:
\newblock Towards energy-, time- and security-aware multi-core coordination.
\newblock In Bliudze, S., Bocchi, L., eds.: 22nd International Conference on
  Coordination Models and Languages (COORDINATION 2020), Malta, LNCS 12134,
  Springer (2020)  57--74

\bibitem{Arbab98}
Arbab, F.:
\newblock What do you mean, coordination?
\newblock Bulletin of the Dutch Association for Theoretical Computer Science
  (NVTI) (1998)  11--22

\bibitem{Arbab04}
Arbab, F.:
\newblock Reo: a channel-based coordination model for component composition.
\newblock Mathematical Structures in Computer Science \textbf{14} (2004)
  329--366

\bibitem{RoedRouxAltm+20b}
Roeder, J., Rouxel, B., Altmeyer, S., Grelck, C.:
\newblock Energy-aware scheduling of multi-version tasks on heterogeneous
  real-time systems.
\newblock In: 36th ACM/SIGAPP Symposium on Applied Computing (SAC 2021), ACM
  (2020)  500--510

\bibitem{RoedRouxGrel21}
Roeder, J., Rouxel, B., Grelck, C.:
\newblock Scheduling {DAGs} of multi-version multi-phase tasks on heterogeneous
  real-time systems.
\newblock In: 14th IEEE International Symposium on Embedded Multicore/Many-core
  Systems-on-Chip (MCSoC 2021), Singapore, IEEE (2021)

\bibitem{RoedPimeGrelAIAI23}
Roeder, J., Pimentel, A., Grelck, C.:
\newblock {GCN}-based reinforcement learning approach for scheduling {DAG}
  applications.
\newblock In: Artificial Intelligence Applications and Innovations, 19th IFIP
  WG 12.5 International Conference, AIAI 2023, Le\'on, Spain. Volume 676 of
  IFIPAICT., Springer (2023)  121--134

\bibitem{RouxAltmGrel21}
Rouxel, B., Altmeyer, S., Grelck, C.:
\newblock {YASMIN}: a real-time middleware for {COTS} heterogeneous platforms.
\newblock In: 22nd ACM/IFIP International Middleware Conference (MIDDLEWARE
  2021), ACM (2021)

\bibitem{WegeNikoNune+23}
Wegener, S., Nikov, K., Nunez-Yanez, J., Eder, K.:
\newblock {EnergyAnalyzer}: Using static {WCET} analysis techniques to estimate
  the energy consumption of embedded applications.
\newblock In W\"agemann, P., ed.: 21st International Workshop on Worst-Case
  Execution Time Analysis (WCET 2023), Vienna, Austria. (2023)

\bibitem{Seewald2019coarse}
Seewald, A., Schultz, U.P., Ebeid, E., Midtiby, H.S.:
\newblock Coarse-grained computation-oriented energy modeling for heterogeneous
  parallel embedded systems.
\newblock International Journal of Parallel Programming \textbf{49} (2021)
  136--157

\bibitem{seewald2019component}
Seewald, A., Schultz, U.P., Roeder, J., Rouxel, B., Grelck, C.:
\newblock Component-based computation-energy modeling for embedded systems.
\newblock In: Proceedings Companion of the 2019 ACM SIGPLAN International
  Conference on Systems, Programming, Languages, and Applications: Software for
  Humanity, ACM (2019)  5--6

\bibitem{CML+18}
Ciatto, G., Mariani, S., Louvel, M., Omicini, A., Zambonelli, F.:
\newblock Twenty years of coordination technologies: State-of-the-art and
  perspectives.
\newblock In: International Conference on Coordination Languages and Models
  (COORDINATION 2018), LNCS 10852, Springer (2018)  51--80

\bibitem{GSS10}
Grelck, C., Scholz, S.B., Shafarenko, A.:
\newblock {Asynchronous Stream Processing with S-Net}.
\newblock International Journal of Parallel Programming \textbf{38} (2010)
  38--67

\bibitem{GrelJulkPencCCGRID12}
Grelck, C., Julku, J., Penczek, F.:
\newblock Distributed {S-Net}: Cluster and grid computing without the hassle.
\newblock In: Cluster, Cloud and Grid Computing (CCGrid'12), 12th IEEE/ACM
  International Conference, Ottawa, Canada, IEEE (2012)

\bibitem{HaM03}
Hammond, K., Michaelson, G.:
\newblock Hume: A domain-specific language for real-time embedded systems.
\newblock In: Generative Programming and Component Engineering, 2nd
  International Conference, GPCE 2003, Erfurt, Germany, LNCS 2830, Springer
  (2003)  37--56

\bibitem{Pradhan15}
Pradhan, S.M., Dubey, A., Gokhale, A., Lehofer, M.:
\newblock {CHARIOT}: A domain-specific language for extensible cyber-physical
  systems.
\newblock In: 15th Workshop on Domain-Specific Modeling (DSM'15), Pittsburgh,
  USA, ACM (2015)  9–16

\bibitem{Berger14}
Berger, C.:
\newblock {SenseDSL}: Automating the integration of sensors for {MCU}-based
  robots and cyber-physical systems.
\newblock In: 14th Workshop on Domain-Specific Modeling (DSM'14), Portland,
  USA, ACM (2014)  41--46

\bibitem{RouxPaghAkes+20}
Rouxel, B., {Pagh Schultz}, U., Akesson, B., Holst, J., Jorgensen, O., Grelck,
  C.:
\newblock {PReGO}: a generative methodology for satisfying real-time
  requirements on cots-based systems: Definition and experience report.
\newblock In: 19th ACM SIGPLAN International Conference on Generative
  Programming: Concepts and Experiences (GPCE 2020), Chicago, USA, ACM (2020)
  70--83

\bibitem{LoevGrel20}
Loeve, W., Grelck, C.:
\newblock Towards facilitating resilience in cyber-physical systems using
  coordination languages.
\newblock In Constantinou, E., ed.: 13th Seminar on Advanced Techniques and
  Tools for Software Evolution (SATToSE 2020), Virtual from Amsterdam,
  Netherlands. Volume 2754., CEUR Workshop Proceedings (2020)

\end{thebibliography}

\begin{thebibliography}{10}
\providecommand{\url}[1]{\texttt{#1}}
\providecommand{\urlprefix}{URL }
\providecommand{\doi}[1]{https://doi.org/#1}

\bibitem{1BIB12}
Acm code of ethics and professional conduct.
  \url{https://www.acm.org/binaries/content/assets/membership/images2/fac-stu-poster-code.pdf},
  accessed: 2023-06-16

\bibitem{1BIB11}
Bcs code of conduct.
  \url{https://www.bcs.org/media/2211/bcs-code-of-conduct.pdf}, accessed:
  2023-06-16

\bibitem{1BIB5}
European green deal.
  \url{https://ec.europa.eu/info/strategy/priorities-2019-2024/european-green-deal\_en},
  accessed: 2023-06-16

\bibitem{1BIB10}
Ieee code of ethics.
  \url{https://www.ieee.org/about/corporate/governance/p7-8.html}, accessed:
  2023-06-16

\bibitem{1BIB8}
Msci – global fossil fuels exclusion indexes.
  \url{https://www.msci.com/our-solutions/indexes/index-categories/esg-indexes/global-fossil-fuels-exclusion-indexes},
  accessed: 2023-06-16

\bibitem{1BIB4}
United nations, sustainable development. \url{https://sdgs.un.org/}, accessed:
  2023-06-16

\bibitem{1BIB2}
United nations framework convention on climate change (unfccc).
  \url{https://unfccc.int/resource/docs/convkp/conveng.pdf} (1992), accessed:
  2023-06-16

\bibitem{1BIB3}
Paris agreement.
  \url{https://unfccc.int/sites/default/files/resource/parisagreement\_publication.pdf}
  (2015), accessed: 2023-06-16

\bibitem{1BIB16}
Sustainability skills | 2015-2016. research into students’ experiences of
  teaching and learning on sustainable development.
  \url{https://s3-eu-west-1.amazonaws.com/nusdigital/document/documents/28740/22840265895e4d786f0a73345e2eb684/NUS\_Sustainability\_Skills\_2016.pdf}
  (2016), accessed: 2023-06-16

\bibitem{1BIB7}
Eu taxonomy for sustainable activities.
  \url{https://ec.europa.eu/info/business-economy-euro/banking-and-finance/sustainable-finance/eu-taxonomy-sustainable-activities\_en}
  (2020), accessed: 2023-06-16

\bibitem{1BIB6}
Iea, net zero by 2050. a roadmap for the global energy sector, flagship report.
  \url{https://www.iea.org/reports/net-zero-by-2050} (2021), accessed:
  2023-06-16

\bibitem{1BIB9}
Environmental, social, and governance criteria, investopedia.
  \url{https://www.investopedia.com/terms/e/environmental-social-and-governance-esg-criteria.asp}
  (2022), accessed: 2023-06-16

\bibitem{1BIB1}
Brooks, I., Laurell~Thorslund, M., Biørn-Hansen, A., Somova, E.: \#tech4bad:
  when do we say no? ITNOW  \textbf{63},  14--15 (2021)

\bibitem{1BIB13}
Drayson, R.: Student attitudes towards, and skills for, sustainable
  development. executive summary: Students.
  \url{https://www.sustainabilityexchange.ac.uk/files/nus\_report\_-\_student\_attitudes\_towards\_and\_skills\_for\_sustainable\_development\_.pdf}
  (2015), accessed: 2023-06-16

\bibitem{1BIB15}
Drayson, R., Bone, E., Agombar, J., Kemp, S.: Student attitudes towards and
  skills for sustainable development.
  \url{https://www.sustainabilityexchange.ac.uk/files/student\_attitudes\_towards\_and\_skills\_for\_sustainable\_development.pdf}
  (2014), accessed: 2023-06-16

\end{thebibliography}

\begin{thebibliography}{99}
\bibitem{similarweb}
Similarweb: App Intelligence; A $360^{\circ}$ View Of The Digital World, \url{https://www.similarweb.com/corp/research/mobile-app-intelligence/}. Last accessed 18
Jun 2023

\bibitem{mdntesting}
MDN Web Docs: Tools and testing, \url{https://developer.mozilla.org/en-US/docs/Learn/Tools_and_testing}. Last accessed 18
Jun 2023

\bibitem{swevolutionbook}
Software evolution and feedback: Theory and practice, Edited by Nazim H. Madhavji ... [et al.]. John Wiley \& Sons Ltd, The Atrium, Southern Gate, Chichester, West Sussex PO19 8SQ, England (2006). ISBN-13: 978-0-470-87180-5

\bibitem{raplzhang}
Zhang, H, Hoffmann, H.: A Quantitative Evaluation of the RAPL Power Control System (2014)

\bibitem{raplinaction}
Khan, K. N., Hirki, M., Niemi, T., Nurminen, J. K., Ou, Z.: RAPL in Action: Experiences in Using RAPL for Power Measurements. ACM Trans. Model. Perform. Eval. Comput. Syst. 3, 2, Article 9 (June 2018), 26 pages. \url{https://doi.org/10.1145/3177754}

\bibitem{architecture}
Smith, S.: Architecting Modern Web Applications with ASP.NET Core and Azure. Microsoft Developer Division, .NET, and Visual Studio product teams, A division of Microsoft Corporation, One Microsoft Way, Redmond, Washington (2023)

\bibitem{balkishanprofiling}
Sharma, B.: Web Front End Profiling Client Side Performance Testing, \url{https://www.linkedin.com/pulse/web-front-end-profiling-client-side-performance-testing-sharma}, Published 29 Dec 2020

\bibitem{unit}
Noureddine, A., Rouvoy, R., Seinturier, L.: Unit Testing of Energy Consumption of Software Libraries. International Symposium On Applied Computing (SAC), March 2014, Gyeongju, South Korea. pp.1200-1205.

\end{thebibliography}
%

\end{document}